\documentclass[twocolumn,twocolappendix]{aastex63}

\accepted{to ApJ on August 11, 2020}

\newcommand{\hd}{HD~142527}
\newcommand{\im}{IM~Lup}
\newcommand{\pfrac}{$P_{\text{frac}}$}
\newcommand{\ceo}{\mbox{C$^{18}$O(2--1)}}
\newcommand{\ttco}{\mbox{$^{13}$CO(2--1)}}
\newcommand{\coto}{\mbox{CO(2--1)}}
\newcommand{\kms}     {km\,s$^{-1}$}

\shorttitle{Line Polarization toward HD~142527 and IM~Lup}
\shortauthors{Stephens et al.}
\begin{document}

\title{Low Level Carbon Monoxide Line Polarization in two Protoplanetary Disks: HD~142527 and IM~Lup}

\author{Ian W. Stephens}
%\affiliation{Harvard-Smithsonian Center for Astrophysics, 60 Garden Street, Cambridge, MA, USA}
\affiliation{Center for Astrophysics $|$ Harvard \& Smithsonian, 60 Garden Street, Cambridge, MA 02138, USA \href{mailto:ian.stephens@cfa.harvard.edu}{ian.stephens@cfa.harvard.edu}} %; 

\author{Manuel Fern\'andez-L\'opez}
\affiliation{Instituto Argentino de Radioastronom\'\i a (CCT-La Plata, CONICET; CICPBA), C.C. No. 5, 1894, Villa Elisa, Buenos Aires, Argentina}
%\email{manferna@gmail.com}

\author{Zhi-Yun Li}
\affiliation{Astronomy Department, University of Virginia, Charlottesville, VA 22904, USA}

\author{Leslie W. Looney}
\affiliation{Department of Astronomy, University of Illinois, 1002 West Green Street, Urbana, IL 61801, USA}

\author{Richard Teague}
\affiliation{Center for Astrophysics $|$ Harvard \& Smithsonian, 60 Garden Street, Cambridge, MA 02138, USA \href{mailto:ian.stephens@cfa.harvard.edu}{ian.stephens@cfa.harvard.edu}} %; 

%\author{}
%\affiliation{}

\begin{abstract}
Magnetic fields are expected to play an important role in accretion processes for circumstellar disks. Measuring the magnetic field morphology is difficult, especially since polarimetric (sub)millimeter continuum observations may not trace fields in most disks. The Goldreich-Kylafis (GK) effect suggests that line polarization is perpendicular or parallel to the magnetic field direction. We attempt to observe \coto, \ttco, and \ceo\ line polarization toward \hd\ and \im, which are large, bright protoplanetary disks. We use spatial averaging and spectral integration to search for signals in both disks, and detect a potential \coto\ Stokes~$Q$ signal toward both disks. The total \coto\ polarization fractions are 1.57\,$\pm$\,0.18\% and 1.01\,$\pm$\,0.10\% for \hd\ and \im, respectively. Our Monte Carlo simulations indicate that these signals are marginal. We also stack Stokes parameters based on the Keplerian rotation, but no signal was found. Across the disk traced by dust of \hd, the 3$\sigma$ upper limits for \pfrac\ at 0.5$\arcsec$ ($\sim$80\,au) resolution are typically less than 3\% for \coto\ and \ttco\ and 4\% for \ceo. For \im\, the 3$\sigma$ upper limits for these three lines are typically less than 3\%, 4\%, and 12\%, respectively. Upper limits based on our stacking technique are up to a factor of $\sim$10 lower, though stacking areas can potentially average out small-scale polarization structure.  We also compare our continuum polarization at 1.3\,mm to observations at 870\,$\mu$m from previous studies. The polarization in the northern dust trap of \hd\ shows a significant change in morphology and an increase in \pfrac\ as compared to 870\,$\mu$m. For \im, the 1.3\,mm polarization may be more azimuthal and has a higher \pfrac\ than at 870\,$\mu$m.
~\\
%{\bf Assuming our stacking technique with Keplerian motion is valid, upper limits can be constrained by a factor of up to $\sim$10 lower.}
\end{abstract}
%The southern part of the disk shows a decrease in \pfrac. 

%\keywords{Star-formation, Polarimetry}

\section{Introduction} \label{sec:intro}

Accretion from circumstellar disks is widely thought to be controlled by magnetohydrodynamic (MHD) turbulence driven by magnetorotational instability \citep[MRI;][]{Balbus1998} and/or by a magnetically driven disk-wind \citep[e.g.,][]{Konigl2000}. Despite their perceived importance, measuring the field morphology and strength in these disks has been quite difficult. The most common way to measure the magnetic field morphology in the interstellar medium is via (sub)millimeter dust polarization observations. Small grains ($\lesssim$10\,$\mu$m) are expected to align their short axes with the magnetic field direction, which causes thermal emission from these grains to be polarized perpendicular with the magnetic field \citep[e.g.,][]{Andersson2015}. However, the larger the grains, the more likely the will align with their long axes perpendicular to the radiation anisotropy \citep{LazarianHoang2007,Tazaki2017}. Moreover, if grains are comparable to the wavelength, efficient dust self-scattering (i.e., emission from one grain scatters off another grain) can cause a polarization signal. High optical depths, such as those estimated for HL Tau at (sub)millimeter wavelengths \citep{Carrasco2016,Carrasco2019}, will also increase the chance that polarization is from scattering rather than from aligned grains \citep{Yang2017}. 

Given that disks often have larger grains and higher optical depths compared to the interstellar medium, (sub)millimeter polarization observations in disks may not show the magnetic field morphologies.  Indeed, polarimetric observations toward disks at 870\,$\mu$m and 1.3\,mm primarily show a mostly uniform polarization morphology parallel with the minor axis \citep[e.g.,][]{Stephens2014b,FL2016,Stephens2017c,Bacciotti2018,Hull2018,Dent2019,Sadavoy2019}, which is a signature of dust scattering from grains that are $\sim$10--200\,$\mu$m in size \citep[e.g.,][]{Kataoka2015,Yang2016a}. Polarization from scattering at longer wavelengths is expected to be minimal since the largest grains are much smaller than the wavelength. Indeed, at 3\,mm for the disks HL~Tau, DG~Tau, and Haro~6--13, the polarization is azimuthal, and thus is probably not from scattering nor from grains aligned with a (commonly expected) toroidal magnetic field \citep{Kataoka2017,Stephens2017c,Harrison2019}. The polarization may be due to alignment with the radiation anisotropy \citep{Tazaki2017}, but polarization from such mechanism for at least HL~Tau has been seriously questioned \citep{Yang2019}. Other studies also show polarization signatures that indicate polarized emission from aligned dust grains even at 870\,$\mu$m \citep[e.g.,][]{Alves2018,Ohashi2018}, but it is uncertain whether or not these grains are truly aligned with the magnetic field.

Magnetic fields also can be probed via polarimetric line observations, either through the Zeeman effect \citep[e.g.,][]{CrutcherKemball2019} or via the Goldreich-Kylafis effect \citep[][henceforth, GK effect]{Goldreich1981}. The Zeeman effect measures the magnetic field strength along the line of sight and requires observations of circular polarization (Stokes~$V$) of spectral lines. Additionally, the Zeeman effect is strongest only for paramagnetic molecular species (i.e., one with an unpaired electron), such as CN, \ion{H}{1}, or OH. While this effect may have been detected toward the innermost region of an accretion disk of FU Orionis via lines in the optical \citep{Donati2005}, it has yet to be detected in the bulk of the disk, despite recent attempts with the CN line at millimeter wavelengths (\citealt{Vlemmings2019}; R. Harrison submitted). 

The GK effect suggests that linear polarization (Stokes~$Q$ and~$U$) from spectral lines measure the magnetic field morphology in the plane of the sky. This effect predicts that a molecule within a magnetic field has its rotational transition split into magnetic sublevels. Depending on the population imbalance of the $\pi$ or $\sigma$ transitions, polarization from the spectral lines are expected to be either parallel or perpendicular to the magnetic field. The fraction of polarized light, \pfrac, is expected to be maximum when the magnetic field and velocity gradient is perpendicular along the line of sight \citep{Deguchi1984}. Line polarization attributed to the GK effect has been detected in numerous star-forming regions, molecular clouds, and protostellar outflows \citep{Glenn1997b,Greaves1999,Girart1999,Greaves2001,Lai2003,Cortes2005,CortesCrutcher2006,Cortes2006,Cortes2008,Forbrich2008,Beuther2010,LiHenning2011,Vlemmings2012,Houde2013,Hezareh2013,LeeCF2014,Ching2016,LeeCF2018b,Hirota2020}, evolved stars \citep{Glenn1997a,Girart2012,Vlemmings2017,Shinnaga2017,Chamma2018,Huang2020}, and possibly a protoplanetary nebula \citep{Sabin2019}.
 Resonant scattering \citep{Houde2013} of the linearly polarized signal can lead to circular polarization (Stokes~$V$). Detections of resonant scattering has been suggested toward molecular clumps and an evolved star \citep{Houde2013,Hezareh2013,Chamma2018}.

A spatially resolved detection of linear or circular polarized light from a molecular rotational transition in a disk has yet to be found. In this paper, we attempt to detect polarized emission from the $J = 2 \rightarrow 1$ rotational transitions of the molecular lines CO, $^{13}$CO, and C$^{18}$O in the disks \hd\ and \im. We primarily attempt to detect linear polarization (Stokes~$Q$ and~$U$) but we also present Stokes~$V$ results. A polarized signal is not immediately obvious from the Stokes images, so we use stacking and spatial averaging over an integrated velocity range to look for a signal. In Section~\ref{sec:obs} we describe the targets and the observations. In Section~\ref{sec:results} we show the 1.3\,mm polarized continuum images and line observations, and discuss the optical depth of each spectral line. In Section~\ref{sec:search} we present our stacking and how we spatial average an integrated velocity range. In Section~\ref{sec:constrain} we give constraints on \pfrac, and in Section~\ref{sec:summary} we summarize the results.
%maybe mention IQUV all looking for detection

\section{Observations} \label{sec:obs}

\subsection{Targets}
We selected the disks \hd\ and \im\ since they are large in the sky and are known to be bright for \coto, \ttco, and \ceo\ transitions \citep{PerezS2015,Cleeves2016}. These two features are important because we need bright lines over a large solid angle to measure the field morphology. These two targets are among the largest and brightest known disks with simple CO morphologies. They are also close enough to each other that they can share the same phase calibrator, and thus both can be efficiently observed during a single observation run. These disks are also no longer embedded in their natal envelopes, so the molecular line emission should come solely from the disk. Moreover, \hd\ is mostly face-on while \im\ is at an intermediate inclination, which allows us to test how inclination affects the polarization.

\hd\ is a close binary system located 157\,pc away \citep{Arun2019}. The primary is an Herbig star with a spectral type of F6--F7IIIe \citep{Malfait1998,vandenAncker1998}, while the companion is an M-dwarf separated by $\sim$0$\farcs$1 \citep{Biller2012,Close2014,Lacour2016}. \hd\ is a well known transition disk with a very large central gap. It is highly asymmetric, with the majority of the dust emission detected toward the northern part of the disk.

\im\ is a K5 Class II T Tauri star located 158\,pc away \citep{Alcala2017,Gaia2018}. High resolution ($\sim$5\,au) 1.3\,mm continuum observations reveal that the disk has spiral substructure \citep{Andrews2018}. 

\subsection{Observing Details and Data Reduction}\label{sec:datareduce}
%Observations were conducted on 2019 April 29 using ALMA Band~6 (1.3\,mm) under the project code 2018.1.01172.S (PI: I. Stephens). We chose Band~6 because this is the only ALMA band where the spectral lines CO, $^{13}$CO, and C$^{18}$O can be observed simultaneously. ALMA has four basebands. For one baseband, we map the dust continuum. This baseband was centered at 234.5\,GHz with a bandwidth of 1.875\,GHz. The other three each had a bandwidth of 59\,GHz and were centered on \coto, \ttco, and \ceo, and the basebands provided spectral resolutions of 0.079, 0.083, and 0.083\,\kms, respectively. J1427--4206 was the flux and bandpass calibrator, J1610--3958 was the phase calibrator, and J1517--2422 was the polarization calibrator.

Observations were conducted on 2019 April 29 using ALMA Band~6 (1.3\,mm) under the project code 2018.1.01172.S (PI: I. Stephens). ALMA was using 46 antennas with baselines ranging between 15 and 704\,m. Weather conditions were good for 1.3\,mm, with the precipitable water vapor column of $\sim$0.85\,mm and a system temperature oscillating between 60 and 120\,K, depending on the spectral baseband and frequency. We chose Band~6 because this is the only ALMA band where the spectral lines CO, $^{13}$CO, and C$^{18}$O can be observed simultaneously.  We tuned four ALMA basebands. One baseband was dedicated for the 1.3\,mm dust continuum emission, and it was centered at 234.5\,GHz and had a bandwidth of 2.0\,GHz. The other three each had a bandwidth of 59\,MHz and were centered on \coto, \ttco, and \ceo. These basebands provided spectral resolutions of 0.079, 0.083, and 0.083\,\kms, respectively. As typical with all ALMA observations, Hanning smoothing was applied by the  correlator to reduce ringing in the spectra. We used the default correlator spectral averaging for each baseband, which is 2 channels averaged for the high spectral resolution basebands and no channels averaged for the continuum baseband. The observations toward \hd\ and \im\ were intertwined with periodic visits to the phase and polarization calibrators every $\sim$8 and $\sim$30~minutes, respectively. The total integrated time over each science target was 1 hour. For \im, the phase center of the observations was errantly located 0$\farcs$8 west from the disk center. 
%a phase rms oscillating between $0\fdg41$ and $0\fdg46$

Delivered data were manually calibrated by the North American ARC staff using the Common Astronomy Software Applications (CASA) package \citep{McMullin2007}. J1427--4206 was used as the flux and bandpass calibrator, J1610--3958 was the phase calibrator, and J1517--2422 was the polarization calibrator. ALMA Band\,6 observations have a typical absolute flux uncertainty of 10\%, and the polarization uncertainties as determined by the D-terms were less than a 5\%. 

We imaged the continuum and spectral lines using CASA v5.6.0. To construct the continuum images, we combined the continuum baseband with the line-free channels of the other three basebands for a total bandwidth of 2.150\,GHz. We ran three phase-only self-calibration iterations on the continuum Stokes~$I$ data. The solution intervals used for the first, second, and third self-calibrations were infinite, 25\,s, and 10\,s, respectively. Self-calibration improved the continuum Stokes~$I$ signal-to-noise ratio for \hd\ and \im\ by a factor of 6 and 3, respectively. The self-calibration solutions from the continuum Stokes~$I$ were then applied to the continuum Stokes~$QUV$ and the spectral line Stokes~$IQUV$. For each spectral line, we subtracted the continuum emission by fitting the continuum using the baseband's line-free channels. To clean the images, we use the CASA task \texttt{tclean} using Briggs weighting with a robust parameter of 0.5. The final continuum and line synthesized beams and noise levels are reported in Table~\ref{tab:obs}. All images in this paper have been corrected for the primary beam. The pixel size for our maps is 0$\farcs$11.
%This table also contains information about images of the CO emission and its isotopologues, obtained after applying the continuum self-calibration solutions and subtracting the continuum emission in the uv-plane 

From the Stokes~$IQU$ images, we constructed polarized intensity ($P_I$), \pfrac, and position angle ($\chi$) maps where:

\begin{equation}
P_I = \sqrt{Q^2 + U^2}
\end{equation}
\begin{equation}
P_{\text{frac}} = P_I/I
\end{equation}
\begin{equation}
\chi = \frac{1}{2}\text{arctan}\left(\frac{U}{Q}\right) .
\end{equation}
Since $P_I$ can only be positive, it has an inherit bias toward positive values. We de-bias the values following \citet{Hull2015}. In our images, we show the polarization as line segments. These segments are typically called ``vectors." While we adopt this custom, we note that they are not true vectors because they have a 180$^\circ$ ambiguity in angle.

\renewcommand{\tabcolsep}{0.1cm}
\begin{deluxetable}{l@{}c@{}c@{}c@{}c}
\label{tab:obs}
\tablewidth{0pt}
\tablecolumns{4}
\tabletypesize{\scriptsize}
\tablecaption{Observational Parameters}
\tablehead{
\colhead{Image} & \colhead{Stokes} & \colhead{rms\tablenotemark{a}} & \colhead{Channel Width} & \colhead{Synthesized beam}   \vspace{-8pt} \\
\colhead{}  & \colhead{} & \colhead{mJy\,bm$^{-1}$} & \colhead{(\kms)}  & \colhead{$\theta_{\text{maj}} \times \theta_{\text{min}}$; PA} 
}
\startdata
\underline{\hd} \\
1.3\,mm cont & $I$ & 0.037 & -- & $0\farcs51\times0\farcs45$; $64.5\degr$ \\
1.3\,mm cont & $QUV$ & 0.022 & -- & $0\farcs51\times0\farcs45$; $64.5\degr$ \\
%1.3\,mm cont & U & 0.022 & -- & $0\farcs51\times0\farcs45$; $64.5\degr$ \\
%1.3\,mm cont & V & 0.022  & --& $0\farcs51\times0\farcs45$; $64.5\degr$ \\
\coto & $IQUV$ & 3.3 & 0.079 & $0\farcs57\times0\farcs54$; $71.3\degr$ \\
\ttco & $IQUV$ & 3.4 & 0.083 & $0\farcs59\times0\farcs56$; $74.2\degr$ \\
\ceo & $IQUV$ & 2.5 & 0.083 & $0\farcs60\times0\farcs57$; $69.7\degr$ \\
\hline 
\underline{\im} \\
1.3\,mm cont & $I$ & 0.022 & -- & $0\farcs51\times0\farcs44$; $69.5\degr$ \\
1.3\,mm cont & $QUV$& 0.022 & -- & $0\farcs51\times0\farcs44$; $69.5\degr$ \\
%1.3\,mm cont & $U$ & 0.021 & -- & $0\farcs51\times0\farcs44$; $69.5\degr$ \\
%1.3\,mm cont & $V$ & 0.021 & -- & $0\farcs51\times0\farcs44$; $69.5\degr$ \\
\coto & $IQUV$ & 3.3 & 0.079 & $0\farcs56\times0\farcs53$; $75.2\degr$ \\
\ttco & $IQUV$ & 3.4 & 0.083 & $0\farcs59\times0\farcs55$; $79.3\degr$ \\
\ceo & $IQUV$ & 2.5 & 0.083 & $0\farcs60\times0\farcs56$; $74.9\degr$ \\
\enddata 
%\tablecomments{}
\tablenotetext{a}{For lines, this is per channel.} %0.079, 0.083, and 0.083
\end{deluxetable}

%\begin{deluxetable}{cccc}
%\label{obs}
%\tablewidth{0pt}
%\tablecolumns{4}
%\tabletypesize{\scriptsize}
%\tablecaption{Observational parameters of the HD\,142527 images}
%\tablehead{
%\colhead{Image} & \colhead{Stokes} & \colhead{rms} & \colhead{Synthesized beam}  \\
%\colhead{}  & \colhead{} & \colhead{mJy} & \colhead{} 
%}
%\startdata
%1.3\,mm continuum & I & 0.037 & $0\farcs51\times0\farcs45$; $64.5\degr$ \\
%1.3\,mm continuum & Q & 0.022 & $0\farcs51\times0\farcs45$; $64.5\degr$ \\
%1.3\,mm continuum & U & 0.022 & $0\farcs51\times0\farcs45$; $64.5\degr$ \\
%1.3\,mm continuum & V & 0.022 & $0\farcs51\times0\farcs45$; $64.5\degr$ \\
%\coto & IQUV & 3.8 & $0\farcs57\times0\farcs54$; $71.3\degr$ \\
%\ttco & IQUV & 3.4 & $0\farcs59\times0\farcs56$; $74.2\degr$ \\
%\ceo & IQUV & 2.5 & $0\farcs60\times0\farcs57$; $69.7\degr$ \\
%\enddata 
%\tablecomments{}
%%\tablenotetext{(a)}{Deconvolved semimajor and semiminor axis of the ring-shaped disk.}
%\label{tobs}
%\end{deluxetable}

\subsection{Stokes~$QUV$ Noise Characterization}\label{sec:characterization}
In this paper, we will do extensive searches for a signal in \coto\ Stokes~$QUV$. Therefore, it is important to understand the noise behavior of these Stokes parameters. We thus analyze the \coto\ Stokes~$QUV$ noise on the maps before the primary beam correction.

For both \hd\ and \im, the noise across the entire \coto\ Stokes~$QUV$ cubes are fit very well with a Gaussian, with an average value of 0.0~$\pm$~3.3\,mJy. We check for any spatial variation in each cube by creating a moment 6 maps (i.e., the root mean square of the spectrum). The average value in these maps are 3.3\,$\pm$\,0.2\,mJy, and the noise is once again Gaussian. The noise characteristics do not change for emission on and off the disk, nor for different velocity bins. As such, it appears the noise characteristics for these cubes are well behaved. 

We checked the autocorrelation of our $QUV$ cubes to search for serial correlations for the noise along the spectral axis of these data. Correlation was found between consecutive channels, but no correlation was detected for the rest of the bandwidth (i.e., there was no periodic noise). This channel-to-channel correlation is due the Hanning smoothing applied by the ALMA correlator. As shown in \citet{Loomis2018}, this Hanning smoothing plus a spectral averaging of 2 (i.e., the default averaging applied by the correlator to these data) provides a channel to channel covariance of 0.3 times the variance measured in a spectra.

\section{Results}\label{sec:results}

\subsection{Continuum}

%hd142527_Debiased_Band6.py
\begin{figure}[ht!]
\begin{center}
\includegraphics[width=1\columnwidth]{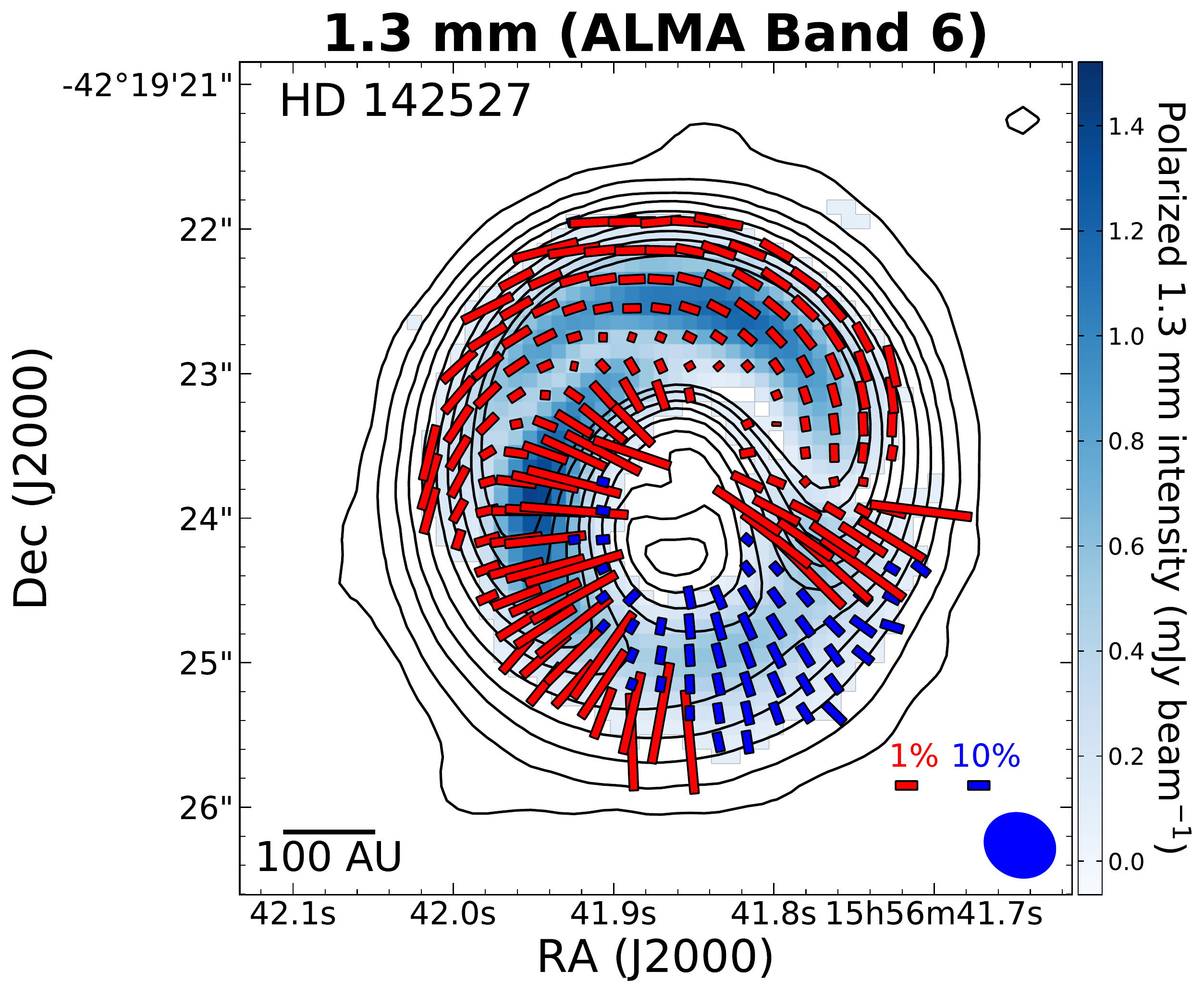}
\end{center}
\caption{1.3\,mm continuum polarimetric map toward \hd. Color-scale shows the polarized intensity $P_I$. Red and blue vectors show the direction of the polarization (i.e., vectors are not rotated). Red and blue vectors are shown for \pfrac\ smaller than and larger than 5\%, respectively (i.e., all blue vectors have higher \pfrac\ than the red). Vectors are shown for signal-to-noise ratio $>$3 for Stokes~$I$ and $P_I$ and for \pfrac~$<$~40\%. Contours show the Stokes~$I$ emission, with levels of [3, 10, 25, 50, 100, 200, 325, 500, 750, 1000]~$\times$~$\sigma_I$, where $\sigma_I$~=~37\,$\mu$Jy\,bm$^{-1}$.
}
\label{fig:hd142527_cont} 
\end{figure}

%imlup_Debiased_Band6.py
\begin{figure}[ht!]
\begin{center}
\includegraphics[width=1\columnwidth]{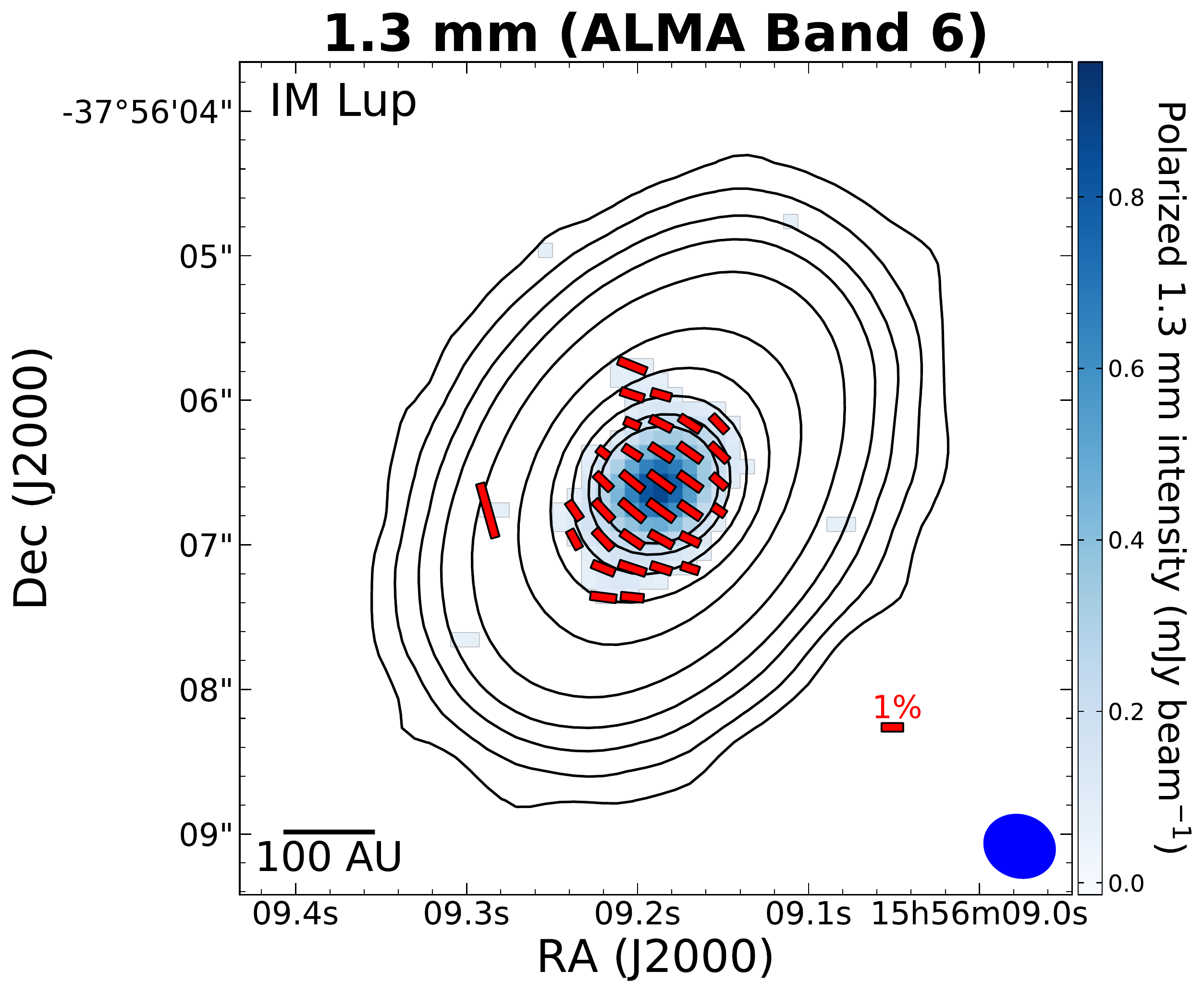}
\end{center}
\caption{Same as Figure~\ref{fig:hd142527_cont}, but now for \im, and contour levels are [3, 10, 25, 50, 100, 200, 325, 500, 750, 1000]~$\times$~$\sigma_I$, where $\sigma_I$~=~22\,$\mu$Jy\,bm$^{-1}$.
}
\label{fig:imlup_cont} 
\end{figure}

%make_comparisons.py
\begin{figure*}[ht!]
\begin{center}
\includegraphics[width=2\columnwidth]{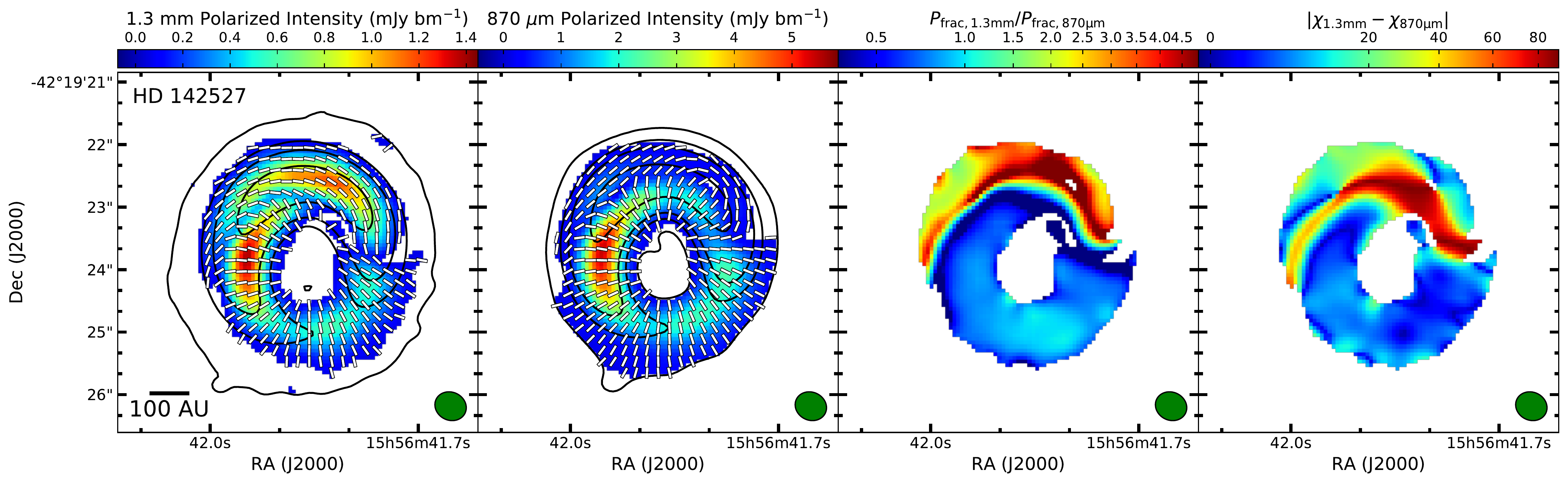}
\end{center}
\caption{Comparisons of the 1.3\,mm and 870\,$\mu$m dust continuum polarization for \hd. 
The first and second panels show the 1.3\,mm and 870\,$\mu$m polarized intensities overlaid with Stokes~$I$ contours and polarization vectors (all vectors are the same size for clarity). Observations at 1.3\,mm have been regridded to the 870\,$\mu$m pixel size and smoothed to the 1.3\,mm resolution (Table~\ref{tab:obs}). 1.3\,mm Stokes~$I$ contours are drawn for [4, 150, 500, 2000]~$\times$~$\sigma_{I,\rm{1.3\,mm}}$, where $\sigma_{I,\rm{1.3\,mm}}$~=~37\,$\mu$Jy\,bm$^{-1}$, and 870\,$\mu$m Stokes~$I$ contours are drawn for [3, 25, 60, 250]~$\times$~$\sigma_{I,\rm{870\,\mu m}}$, where $\sigma_{I,\rm{870\,\mu m}}$~=~80\,$\mu$Jy\,bm$^{-1}$. The third panel shows the ratio between the polarization fractions at 1.3\,mm and 870\,$\mu$m. The fourth panel shows the difference between the 1.3\,mm and 870\,$\mu$m polarization angles. The latter two panels have a square root stretch for the color scale.
}
\label{fig:hd142527_comparisons} 
\end{figure*}

%make_comparisons.py

\begin{figure*}[ht!]
\begin{center}
\includegraphics[width=2\columnwidth]{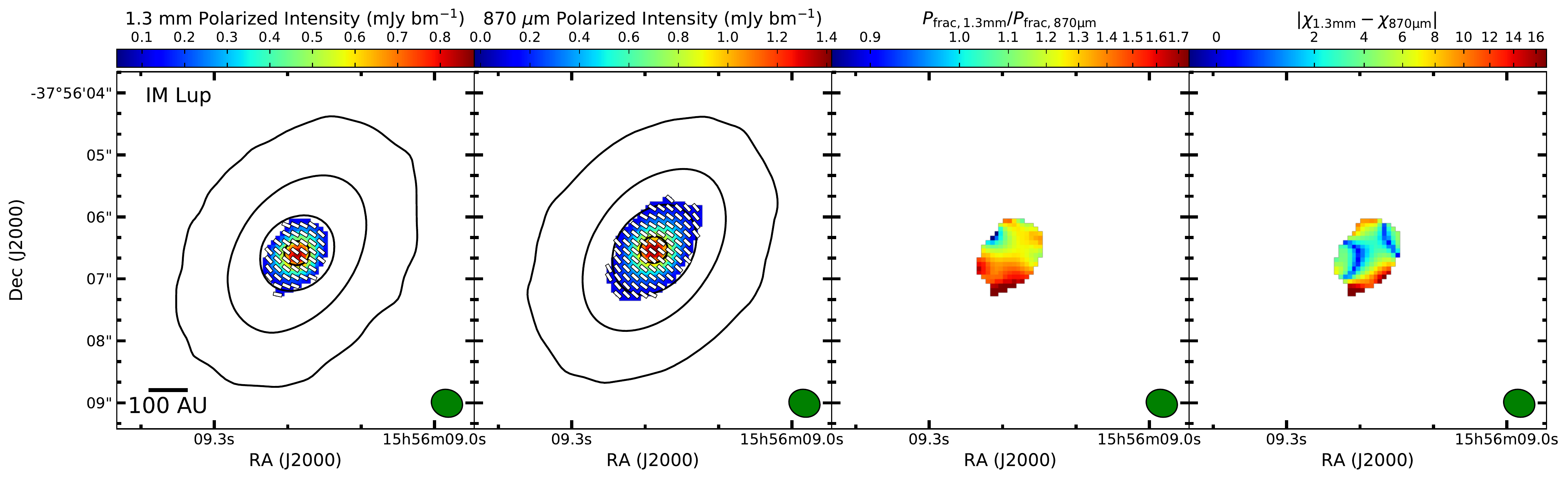}
\end{center}
\caption{Same as Figure~\ref{fig:hd142527_comparisons}, but now for \im. 1.3\,mm Stokes~$I$ contours are drawn for [4, 150, 500, 2000]~$\times$~$\sigma_{I,\rm{1.3\,mm}}$, where $\sigma_{I,\rm{1.3\,mm}}$~=~22\,$\mu$Jy\,bm$^{-1}$, and 870\,$\mu$m Stokes~$I$ contours are drawn for [3, 100, 250, 900]~$\times$~$\sigma_{I,\rm{870\,\mu m}}$, where $\sigma_{I,\rm{870\,\mu m}}$~=~100\,$\mu$Jy\,bm$^{-1}$.
}
\label{fig:imlup_comparisons} 
\end{figure*}

We present the 1.3\,mm continuum for \hd\ and \im\ in Figures~\ref{fig:hd142527_cont} and~\ref{fig:imlup_cont}, respectively. ALMA polarization observations at 870\,$\mu$m were previously presented for \hd\ in \citet{Kataoka2016b} and \citet{Ohashi2018} and for \im\ in \citet{Hull2018}. Since the focus of this paper is on the line polarization, detailed multi-wavelength analysis is beyond the scope of this paper. We briefly comment on the results in this subsection.

For \hd, an analysis by \citet{Ohashi2018} suggested that two different polarization mechanisms create the polarization pattern observed at 870\,$\mu$m. The northern part of the disk is optically thick, and the polarization morphologies show signatures of scattering (also see \citealt{Kataoka2016b}). The southern part of the disk is optically thin, and the morphology and high values of \pfrac\ are suggestive of polarization due to aligned grains \citep{Ohashi2018}. Since a radial polarization morphology is expected for a toroidal field (grains are expected to align with their short axes parallel with the magnetic field), \citet{Ohashi2018} attributes the southern morphology to grains aligned with a toroidal magnetic field. 

To compare the dust continuum polarization observations for \hd\ at 1.3\,mm to those of archival 870\,$\mu$m observations, we requested and downloaded calibrated data via the ALMA helpdesk (project code 2015.1.00425.S, PI: A. Kataoka). We re-imaged the calibrated data in the same manner as discussed in Section~\ref{sec:datareduce}. The final resolution was 0$\farcs$27~$\times$~0$\farcs$24. We regridded the 1.3\,mm observations to have the same pixel size used at 870\,$\mu$m (0$\farcs$06), and we smoothed the 870\,$\mu$m to have the same resolution as the 1.3\,mm observations.

Figure~\ref{fig:hd142527_comparisons} shows the comparisons of the polarization morphologies at the two wavelengths. The first two panels show the polarized intensity for each wavelength. Both have a strong peak toward the east. However, the 1.3\,mm polarized intensity shows a strong peak toward the north, which is not seen at 870\,$\mu$m. Moreover, the polarization vector direction changes substantially toward this northern region (i.e., toward the Stokes~$I$ peak). The third panel shows the ratio of \pfrac\ between 1.3\,mm and 870\,$\mu$m, and the fourth shows the polarization position angle differences between 1.3\,mm and 870\,$\mu$m. 

For both wavelengths, \pfrac\ and position angles toward the southern part of the disk are roughly identical, which is consistent with the aligned grain interpretation advocated by \citep{Ohashi2018}. However, the ratio $P_{\rm{frac,1.3mm}}/P_{\rm{frac,870\mu m}}$ is slightly less than unity, which is unexpected since the emission at 1.3 mm is more optically thin than that at 870\,$\mu$m and the polarization fraction is expected to increase with decreasing optical depth \citep{Yang2017}, as seen with BHB07-11 \citep{Alves2018}.

For the northern part of the disk, the polarization between wavelengths is substantially different, with $P_{\rm{frac,1.3mm}}/P_{\rm{frac,870\mu m}}$ reaching values as high as $\sim $5, and the difference in polarization angle is as large as $90^\circ$ (i.e., perpendicular). The Band 7 polarization was interpreted as coming from dust scattering, mainly because of the flip of the polarization orientations (from radial to azimuthal direction) across the horseshoe-shaped dust trap \citep{Ohashi2018}.  The flip is less prominent in Band~6 but is still present; it occurs at smaller radii compared to the Band~7 case. Whether it is still consistent with the scattering interpretation remains to be determined, and may require consideration of non-ideal polarization effects, such as large non-spherical grains polarizing emission beyond the Rayleigh regime \citep[e.g.,][]{Kirchschlager2020}. Detailed modeling is needed but is beyond the scope of this paper.

%of $\sim$5. The position angles of many vectors in this region change to mostly perpendicular (i.e., $|\chi_{\rm{1.3mm}} - \chi_{\rm{870\mu m}}|$ = 90$^\circ$). The rapid change of polarization fraction with wavelength is only expected for scattering. However, the flip in the polarization angle is unexpected for pure scattering. This feature may indicate multiple polarization mechanisms at play toward this region, or other non-ideal polarization effects \citep[e.g.,][]{Kirchschlager2020}. Detailed modeling is needed beyond the scope of this paper.}

For \im, \citet[][ALMA project code 2016.1.00712.S, PI: C. Hull]{Hull2018} found that the detected 870\,$\mu$m continuum polarization was mostly (but not completely) aligned with the minor axis, which is expected for scattering. To compare with these observations, we downloaded the polarization images provided in the online version of \citet{Hull2018}, which have a resolution of  0$\farcs$50~$\times$~0$\farcs$40. As above, we regridded the 1.3\,mm observations to have the same pixel size that we used at 870\,$\mu$m (0$\farcs$1), and we smoothed the 870\,$\mu$m to have the same resolution as the 1.3\,mm observations. We show the comparison in Figure~\ref{fig:imlup_comparisons}.

Along the major axis, \citet{Hull2018} detected polarization over an extent of $\sim$2$\arcsec$, while we detect it over $\sim$1.5$\arcsec$. Nevertheless, the 1.3\,mm results show vectors less aligned with the minor axis than that found in \citet{Hull2018}. The 1.3\,mm polarization angles are suggestive of a more azimuthal morphology along the edges of the detected polarized emission, with the main differences at the northern and southern parts of the disk (right panel, Figure~\ref{fig:imlup_comparisons}). This possible change in polarization morphology between the wavelengths of 870\,$\mu$m and 1.3\,mm was also found between these wavelengths for HL~Tau \citep{Stephens2017c}. This feature could possibly indicate two competing polarization mechanisms, such as scattering and alignment with the radiation anisotropy. Nevertheless, pure scattering models show that, while very optically thick regions show uniform vectors aligned with the minor axis, optically thinner regions have a significant azimuthal component \citep{Yang2017}. Indeed, 1.3\,mm dust emission is optically thinner than emission at 870\,$\mu$m, so it is possible that the morphologies at both wavelengths can be explained by dust scattering.
%We calculate the standard deviation of the distribution of 1.3\,mm vectors shown in Figure~\ref{fig:imlup_cont} and the 870\,$\mu$m vectors shown in \citet{Hull2018}, and find them to be 7$^\circ$ and 14$^\circ$, respectively. 

Additionally $P_{\rm{frac,1.3mm}}/P_{\rm{frac,870\mu m}}$ is primarily larger than 1 for \im. This feature is unexpected for optically thin dust grains of sizes $\sim$100\,$\mu$m, as originally proposed by \citet{Hull2018}. However, this might be explained via optical depth and beam smearing effects \citep{Lin2020b} or by grain sizes that are a few 100s of $\mu$m \citep[e.g.,][]{Kataoka2015,Kataoka2017}.

\subsection{Line Observations}

%hd142527_co_21_mom0.py
%hd142527_13co21_mom0.py
%hd142527_c18o21_mom0.py
\begin{figure*}[ht!]
\begin{center}
\includegraphics[width=0.66\columnwidth]{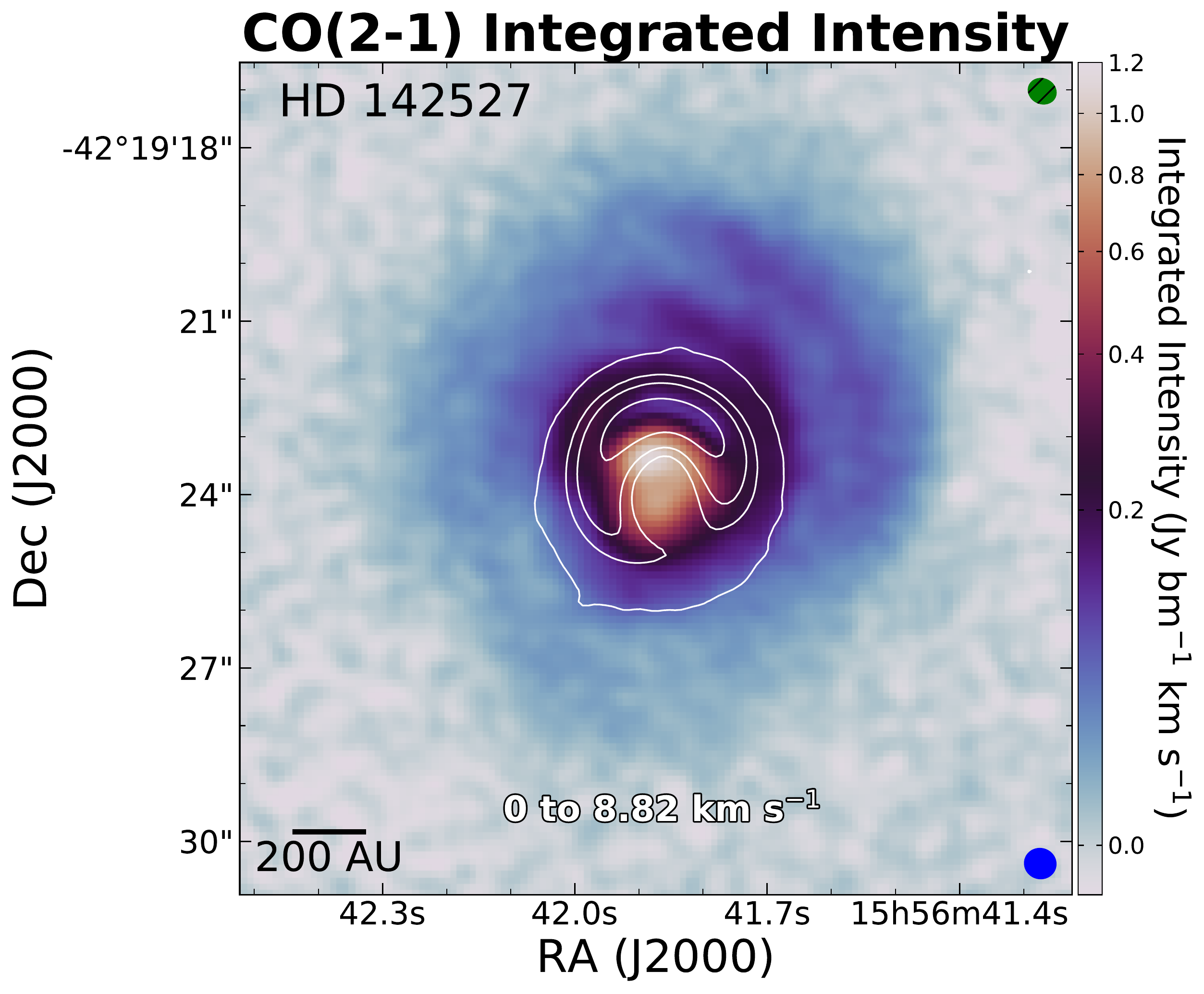}
\includegraphics[width=0.66\columnwidth]{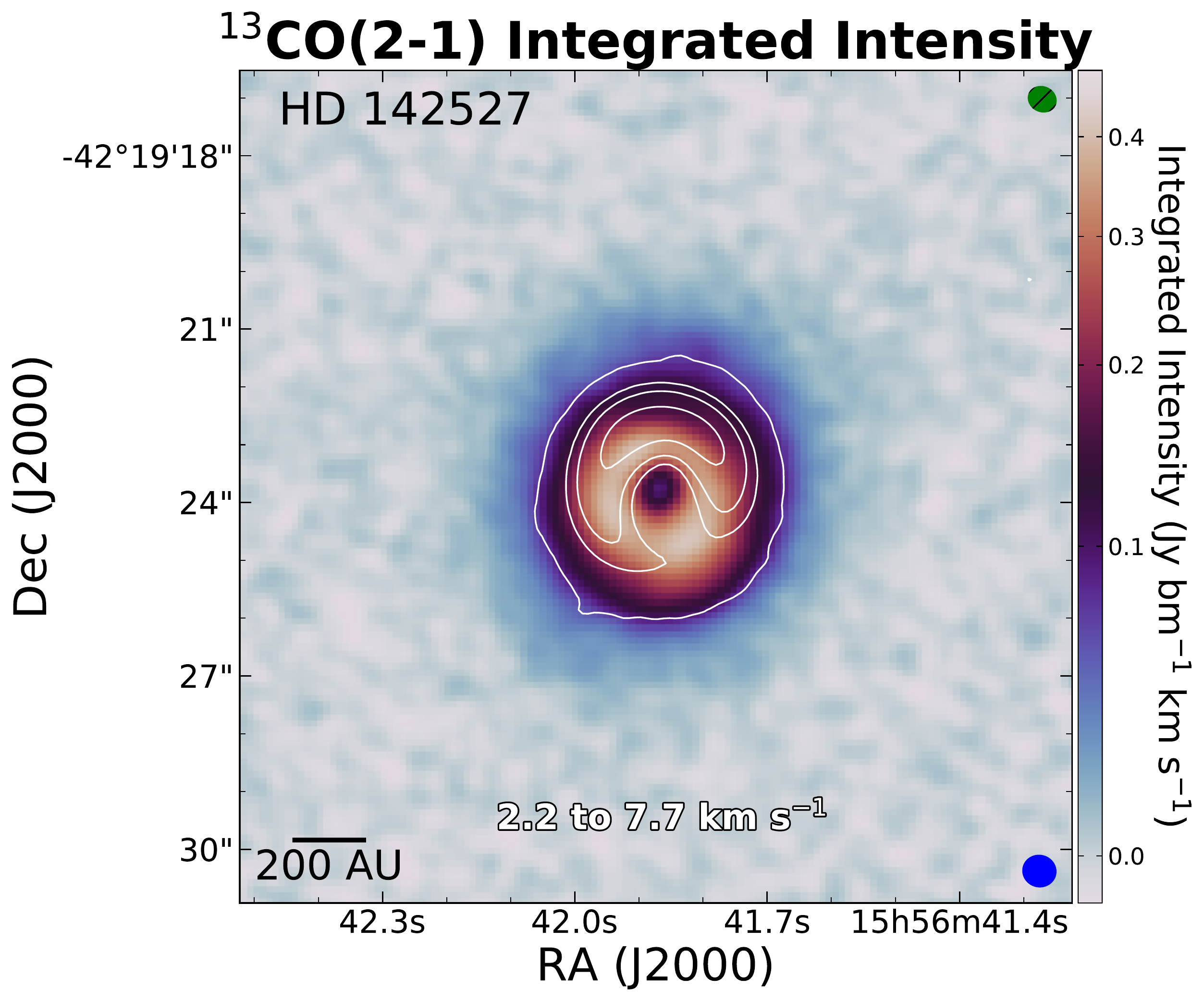}
\includegraphics[width=0.66\columnwidth]{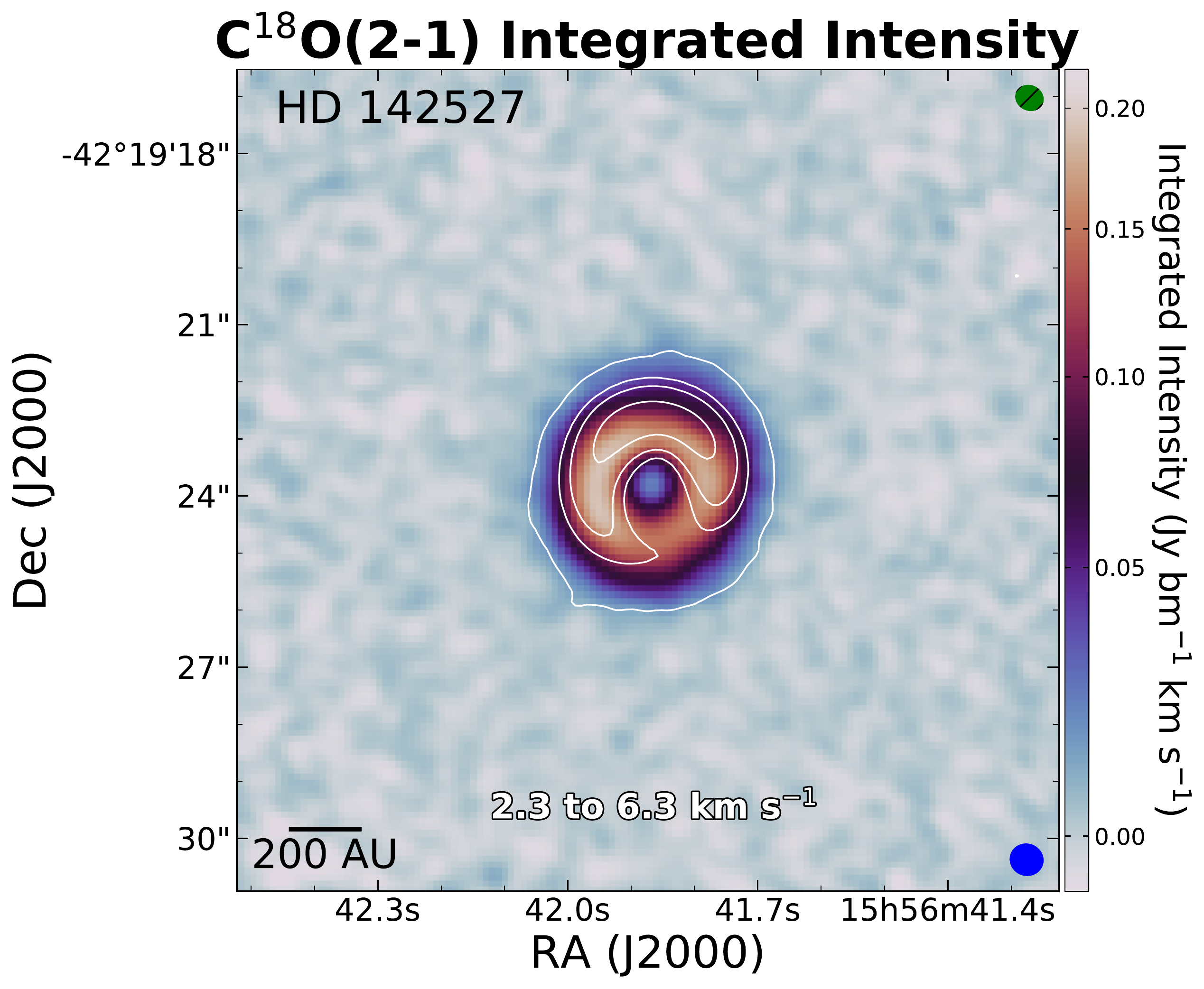}
\end{center}
\caption{From left to right, \hd\ colormaps of the integrated intensity (moment~0) for \coto, \ttco, and \ceo. The integrated velocity range is labeled at the bottom of each figure. White contours show the continuum emission, with contours levels at [4, 150, 500, 2000]~$\times$~$\sigma_I$, where $\sigma_I$~=~37\,$\mu$Jy\,bm$^{-1}$. The green ellipse at the top right is the synthesized beam for the continuum, while the blue ellipse at the bottom right is for the spectral line. %Images have been corrected for the primary beam.
%4*I_rms,150*I_rms,500*I_rms,2000*I_rms
}
\label{fig:hd142527_mom0} 
\end{figure*}

\begin{figure*}[ht!]
\begin{center}
\includegraphics[width=0.66\columnwidth]{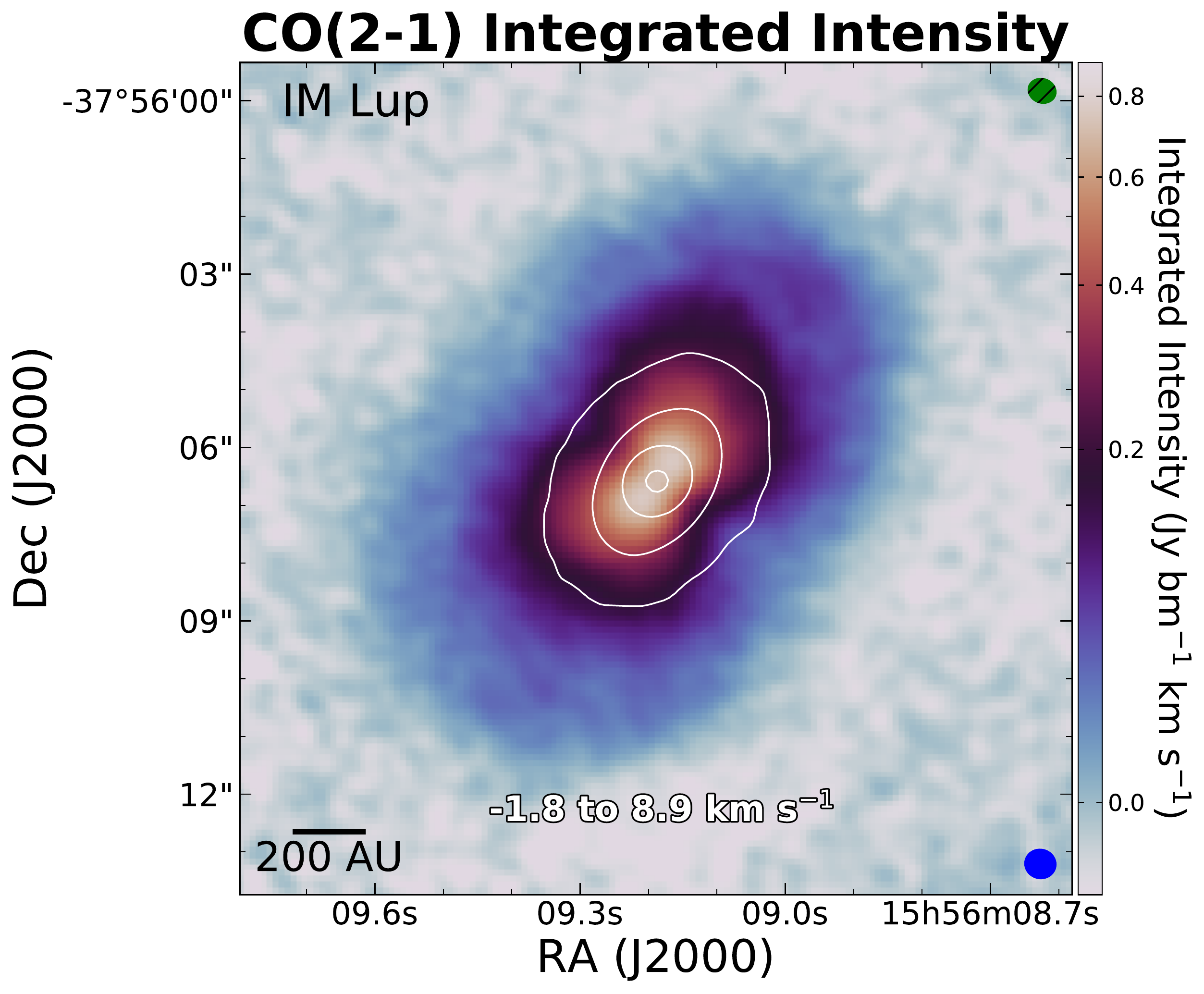}
\includegraphics[width=0.66\columnwidth]{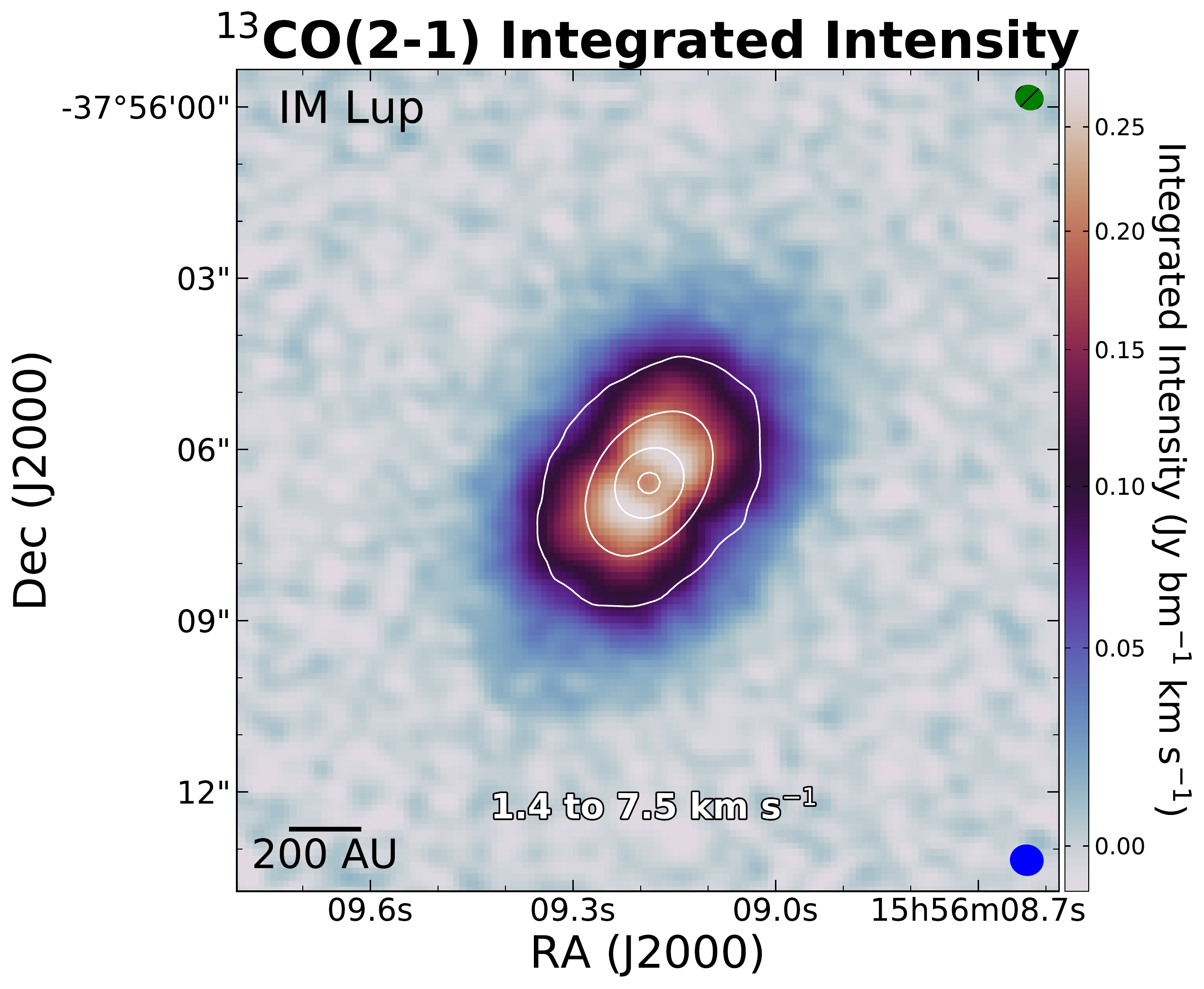}
\includegraphics[width=0.66\columnwidth]{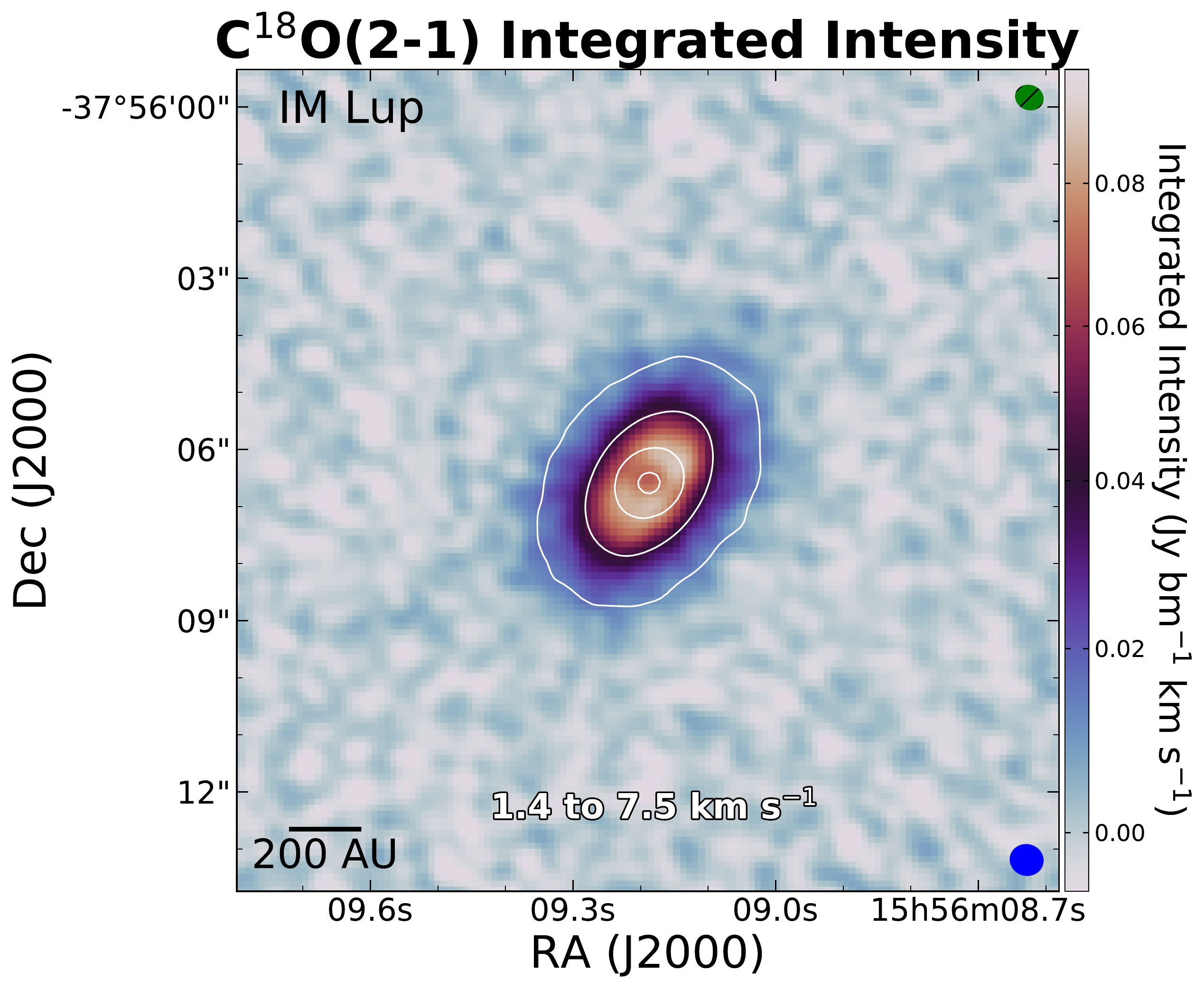}
\end{center}
\caption{Same as Figure~\ref{fig:hd142527_mom0}, but now for \im\ and  $\sigma_I$~=~22\,$\mu$Jy\,bm$^{-1}$.
}
\label{fig:imlup_mom0} 
\end{figure*}

We show integrated intensity (moment~0) maps of \coto, \ttco, and \ceo\ for \hd\ and \im\ in Figures~\ref{fig:hd142527_mom0} and~\ref{fig:imlup_mom0}, respectively. Each image is overlaid with contours of the continuum emission. Compared to the continuum for both disks, \coto\ and \ttco\ lines are detected over a larger extent than that seen for the continuum, while the \ceo\ line is detected over roughly the same area. The measured dimensions of the major and minor axes over which \coto\ emission is detected are 11.2$\arcsec$~$\times$~9$\farcs$6 (1760~$\times$~1510\,au) for \hd\ and 12.0$\arcsec$~$\times$~8$\farcs$7 (1900~$\times$~1370\,au) for \im. 

For \hd, the \coto\ moment~0 map (Figure~\ref{fig:hd142527_mom0}, left) indicates a spiral-like structure. The spiral structure is well known for this disk, as it was detected in the infrared \citep{Casassus2012,Rameau2012}, in Near-IR $H$- and $K_{\text{s}}$-band polarimetric observations \citep{Canovas2013,Avenhaus2014}, and in previous ALMA \mbox{CO(2--1)} and \mbox{CO(3--2)} observations \citep{Christiaens2014}. The \ttco\ and \ceo\ emission is ring-like, with a local minimum of emission toward the center of the disk.

For \im, a pinched pattern is seen in the \coto\ and \ttco\ moment~0 maps, which indicates emission from gas above the midplane. \coto\ appears to be slightly asymmetric toward the northwest part of the disk. Like \hd, \ttco\ and \ceo\ line emission also have a local minimum toward the center of the disk. All of the above features were also mentioned in \citet{Cleeves2016}.

\subsection{Optical Depth}\label{sec:optical}
Linear polarization due to the GK effect is largely affected by the optical depth of the spectral line \citep[e.g.,][]{Deguchi1984}. The optical depth can be estimated from the ratio of brightness temperatures \citep[e.g.,][]{Lyo2011}. Given the brightness temperature for the first and second line of $T_{B1}$ and $T_{B2}$, we can solve for the optical depth using the equation

\begin{equation}\label{eq:opticaldepth}
R = \frac{T_{B1}}{T_{B2}} = \frac{1-e^{-\tau_1}}{1-e^{-\tau_2}}  = \frac{1-e^{-\tau_1}}{1-e^{-\tau_1/X}} .
\end{equation}

Since the conversion factor between flux and brightness temperature for each line in this study are roughly identical (vary by less than 1\%), we instead use R = $F_{\nu 1}/F_{\nu 2}$. $X$ is the isotopic line ratio, which we adopt as [$^{12}$CO]/[$^{13}$CO] = 70 and [$^{12}$CO]/[C$^{18}$O] = 500. The optical depth was solved for numerically. We made optical depth maps (not shown) both channel by channel and from the integrated intensity maps of each line. 

For \hd, the calculated optical depths at some locations are either highly uncertain or cannot be computed (i.e., the less abundant species is brighter, making Equation~\ref{eq:opticaldepth} unsolvable), given that \coto\ emission has features of self-absorption and absorption against the continuum. We find that \coto\ optical depths are $>$30 for all locations except the center of the disk, where the optical depth is as low as 14. The optical depths of \ttco\ across the disk, as calculated from its flux ratio with \ceo, range from $\sim$1 to 7, and the optical depth for \ceo\ is between 0.1 and 1. When looking at optical depth maps channel by channel, \coto\ is optically thick at all velocities where both \coto\ and \ttco\ are detected. At the highest velocities from systemic (at about 1 and 6.5\,\kms), the \coto\ optical depths are $\sim$8--10.
%(1) \coto\ optical depths are $>30$ in all locations except for the center of the disk, where the optical depth is as low as 14. (2) The optical depth of \ttco\, as calculated from \ceo, range from $\sim$1 to 7, (3) given the assumed abundance ratio, the optical depth for \ceo\ is between 0.1 and 1. When looking at optical depth maps channel by channel, \coto\ is optically thick at all velocities where both \coto\ and \ttco\ are detected. At the highest velocities from systemic (at about 1 and 6.5\,\kms), the \coto\ optical depths are $\sim$8--10.

For \im, the optical depths could be calculated at all locations, though we note that there is likely some absorption against the continuum toward the center of the disk \citep{Cleeves2016}. Based on the integrated intensity maps, we find that \coto\ optical depths are typically about 40, but range from about 10--100. Given the abundance ratios, \ttco\ is just around the optically thick ($\tau$~=~1) regime, and \ceo\ is optically thin. Again, when looking at optical depth maps channel by channel, \coto\ is optically thick at all velocities where both \coto\ and \ttco\ are detected. At the highest velocities from systemic (at about 1.5 and 7.5\,\kms), the \coto\ optical depths are $\sim$25.
 
%The most detailed models of the GK Effect simply consider molecules placed within a magnetic field that may or may not have an additional external source of radiation. Without an external source of radiation, the GK effect is expected to be strongest at $\tau$~$\approx$~1, and significantly reduced for high ($\gtrsim$10) and low ($\lesssim$0.1) optical depths \citep[e.g.,][]{Deguchi1984}. In such a case, \ttco\ is in the optimal optical depth range for a polarized signal detection, \coto\ is suboptimal, and \ceo\ is mostly suboptimal (only toward select areas of \hd). However, when an external source is added (due to, for example, compact dust), CO emission is expected to be strongly polarized even at low optical depths and will be perpendicular to the polarization observed for $\tau$~$\approx$~1 \citep{Cortes2005}. Optically thick emission will again have very low values of \pfrac. Given this model, we would expect the GK effect to be easier to detect for \ttco\ and \ceo\ rather than \coto. However, as will be shown later in this paper, we only detect polarization for \coto, albeit over a limited velocity range and area of the disk. 

Comprehensive models that show in detail how the values of \pfrac\ change with optical depth do not consider disks (the GK effect in disks is modeled in \citealt{Lankhaar2020}, which we will discuss in Section~\ref{sec:summary}). These models simply consider molecules placed within a magnetic field, and may or may not have an additional external source of radiation. Without an external source of radiation, the GK effect is expected to be strongest at $\tau$~$\approx$~1, and significantly reduced for high ($\gtrsim$10) and low ($\lesssim$0.1) optical depths \citep[e.g.,][]{Deguchi1984}. In such a case, \ttco\ is in the optimal optical depth range for a polarized signal detection, \coto\ is suboptimal, and \ceo\ is mostly suboptimal (only toward select areas of \hd). However, when an external source is added (due to, for example, compact dust), CO emission is expected to be strongly polarized even at low optical depths and will be perpendicular to the polarization observed for $\tau$~$\approx$~1 \citep{Cortes2005}. Optically thick emission will again have very low values of \pfrac. Given this model, we would expect the GK effect to be easier to detect for \ttco\ and \ceo\ rather than \coto. However, as will be shown later in this paper, we only detect polarization for \coto, albeit over a limited velocity range and area of the disk.

We stress that a disk is more complex than these models, as additional factors must be considered such as the disk's inclination, density profile, and temperature profile. While these features are taken into account in the PORTAL software \citep{Lankhaar2020}, the effect of optical depth on \pfrac\ was not discussed in detail, though it is mentioned that \pfrac\ is expected to be near 0 for high optical depths. Using PORTAL to show how optical depth changes the polarization of lines requires a detailed study that is beyond the scope of this paper.

Finally, it is worth mentioning that given the estimated densities for \hd\ and \im\ \citep{PerezS2015,Cleeves2016}, the lines are certainly thermalized in the midplane. As such, collisions will dominate over the radiative rate, which causes no polarization from these spectral lines. However, polarized emission could potentially be emitted from less dense regions above the midplane.

\section{Searching for a Polarized Signal in Spectral Lines}\label{sec:search}
In this section, we attempt a thorough search for the polarization in the CO isotopologues. When we mention Stokes parameters in this section, we specifically refer to the isotopologues and not the continuum.

For \hd\ and \im, Stokes~$QUV$ maps appear to be pure noise if we integrate over the same velocity ranges as their respective Stokes~$I$ images (i.e., those shown in Figures~\ref{fig:hd142527_mom0} and~\ref{fig:imlup_mom0}). Additionally, if we step through the Stokes~$QUV$ images channel by channel, we see no strongly polarized signals at any velocity. In other words, any sort of polarization signal is not obvious for any of the observed spectral lines in both disks. We decided to do a deeper search for a polarized signal via spectral stacking and via spatial averaging over an integrated velocity range. We first motivate the stacking technique using a simple morphological model.

\subsection{Morphological Model}\label{sec:model}
To motivate possible ways for stacking, we create model images of the expected Stokes~$Q$ and $U$ maps for purely elliptical polarization for \hd\ and \im, which may be expected for a toroidal field. Along with measured quantities of the disks (e.g., inclinations and position angles), the only assumptions that go into the model is a surface density profile and a polarization morphology. This notably is not a model of the GK effect, as it ignores very important parameters such as the optical depth. The point of the model is to simply describe which quadrants of the disk we expect to see positive and negative values for Stokes~$Q$ and $U$ based on an elliptical (and eventually radial; see below) polarization morphology.

Following \citet{Yang2019}, we create an image of an elliptical polarization pattern for a disk with inclination, $i$. The polarization pattern is then rotated according to the position angle of the disk, PA. Each pixel of the image now has a polarization angle, $\theta_{\text{ell}}$, which is measured counterclockwise from north.

%the position angle (measured counterclockwise from the major axis) tangent to an ellipse ($\theta_{\text{ell}}$)

%\begin{equation}
%\theta_{\text{ell}} = \text{arctan}\left(-\frac{\text{cos}^2(i)}{\text{tan}(\theta_{\text{sky}})}\right),
%\end{equation}
%where $i$ is the inclination and and $\theta_{\text{sky}}$ is the azimuthal angle measured clockwise from the major axis of the disk.

\begin{figure}[ht!]
\begin{center}
\includegraphics[width=1\columnwidth]{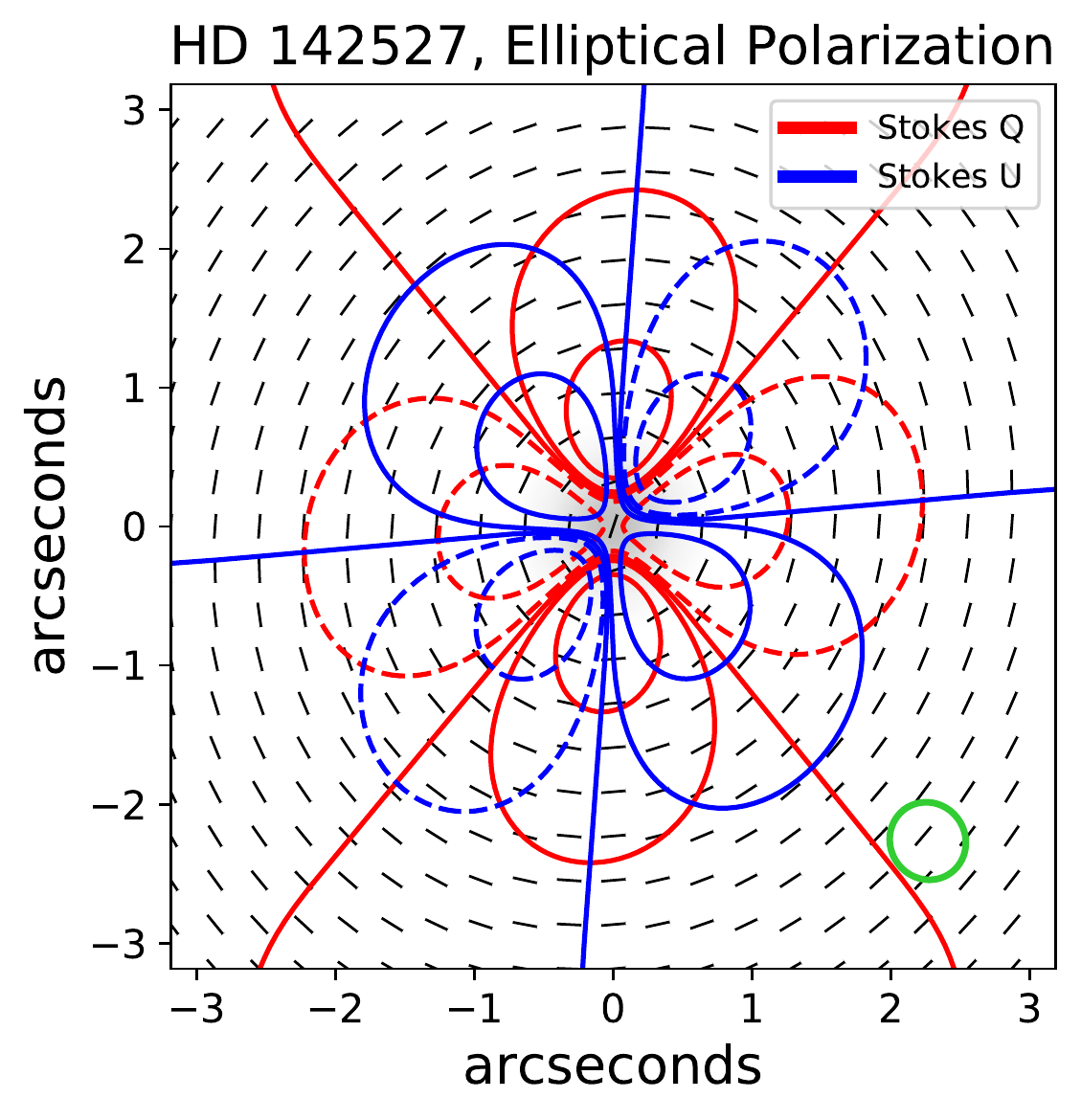}
\end{center}
\caption{Simple model for \hd\ of elliptical polarization position angles (vectors) and the resulting Stokes~$Q$ and $U$ parameters, which are shown as red and blue contours, respectively. Stokes~$Q$ and $U$ maps are normalized so that the maximum value of $U$ is $U_{\text{max}}$~=~1, and contours for each Stokes parameter are drawn for [--0.4, --0.1, 0, 0.1, 0.4] $\times$ $U_{\text{max}}$. Solid and dashed contours show positive and negative levels, respectively. The model was smoothed to the same resolution as the observations, which is shown as a green ellipse at the bottom right.
}
\label{fig:hd142527_model} 
\end{figure}

\begin{figure}[ht!]
\begin{center}
\includegraphics[width=1\columnwidth]{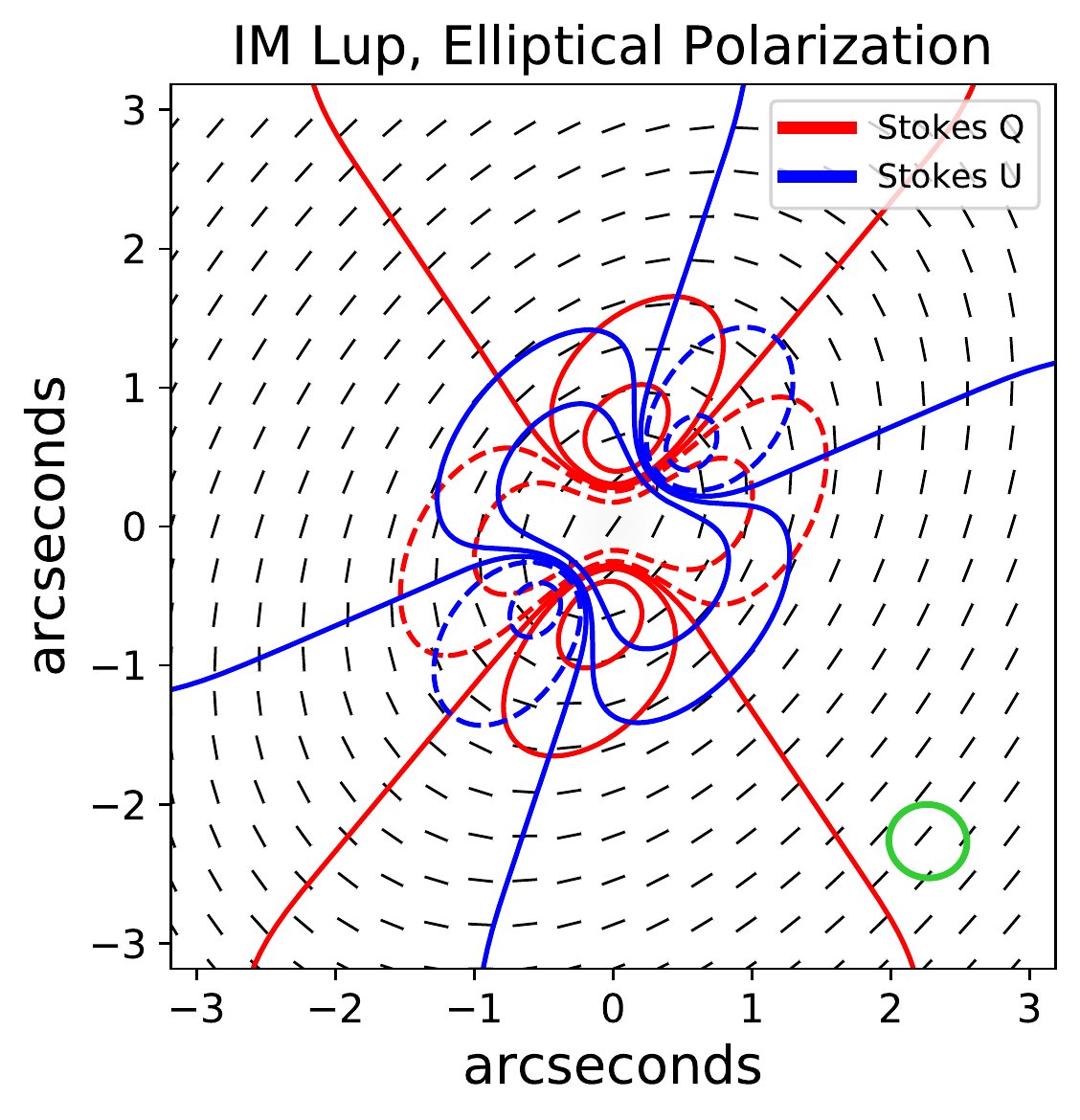}
\end{center}
\caption{Same as Figure~\ref{fig:hd142527_model}, but for \im.
}
\label{fig:imlup_model} 
\end{figure}

For \hd\ we assume PA~=~160$^\circ$ and $i$~=~28$^\circ$ \citep{PerezS2015}, and for \im\ we assume PA~=~144$^\circ$ and $i$~=~48$^\circ$ \citep{Cleeves2016}. We assume a surface density profile that follows the form in \citet{Andrews2011}, and we assume the surface density is directly proportional to the Stokes~$I$ flux such that
\begin{equation}\label{eq:surface}
I \propto \left(\frac{R}{R_c}\right)^{-\gamma} \text{exp}\left[-\left(\frac{R}{R_c}\right)^{2-\gamma}\right] ,
\end{equation}
where $R$ is the radius of the disk, $R_c$ is the characteristic radius, and $\gamma$ is the gas surface density exponent, which we assume to be 1. For \hd\ and \im, we assume $R_c$ is 200 and 100\,au, respectively \citep{PerezS2015,Cleeves2016}. From the Stokes~$I$ and $\theta_{\text{ell}}$ maps, we can determine the Stokes parameters $Q$ and $U$ via

\begin{equation}
Q =  IP_{\text{frac}}\,\text{cos}(2\theta_{\text{ell}})
\end{equation}
\begin{equation}
U = IP_{\text{frac}}\,\text{sin}(2\theta_{\text{ell}}).
\end{equation}
For this model, we assume \pfrac\ to be constant across the disk. Then, \pfrac\ and the constant needed for the proportionality in Equation~\ref{eq:surface} become a single constant, which we call $K$. We then smoothed each map to have the same resolution as our observations. For any value $K$ of these two disks, the Stokes~$U$ peak ($U_{\text{max}}$) happens to be larger than that of Stokes~$Q$. We then normalize both the Stokes~$Q$ and $U$ maps by dividing by $U_{\text{max}}$ so that we can draw contours with respect to the maximum value in the Stokes images. Modeled polarization angles and Stokes~$Q$ and $U$ contours for \hd\ and \im\ are shown in Figure~\ref{fig:hd142527_model} and~\ref{fig:imlup_model}, respectively.%distance = 157\,pc \citep{Alcala2017}

Since the GK effect has an ambiguity with being perpendicular or parallel with the magnetic field, a radial polarization pattern is also possible for a toroidal field. In such a case, Figure~\ref{fig:hd142527_model} and~\ref{fig:imlup_model} would have to be modified by rotating all angles by 90$^\circ$ and by switching the Stokes~$Q$ and $U$ contours with each other. Indeed, the GK disk model in \citet{Lankhaar2020} suggests that radial polarization would be seen for a face-on disk for both toroidal and radial magnetic fields.

It is also important to note that the angular extent of the gaseous disk goes substantially beyond what is shown in these models (Figures~\ref{fig:hd142527_mom0} and~\ref{fig:imlup_mom0}). The modeled Stokes~$Q$ and~$U$ emission declines rapidly with radius due to the exponential part of the assumed intensity profile (Equation~\ref{eq:surface}). We note that the real Stokes~$Q$ and~$U$ contours levels could extend farther out if the polarized fraction is higher in the outer disk. A higher \pfrac\ in the outer disk indeed would be expected for \coto\ and \ttco, as the outer disk is less optically thick than the inner disk. Regardless of the assumptions about \pfrac, we still expect the 4-petal daisy-like pattern in Stokes~$Q$ and $U$ for elliptical and radial polarization patterns.

The 4-petal daisy-like pattern for the elliptical polarization model suggest four distinct disk wedges, where two wedges are positive and two wedges are negative. For Stokes~$Q$, the intensity switches signs (e.g., from positive to negative) at around the major and minor axes of the disks. The four sign switches for Stokes~$U$ are each located between two different sign switches for Stokes~$Q$ (the $Q$ and $U$ maps are both 4-petal daisy-like patterns offset by $\sim$45$^\circ$). These transitions are at $\sim$45$^\circ$ from the major and minor axes when deprojected to the disk plane. Again, for a radial polarization pattern, Stokes~$Q$ and $U$ contours are switched, so the separation between wedges would be at the same locations.

Since we want to make sure we avoid stacking areas of the disk with positive Stokes~$Q$ and~$U$ emission with those areas that are negative, these models motivate us to stack Stokes parameters spatially based on disk quadrants. These areas will be the basis of some of the stacking used in Section~\ref{sec:gofish}. We first stack based on deprojected disk quadrants starting at the major axis and binning for every 90$^\circ$. We then do 90$^\circ$ deprojected quadrants again, but this time starting at a deprojected angle 45$^\circ$ from the major axes. Moreover, for certain quadrants we cut out the central part of the disk to make sure we do not add positive with negative emission, which will be discussed in the next subsection. 

We are binning based on 90$^\circ$ quadrants and start the binning for two separate position angles, which provides a fine probe in many areas of the disk. As such, the stacking method is likely to find polarization morphologies that are not elliptical or radial, if they exist.
 %This is a good approximation for \hd, but a little less-so for \im. However, we use this approximation because it helps us better probe possible polarization morphologies that are non-elliptical and non-radial. 
%Nevertheless, we also tried stacking that closely matches the Stokes~$U$ transitions, and the stacking still looked appeared as noise.

\subsection{Stacking via \texttt{GoFish}}\label{sec:gofish}
Spectral lines are Doppler shifted due to the rotational profile of the disk. A signal can be stacked for the Stokes parameters by aligning their spectra to common centroid velocities based on a disk's deprojected Keplerian rotational profile. We do this by using the software package \texttt{GoFish} \citep{GoFish}.

\texttt{GoFish} takes in a number of parameters to estimate the three-dimensional Keplerian rotational profile used for stacking. This includes the stellar mass, $M_\star$, the position angle of the major axis pointing to the most red-shifted side, PA$_{\text{red}}$, the inclination of the source, $i$, the distance to the source, $d$, and the emission scale height at 1$\arcsec$, $z_0$. We find estimates for most these parameters from disk models from \citet[][\hd]{PerezS2015} and \citet[][\im]{Cleeves2016}. For both disks we assume $z_0$~=~0$\farcs$1. For \hd, we assume $M_\star$~=~2.8\,$M_\odot$, PA$_{\text{red}}$ = 160$^\circ$, $i$~=~28$^\circ$, and $d$~=~157\,pc. For \im, we assume $M_\star$~=~1\,$M_\odot$, PA$_{\text{red}}$ = 144$^\circ$, $i$~=~48$^\circ$, and $d$~=~158\,pc. For \im, we also had to specify an x-offset of 0.8$\arcsec$ since the phase center was errantly offset to the disk center by this amount. Other input parameters for \texttt{GoFish} were all assumed to be their defaults.

\begin{figure}[ht!]
\begin{center}
\includegraphics[width=1\columnwidth]{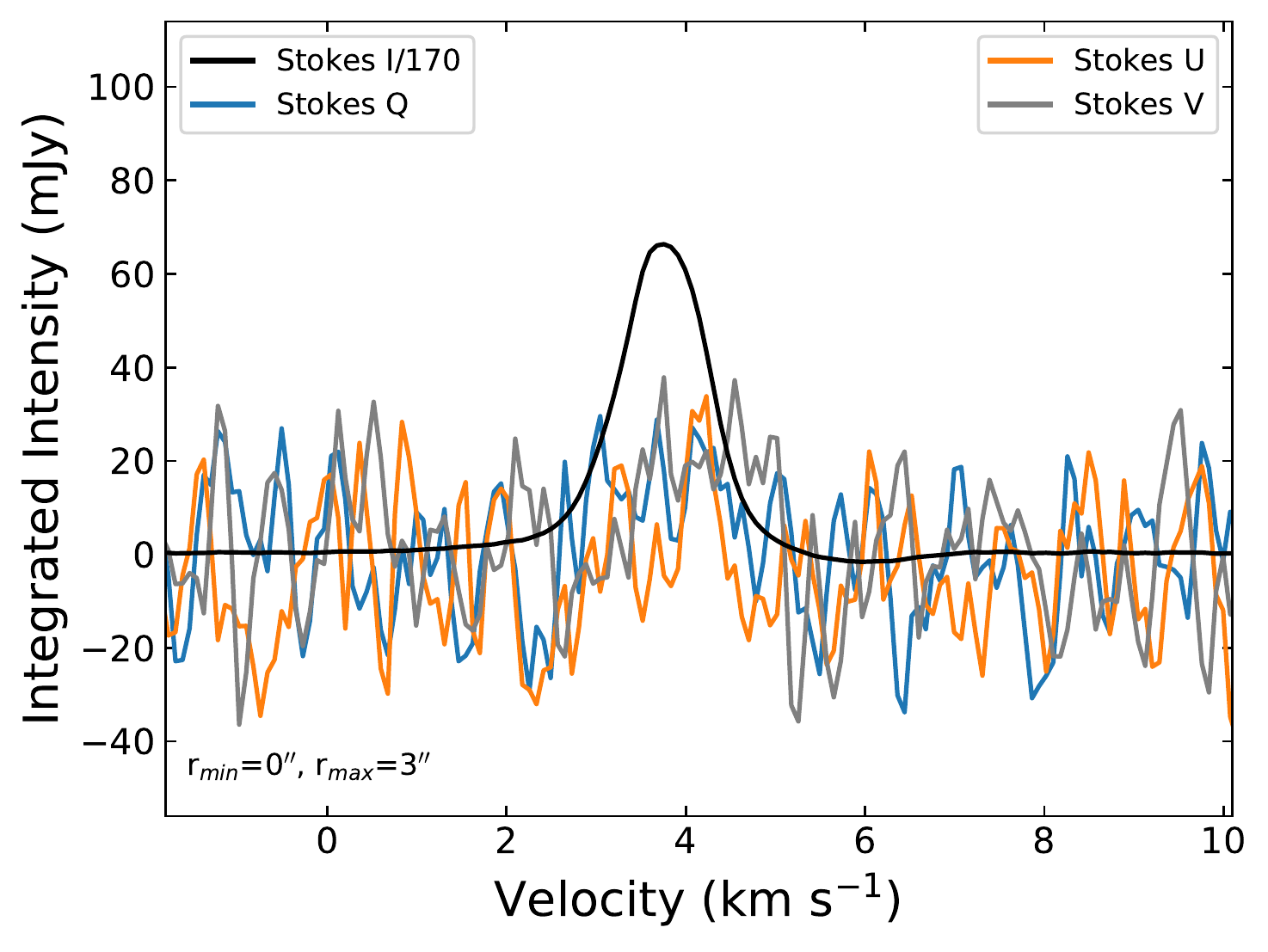}
\end{center}
\caption{\hd\ $IQUV$ \coto\ spectra, stacking based on the Keplerian profile of the disk via \texttt{GoFish} across the entire disk out to a radius of 3$\arcsec$. The stacked Stokes~$I$ signal has been divided by 170.
}
\label{fig:hd142527_gofish_all} 
\end{figure}

\begin{figure}[ht!]
\begin{center}
\includegraphics[width=1\columnwidth]{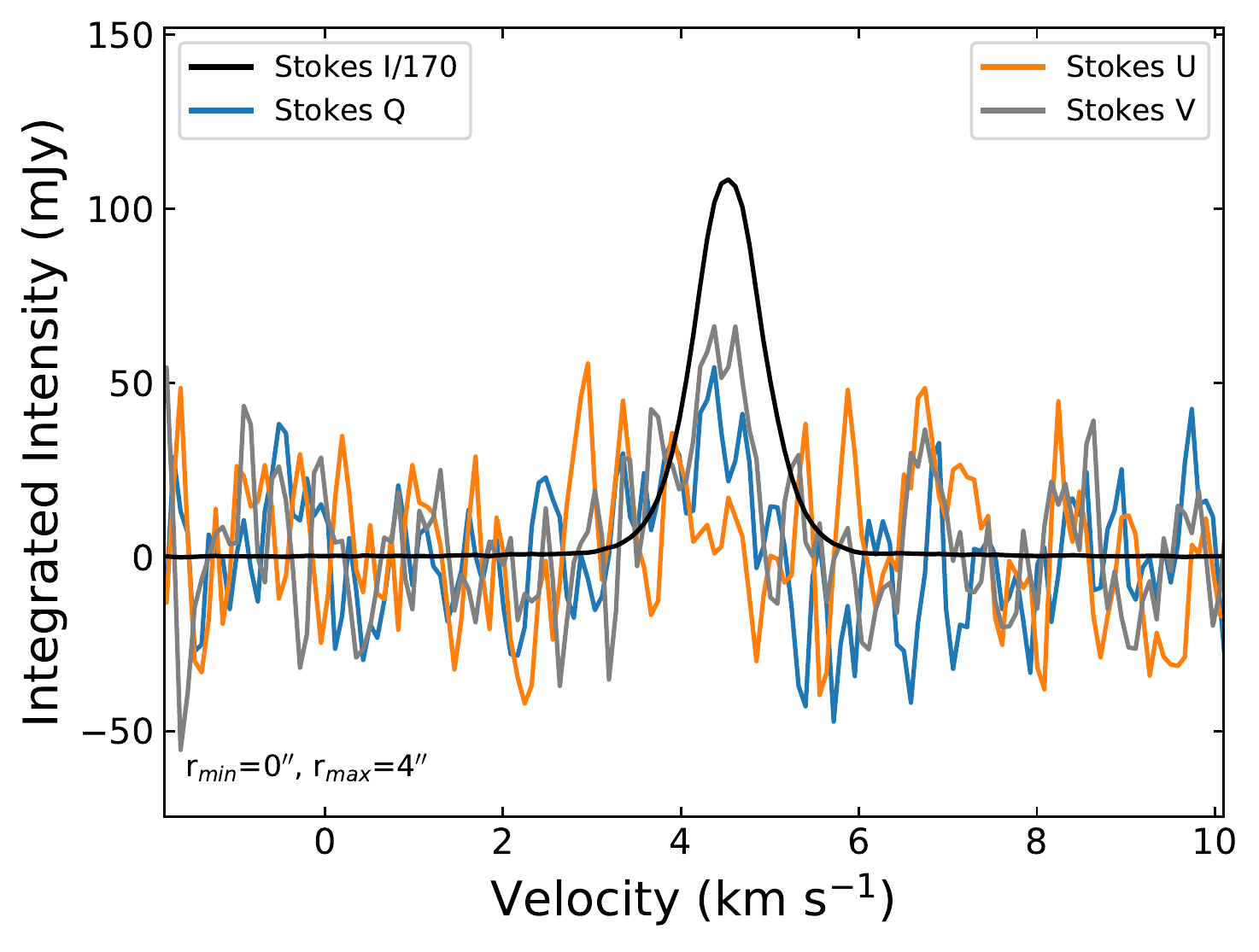}
\end{center}
\caption{\im\ $IQUV$ \coto\ spectra, stacking based on the Keplerian profile of the disk via \texttt{GoFish} across the entire disk out to a radius of 4$\arcsec$. The stacked Stokes~$I$ signal has been divided by 170.
}
\label{fig:imlup_gofish_all} 
\end{figure}

\texttt{GoFish} can also take in a minimum and maximum radius to apply the technique and can also integrate over user-specified disk wedges. We tried a vast mixture of wedges and radii for all 3 spectral lines. Nevertheless, we could not find a significant signal in Stokes~$Q$, $U$, and $V$ via stacking. Here we present the stacking technique by first stacking each Stokes parameter across the entire disk, and then by stacking by the quadrants mentioned in Section~\ref{sec:model}. We first select the radius in the disk plane, $r_{\text{max}}$, over which we do the stacking. We choose a value where the Stokes~$I$ flux drops to 90\% of the peak value of the Stokes~$I$ moment~0 map. Rounded to the nearest arcsecond for \coto, $r_{\text{max}}$ is 3$\arcsec$ for \hd\ and 4$\arcsec$ for \im. The \coto\ results for \hd\ and \im\ are shown in Figures~\ref{fig:hd142527_gofish_all} and~\ref{fig:imlup_gofish_all}. The results for \ttco\ and \ceo\ are shown in the Appendix.

%\begin{figure*}[ht!]
%\begin{center}
%\includegraphics[width=2\columnwidth]{{hd142527_line=co_21_panels_rmin=0.17_rmax=3}.pdf}
%\end{center}
%\caption{\hd\
%}
%\label{fig:hd142527_gofish_all} 
%\end{figure*}
%
%
%\begin{figure*}[ht!]
%\begin{center}
%\includegraphics[width=2\columnwidth]{{imlup_line=co_21_panels_rmin=0.5_rmax=4}.pdf}
%\end{center}
%\caption{\im\
%}
%\label{fig:imlup_gofish_all} 
%\end{figure*}

\begin{figure*}[ht!]
\begin{center}
\includegraphics[width=0.49\columnwidth]{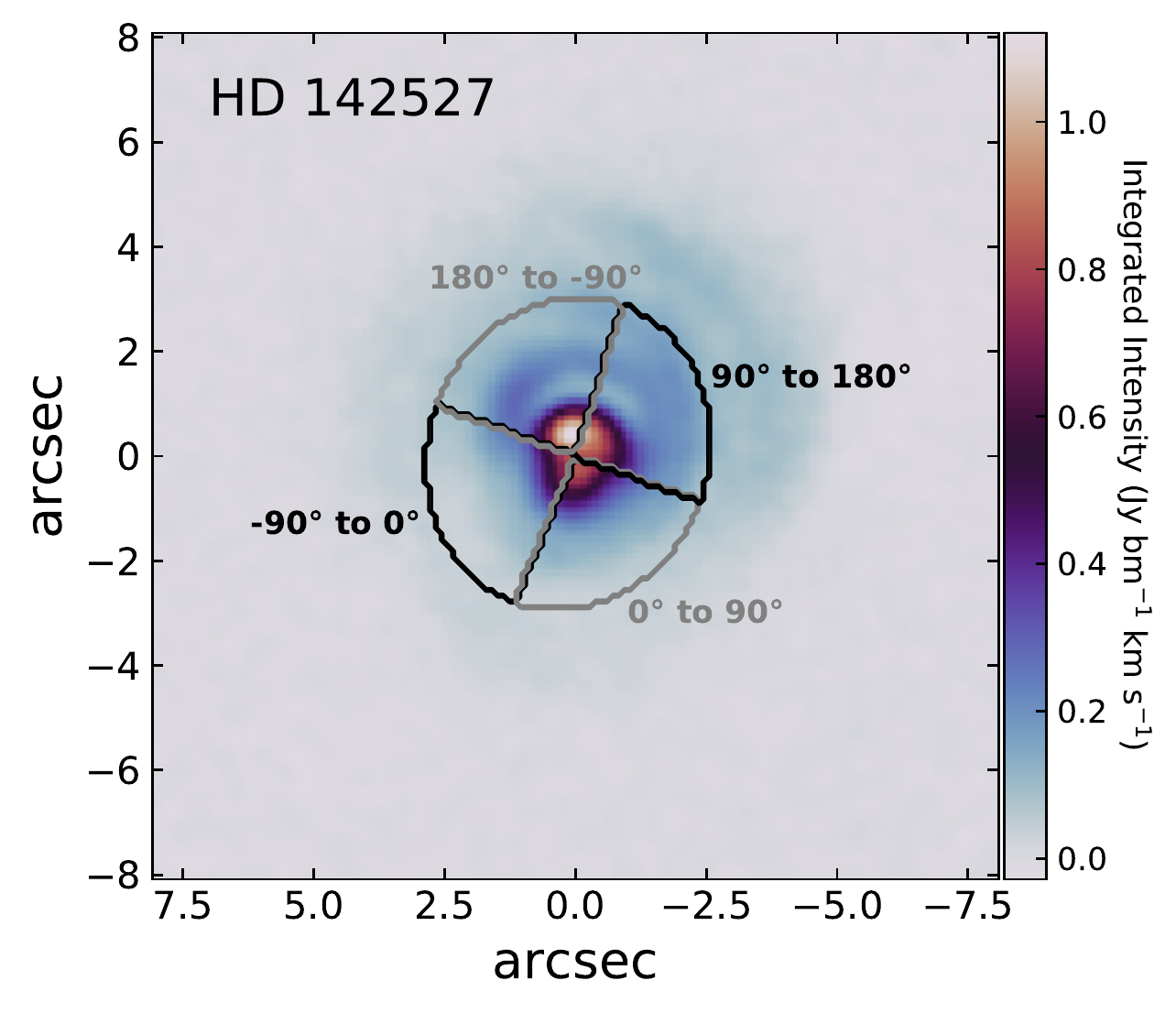}
\includegraphics[width=0.49\columnwidth]{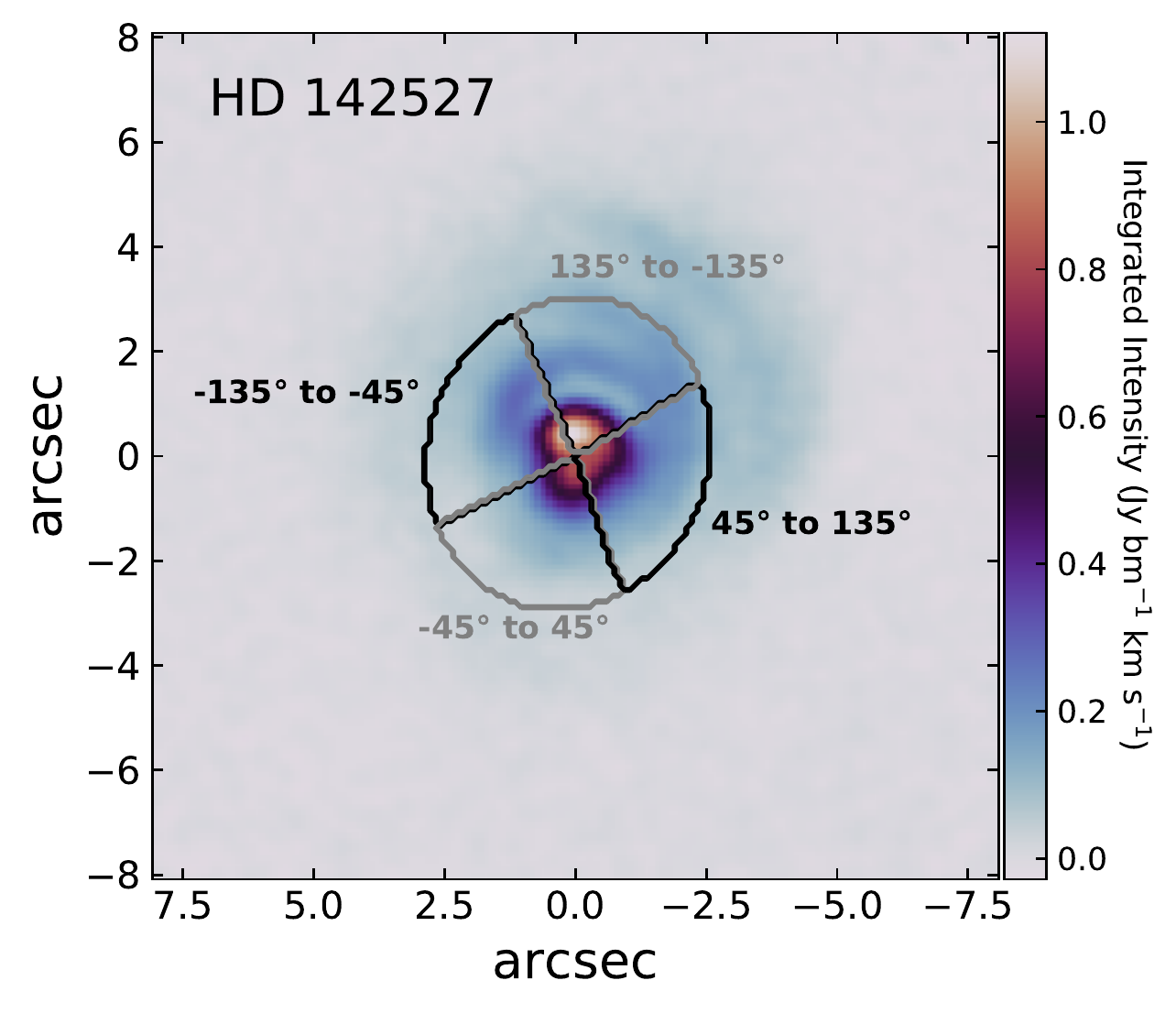}
\includegraphics[width=0.49\columnwidth]{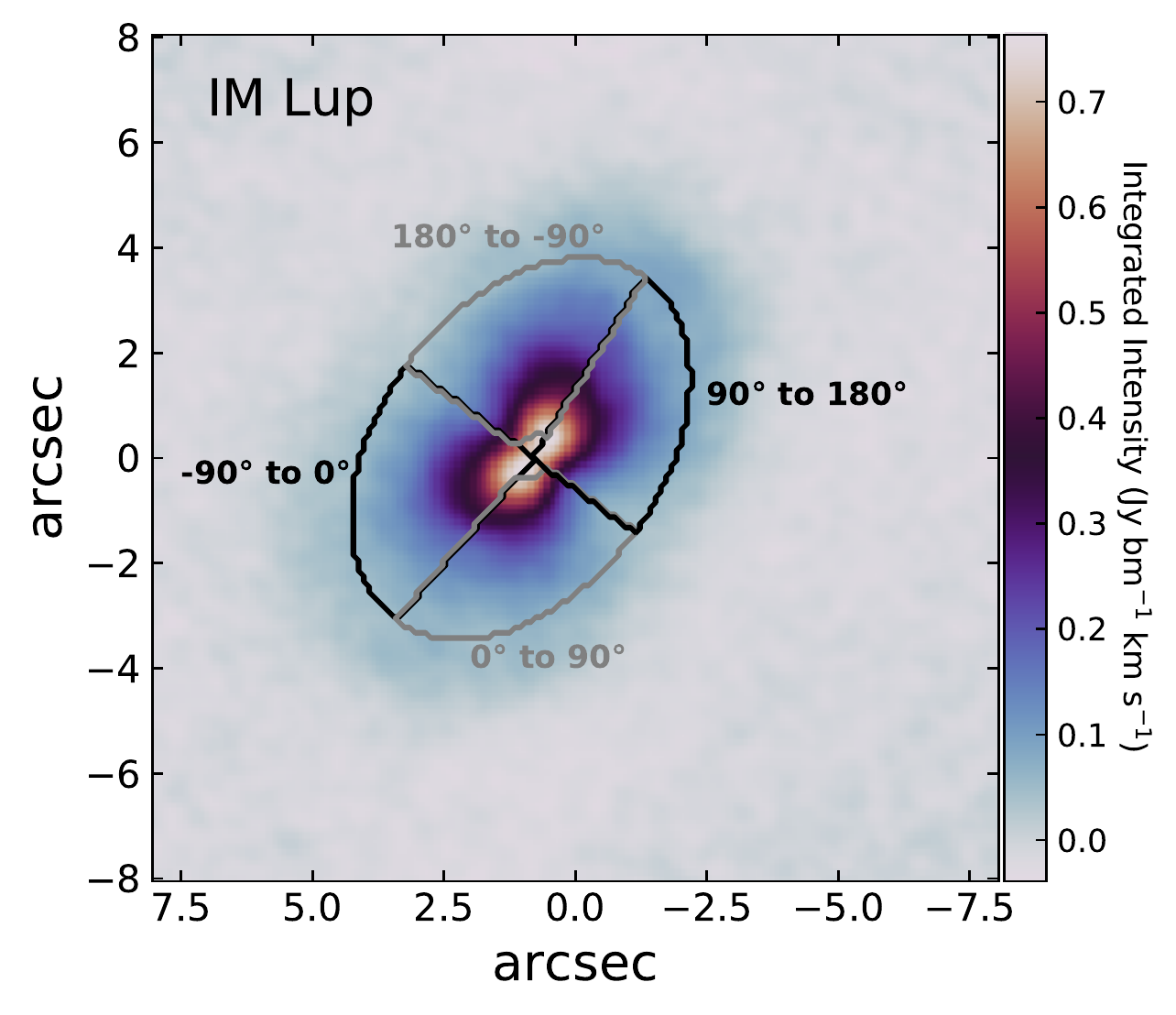}
\includegraphics[width=0.49\columnwidth]{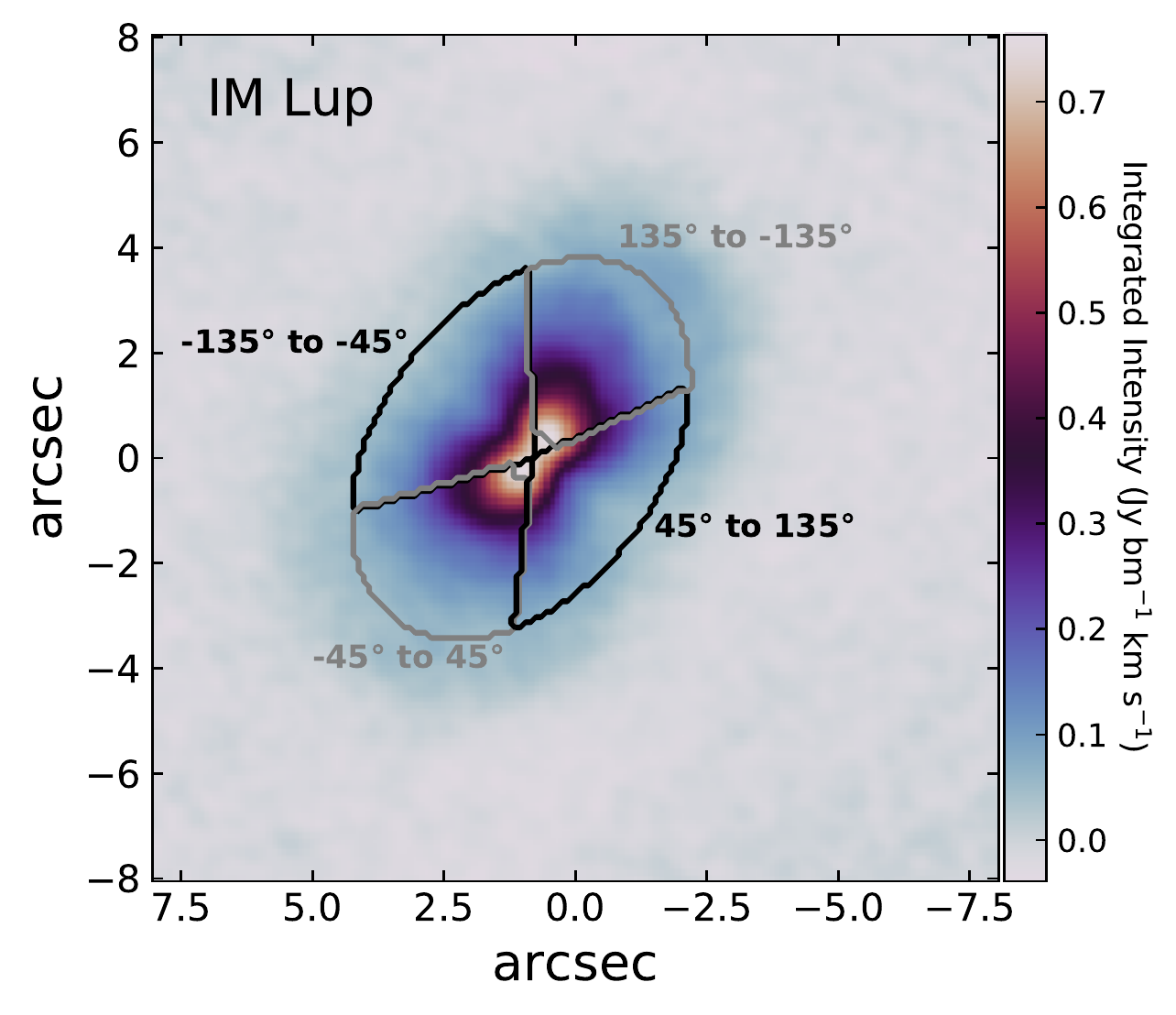}
\includegraphics[width=2\columnwidth]{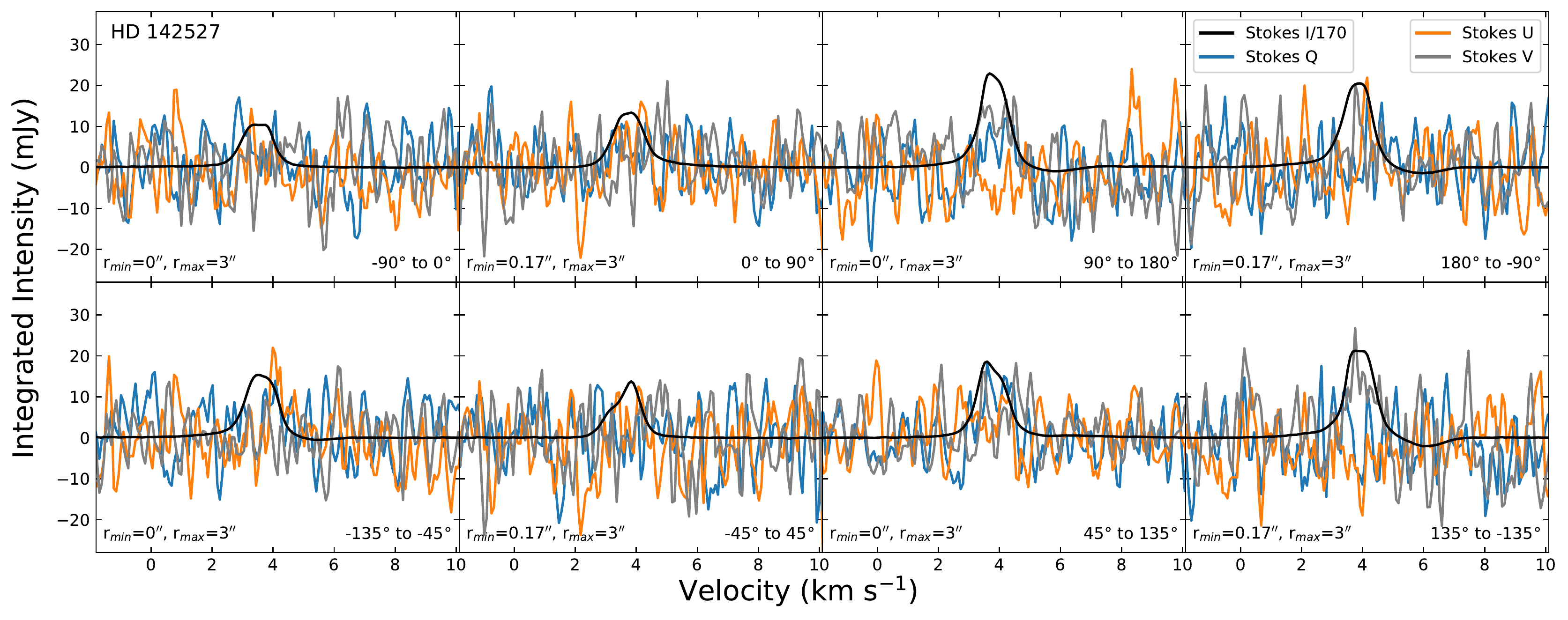}
\includegraphics[width=2\columnwidth]{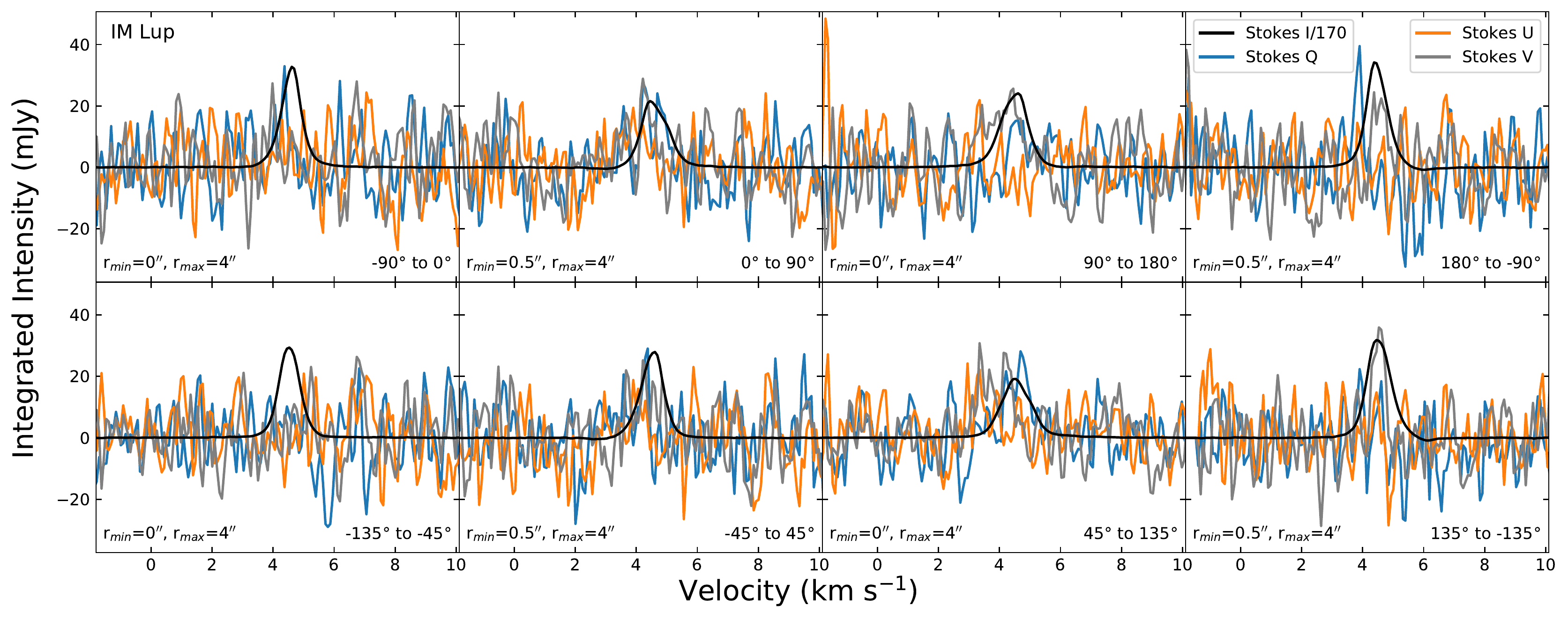}
\end{center}
\caption{Stacking based on Keplerian rotation of selected quadrants (90$^\circ$ wedges in disk plane) via \texttt{GoFish}. The top row shows selected quadrants overlaid on the moment~0 maps (integrated over same velocity interval as Figures~\ref{fig:hd142527_mom0} and~\ref{fig:imlup_mom0}). The left two images show the quadrants for \hd, and the right two images show quadrants for \im. The number range next to each quadrant correspond to the quadrant's angle range from the disk's axis, starting from PA$_{\text{red}}$. The middle row shows 8 panels where we stack for the quadrants shown in the top row for \hd. The top four of these panels correspond to the first \hd\ image, while the bottom four correspond to the second image. The bottom left of each of the 8 panels shows the minimum and maximum radii, $r_{\text{min}}$ and $r_{\text{max}}$, used for the quadrant. The bottom right of each panel shows the angle range of the quadrants indicated in the top panels. The 8 panels on the bottom row are like the middle row, except for \im.
}
\label{fig:gofish_everything} 
\end{figure*}

As mentioned in Section~\ref{sec:model}, for elliptical or radial polarization patterns, averaging polarization over the entire disk causes the fluxes of the Stokes~$Q$ and~$U$ parameters to be averaged out, as there are both positive and negative components (Figures~\ref{fig:hd142527_model} and~\ref{fig:imlup_model}). Therefore, we also use \texttt{GoFish} to stack for different quadrants (90$^\circ$ wedges) for each disk, as shown for \coto\ in Figure~\ref{fig:gofish_everything}. We choose four 90$^\circ$ quadrants, where each 90$^\circ$ quadrant is in reference to the disk's axis rather than in projection. Since the Stokes~$Q$ and~$U$ contours each draw a 4-petal daisy-like pattern that offset from each other by $\sim$45$^\circ$ (Figure~\ref{fig:hd142527_model} and~\ref{fig:imlup_model}), we select four 90$^\circ$ quadrants where we start at PA$_{\text{red}}$, and another four where we start at PA$_{\text{red}}$~$-$~45$^\circ$. Also, for elliptical polarization, the central region is expected to be negative for Stokes~$Q$ and positive for Stokes~$U$, which is especially evident for the inner 0$\farcs$5 for \im\ (Figure~\ref{fig:imlup_model}). As such, for alternating quadrants, we shift between $r_{\text{min}}$~=~0$\arcsec$ and 0$\farcs$17 for \hd, and $r_{\text{min}}$~=~0$\arcsec$ and 0$\farcs$5 for \im. The quadrants with a non-zero $r_{\text{min}}$ are drawn in gray in Figure~\ref{fig:gofish_everything}. Once again, we select $r_{\text{max}}$ of 3$\arcsec$ for \hd\ and 4$\arcsec$ for \im. 

No significant detection was found for any quadrant. Our attempts using other many other values for $r_{\text{min}}$ and $r_{\text{max}}$ (not shown) also did not detect any significant signal in any of the Stokes parameters. We also show the resulting non-detections for \ttco\ and \ceo\ in the Appendix.  We discuss the constraints we can place on \pfrac\ from stacking (for the entire disk and for quadrants) in Section~\ref{sec:constrain}.

%hd142527
%	distance = 157 #+-1 pc, Arun+2019. gaia
%	mass = 2.2*(distance/140.)**2 #Perez+2015
%	red_PA = 160 #Perez+2015 assumes minor axis PA of 70. Red is in quadrant 3.
%	inclination = 28 #Perez+2015
%	odd_rmin=0
%	even_rmin=0.17
%	rmax=5
%	z0=0.1
%	psi=1
%	vguess = 3.6

%imlup
%	distance = 158 #+-3 pc, Andrews+2018. Quotes a paper i cannot find
%	mass = 1 #Andrews2018 is 10^-0.05
%	red_PA = 144 #Cleeves+2016, red major.
%	inclination = 48 #Cleeves+2016
%	odd_rmin=0
%	even_rmin=0.5
%	rmax=5
%	cubeI = imagecube('imlup/imlup.'+line+'.StokesI.image.pbcor.fits',clip=8)	
%	cubeQ = imagecube('imlup/imlup.'+line+'.StokesQ.image.pbcor.fits',clip=8)	
%	cubeU = imagecube('imlup/imlup.'+line+'.StokesU.image.pbcor.fits',clip=8)
%	cubeV = imagecube('imlup/imlup.'+line+'.StokesV.image.pbcor.fits',clip=8)
%	z0=0.1 #Cleeves=2016 find 12 au at 100 au, so kind of similar
%	psi=1 #Cleeves+2016
%	vguess = 4.5

\subsection{Searching for Signal within an Aperture}\label{sec:searchap}
For each spectral line, we also manually inspected the Stokes~$Q$, $U$, and $V$ cubes for a signal. This search was done by looking at the average spectra within a circular aperture. The size and position of the circular aperture was changed, allowing us to carefully inspect Stokes~$Q$, $U$, and $V$ spectra across the entire disk in which there is Stokes~$I$ signal. Using this method, we found that for \coto, there appears to be a signal in the Stokes~$Q$ cube toward both disks. 

We then developed a script that alters the size and position of the circular aperture to find the location with the highest signal to noise. Specifically, we found the location where SNR$_Q$~=~$| \Sigma Q |$/$\sigma_{\Sigma Q}$ was maximum, where $| \Sigma Q |$ is the magnitude of the sum of Stokes~$Q$ for $N$ consecutive channels and $\sigma_{\Sigma Q}$ is the error on the sum. As stated in Section~\ref{sec:characterization}, there exists a covariance between consecutive channels that is 0.3 times the variance in the spectrum. When accounting for these covariances, $\sigma_{\Sigma Q} = \sigma_{Q,ap} \sqrt{1.6N - 0.6}$, where $\sigma_{Q,ap}$ is the noise of Stokes~$Q$ in a single channel for the spectra for an aperture. When searching for the highest SNR$_Q$ within a circular aperture, the aperture was centered on a pixel and radii had sizes that were integers.

%Q_file="hd142527.co_21.StokesQ.real.updatedstdQ.huge.SNR.fits"
%from astropy.io import fits
%from astropy.stats import sigma_clipped_stats
%hdulistQ = fits.open(Q_file)
%print(sigma_clipped_stats(hdulistQ[0].data))
%## -- End pasted text --
%WARNING: Input data contains invalid values (NaNs or infs), which were automatically clipped. [astropy.stats.sigma_clipping]
%(5.4689603, 5.3031383, 0.9816522)

%hd142527_detected_spectra.py
%technically cov, but figure the same....
\begin{figure}[ht!]
\begin{center}
\includegraphics[width=1\columnwidth]{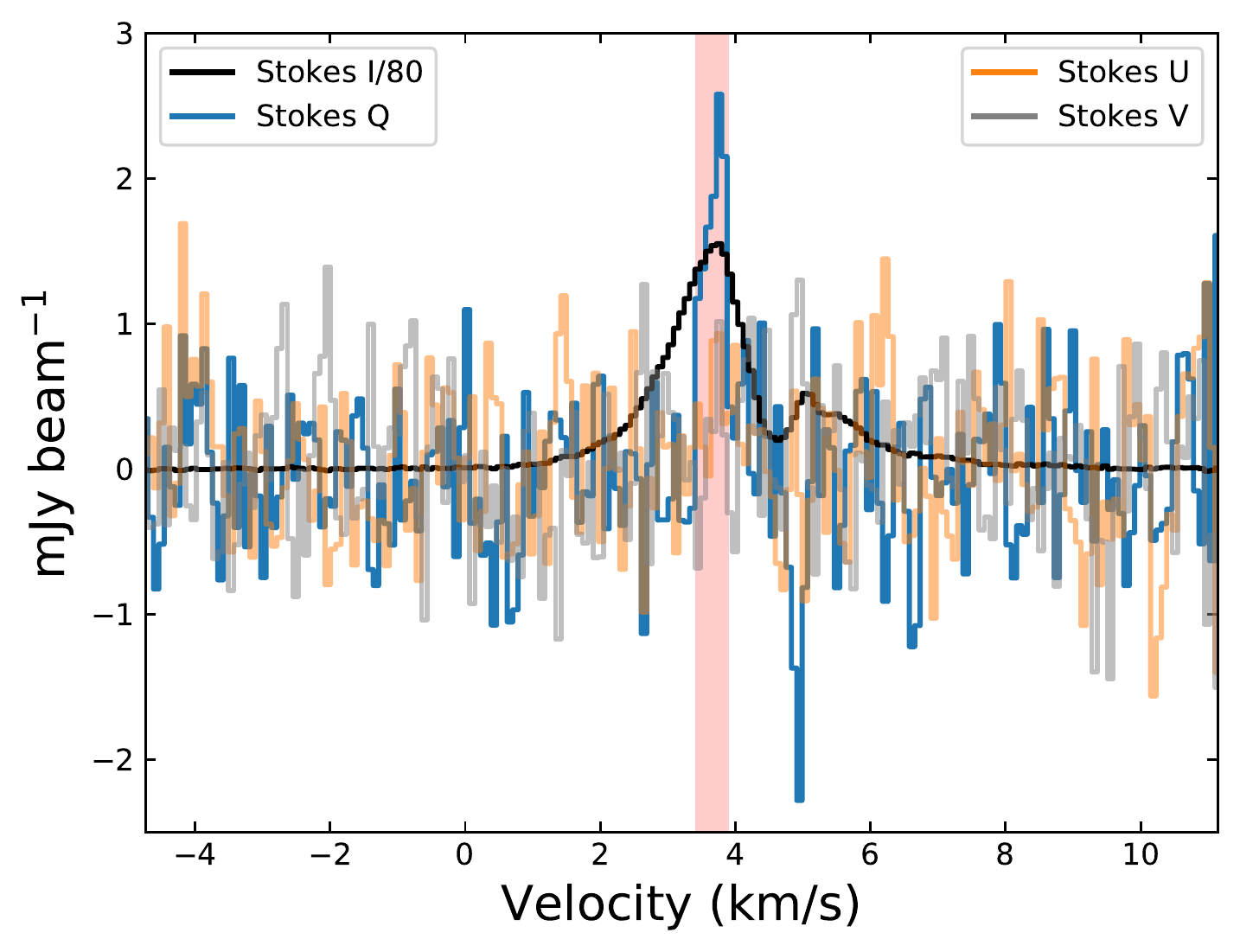}
\end{center}
\caption{Average \hd\ \coto\ Stokes~$IQUV$ spectra for the circle shown in Figure~\ref{fig:hd142527_spectra_detected_IQUV}. Stokes~$I$ has been divided 80. The pink shaded area shows the velocity range in which we integrated the signals.
}
\label{fig:hd142527_spectra_detected} 
\end{figure}

\begin{figure}[ht!]
\begin{center}
\includegraphics[width=1\columnwidth]{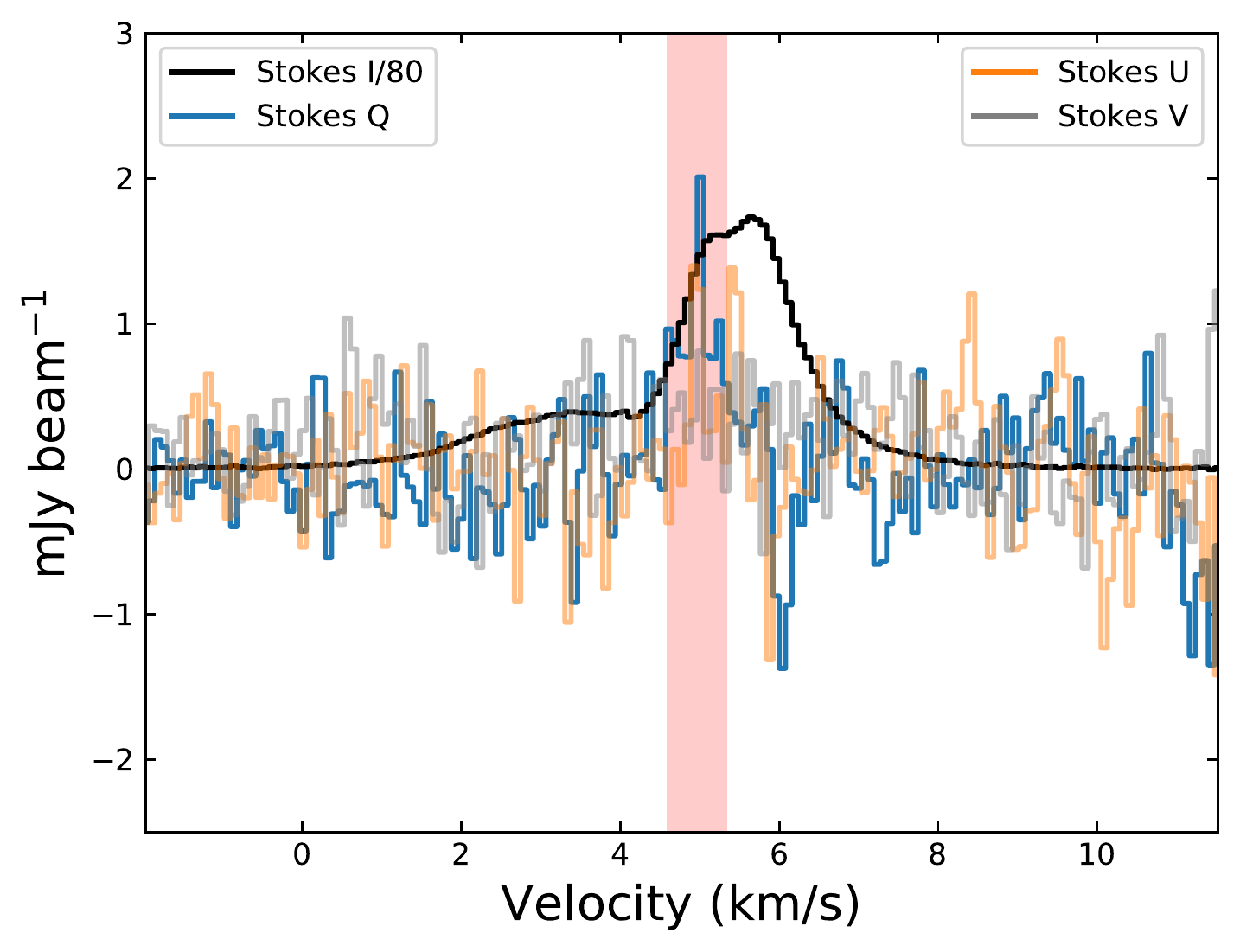}
\end{center}
\caption{Average \im\ \coto\ Stokes~$IQUV$ spectra for the circle shown in Figure~\ref{fig:imlup_spectra_detected_IQUV}. Stokes~$I$ has been divided 80. The pink shaded area shows the velocity range in which we integrated the signals.
}
\label{fig:imlup_spectra_detected} 
\end{figure}

%hd142527_IQUV_mom0cut.py
\begin{figure*}[ht!]
\begin{center}
\includegraphics[width=2\columnwidth]{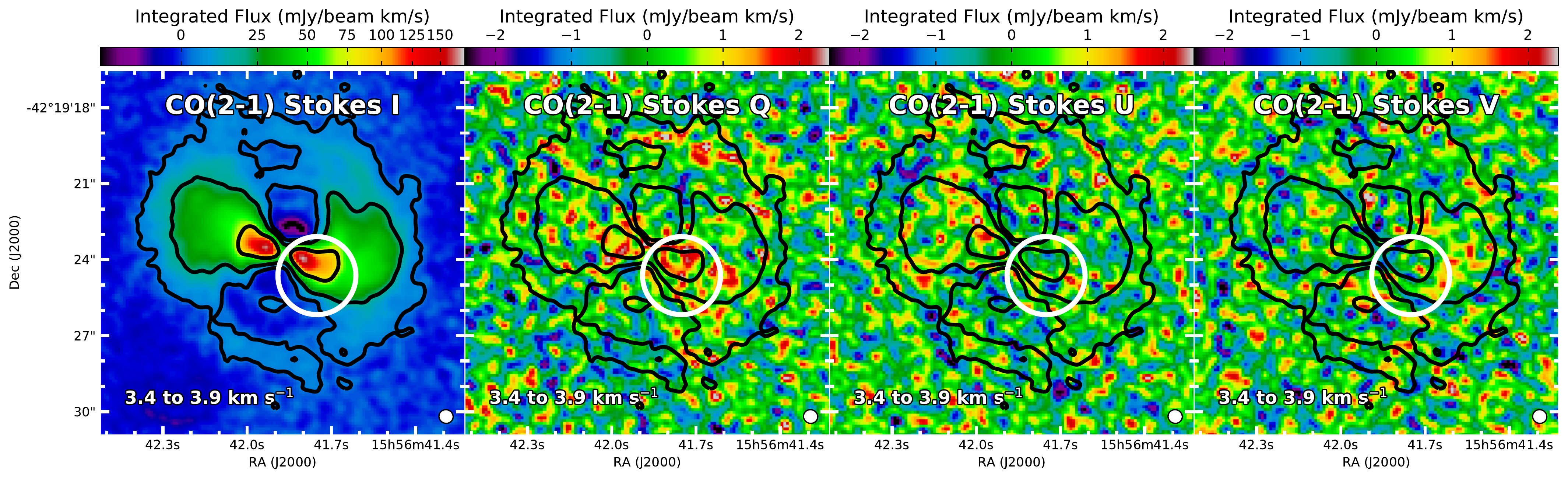}
\end{center}
\caption{\hd\ $IQUV$ moment~0 maps over the velocity range shown in the bottom left of the panel, which is the pink shaded area shown in Figure~\ref{fig:hd142527_spectra_detected}. The spectra in Figure~\ref{fig:hd142527_spectra_detected} is the average spectra taken at the location of the white circle.}
\label{fig:hd142527_spectra_detected_IQUV} 
\end{figure*}

%imlup_IQUV_mom0cut.py
\begin{figure*}[ht!]
\begin{center}
\includegraphics[width=2\columnwidth]{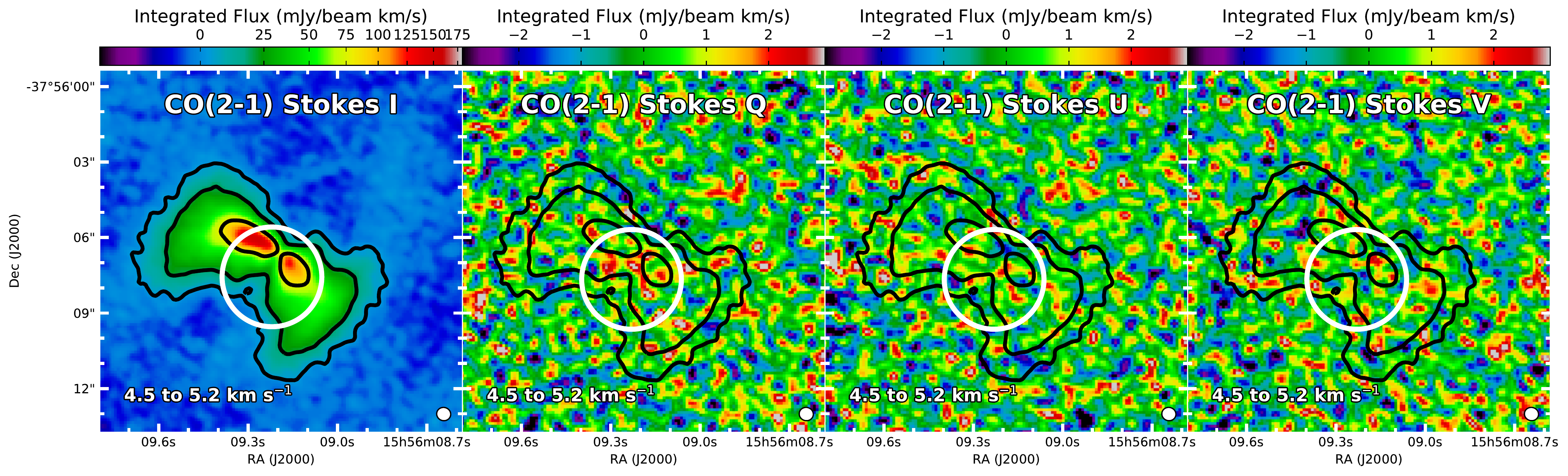}
\end{center}
\caption{\im\ $IQUV$ moment~0 maps over the velocity range shown in the bottom left of the panel, which is the pink shaded area shown in Figure~\ref{fig:imlup_spectra_detected}. The spectra in Figure~\ref{fig:imlup_spectra_detected} is the average spectra taken at the location of the white circle.
}
\label{fig:imlup_spectra_detected_IQUV} 
\end{figure*}

Figures~\ref{fig:hd142527_spectra_detected} and~\ref{fig:imlup_spectra_detected} show the spectra of the four Stokes parameters where SNR$_Q$ is maximum for \hd\ and \im, respectively. These displayed spectra are the average spectra within the circular aperture, and the channels that were integrated to obtain the maximum SNR$_Q$ are shown in pink. For \hd\ and \im, the values for $\Sigma Q$ over the pink areas in Figures~\ref{fig:hd142527_spectra_detected} and~\ref{fig:imlup_spectra_detected} are 10.8\,$\pm$\,1.2\,mJy\,bm$^{-1}$ (SNR$_Q$~=~8.7) and 9.9\,$\pm$\,1.0\,mJy\,bm$^{-1}$ (SNR$_Q$~=~9.6), respectively. For \hd\ and \im, the values for $\Sigma U$ in the same areas are 2.7\,$\pm$\,1.5\,mJy\,bm$^{-1}$ and 3.7\,$\pm$\,1.2\,mJy\,bm$^{-1}$, respectively. At these locations for \hd\ and \im, the de-biased \pfrac\ is 1.56\,$\pm$\,0.18\% and 1.01\,$\pm$\,0.10\%, respectively. The position angles $\chi$ are 6$\fdg$9\,$\pm$\,3\fdg9 and 10$\fdg$1\,$\pm$\,3$\fdg$2, respectively.

We create moment~0 maps for each of the four \coto\ Stokes parameters integrated about the pink shaded velocity ranges. These maps are shown in Figures~\ref{fig:hd142527_spectra_detected_IQUV} and~\ref{fig:imlup_spectra_detected_IQUV} for \hd\ and \im, respectively. The circular aperture indicates the area where we show the average spectra in Figures~\ref{fig:hd142527_spectra_detected} and~\ref{fig:imlup_spectra_detected}. For \hd, there is elevated emission across the major axis of the Stokes~$Q$ image; given the sensitivity in the $QUV$ moment~0 maps is $\sigma$~=~0.78\,mJy\,bm$^{-1}$\,\kms, anything that is red-colored is above 2$\sigma$ and white-colored is above 3$\sigma$. While the $>$2$\sigma$ pixels are blotchy rather than a smooth cohesive structure, we note that for an image that is purely noise, only 2.3\% of the area is expected to be above the 2$\sigma$ value. Since the Stokes~$Q$ image of \hd\ has a region where substantially more than 2.3\% of the area is above 2$\sigma$, this is likely a robust detection. Similarly, the red and white colors for \im\ (Figure~\ref{fig:imlup_spectra_detected}) is above 2$\sigma$ and 3$\sigma$, respectively, as the $QUV$ moment~0 maps is $\sigma$~=~0.96\,mJy\,bm$^{-1}$\,\kms.

%plot_SNR_plots.py
\begin{figure*}[ht!]
\begin{center}
\includegraphics[width=0.9\columnwidth]{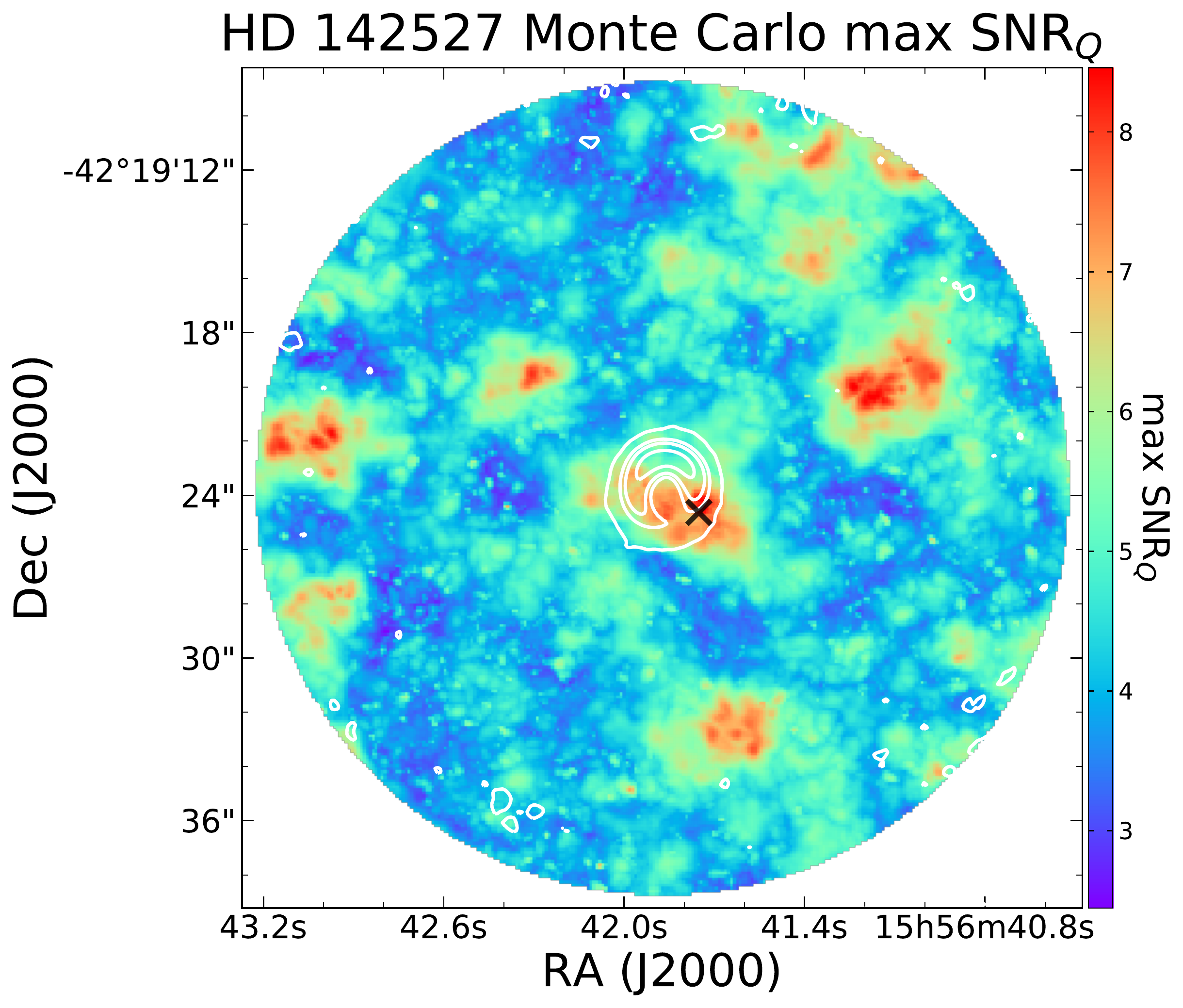}~~~~~~~
\includegraphics[width=0.9\columnwidth]{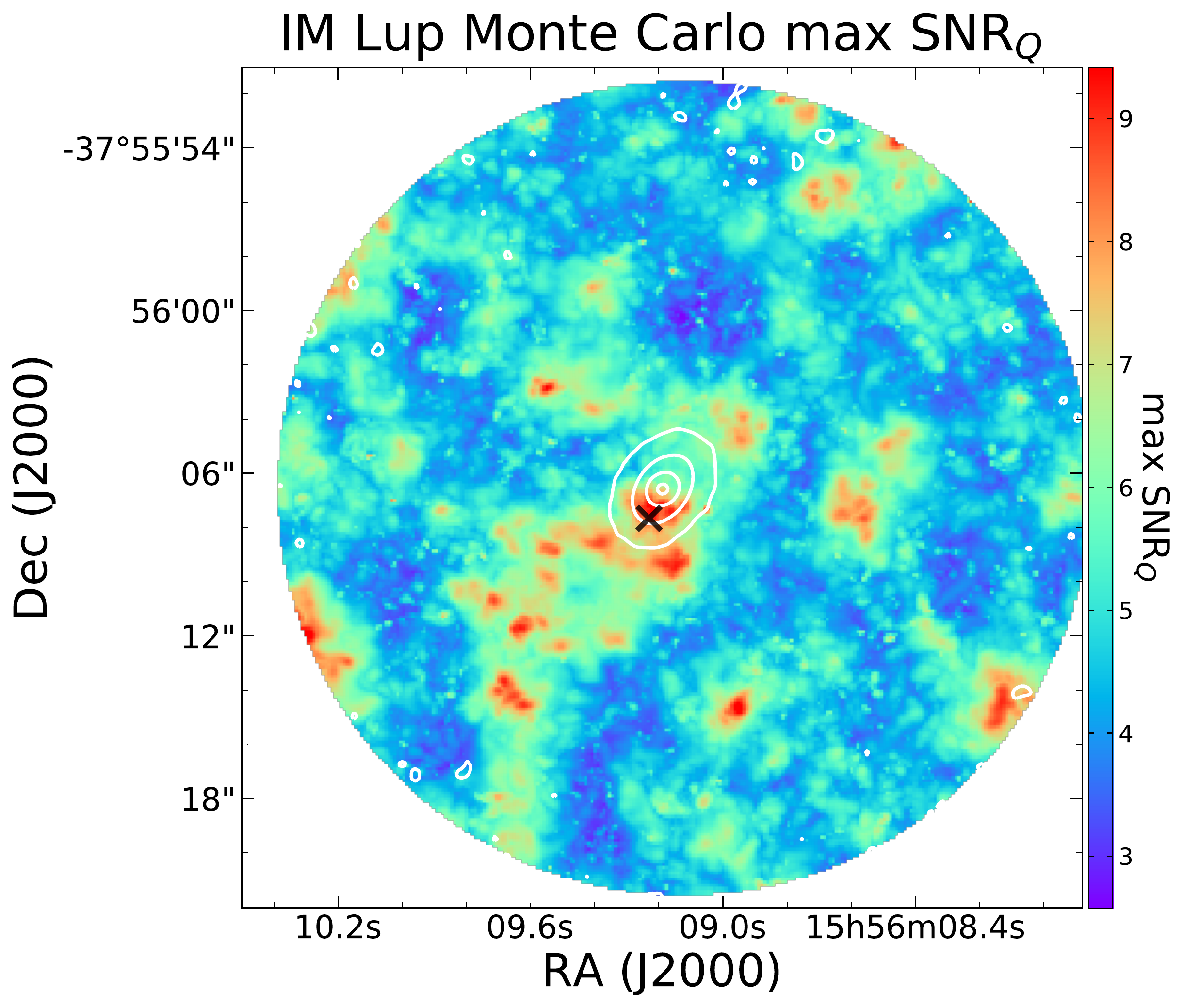}
\end{center}
\caption{Monte Carlo simulation results showing the maximum \coto\ SNR$_Q$ detected for each pixel for \hd\ (left) and \im\ (right). To determine these values, at each pixel we spatial averaged spectra within circular apertures between 0$\farcs$11~$\times$~3$\arcsec$ and considered spectral window sizes between $\sim$0.16 and 2.5\,\kms. Of all possible combinations, we picked the largest value, i.e., the maximum SNR$_Q$. See Section~\ref{sec:searchap} for more details. White contours show the 1.3\,mm Stokes~$I$ contours, with levels of  [4, 150, 500, 2000]~$\times$~$\sigma_{I,\rm{1.3\,mm}}$, where $\sigma_{I,\rm{1.3\,mm}}$~=~37\,$\mu$Jy\,bm$^{-1}$ for \hd\ and 22\,$\mu$Jy\,bm$^{-1}$ for \im. The black crosses in each panel show the the location of the maximum SNR$_Q$ detected; these spectra are shown for \hd\ and \im\ in Figures~\ref{fig:hd142527_spectra_detected} and~\ref{fig:imlup_spectra_detected}, respectively. }
\label{fig:SNRQ} 
\end{figure*}

Since we used an algorithm that optimizes a circular aperture to have the best SNR as possible, we wanted to confirm that this algorithm does not create a false signal. We used a Monte Carlo simulation to find the typical SNR that one would expect at each pixel of our \coto\ Stokes~$Q$ maps. As above, for each 0$\farcs$11 pixel, we searched for the highest SNR$_Q$ possible for a range of circular apertures and channels. We focused on pixels within a radius of 15$\arcsec$ from the center of the map, which allows us to probe well beyond the disks where Stokes~$Q$ is expected to be pure noise. We tried circular apertures with integer radii sizes between 1 and 27~pixels (0$\farcs$11~$\times$~3$\arcsec$). For every circular aperture, we looked for the highest SNR$_Q$ possible for 2 to 32 consecutive channels (i.e., a window of sizes between $\sim$0.16 and 2.5\,\kms), restricting the search for a signal within a $v_{\text{lsr}}$ velocity range of 0~to~7\,\kms\ for \hd\ and 1~to~8\,\kms\ for \im. We chose a maximum window of 2.5\,\kms\ because this captures the majority of Stokes~$I$ for any pixel, and we chose the $v_{\text{lsr}}$ ranges because they capture the vast majority of the Stokes~$I$ emission across the entire map.

Figure~\ref{fig:SNRQ} shows the maximum SNR$_Q$ at each pixel based on these Monte Carlo simulations. For \hd, the mean SNR$_Q$ for each pixel is 4.7 with a standard deviation of 0.9. The signal shown in Figure~\ref{fig:hd142527_spectra_detected} is indeed the pixel with the highest SNR$_Q$ found (SNR$_Q$~=~8.7). However, there is a signal off the disk with comparable signal (SNR$_Q$~=~8.6). For \im, the mean SNR$_Q$ for each pixel is 5.0 with a standard deviation of 1.0, and the signal shown in Figure~\ref{fig:imlup_spectra_detected} (SNR$_Q$~=~9.6) is not the pixel with the highest SNR$_Q$. Instead, a few pixels off of the disk have slightly higher values of SNR$_Q$, with values up to 10.1. We note that the distribution of the maximum SNR$_Q$ for all pixels is not Gaussian, but rather positively skewed. Figure~\ref{fig:SNRQ} shows that there exists high SNRs well outside of both disks.

Based on these results, we consider these Stokes~$Q$ detections as marginal. While the signals in Figures~\ref{fig:hd142527_spectra_detected} and~\ref{fig:imlup_spectra_detected} appear to be robust by eye, and we have characterized the noise as well-behaved (Section~\ref{sec:characterization}), our Monte Carlo simulations show that our methodology of finding this signal can find signals that are almost certainly false as well. If we repeat the same Monte Carlo simulations toward these disks for Stokes~$U$, the highest values of SNR$_U$ within the areas of the gaseous disks are 6.9 and 8.0 for \hd\ and \im\ respectively. Since \coto\ Stokes~$U$ maps share the same noise characteristics as Stokes~$Q$ (Section~\ref{sec:characterization}), these lower values indicate that the marginal detections of Stokes~$Q$ could indeed be real. 
{\bf Nevertheless, finding a marginal signal for each disk only for \coto\ Stokes~$Q$ is peculiar and may indicate a more complex noise behavior that could not be found with our tests.}
%Nevertheless, the fact that these marginal detections for each disk are only found for \coto\ Stokes~$Q$ is peculiar

As seen in Figures~\ref{fig:hd142527_spectra_detected} and~\ref{fig:imlup_spectra_detected}, for both \hd\ and \im, there is also a negative dip in the Stokes~$Q$ spectra at a velocity $\sim$1~\kms\ higher than the pink shaded area. If we again search the nearby vicinity for the maximum SNR$_Q$ for different apertures at these negative dips, we find the maximum SNR$_Q$ for \hd\ and \im\ are 5.8 and 8.2 respectively. Based on our Monte Carlo simulations, these are marginal at best. However, the fact that the Stokes~$Q$ spectra for both of these disks show a negative dip at the same relative velocity is intriguing, and perhaps indicates a flip in polarization angle at the higher velocity. Nevertheless, given these data, we cannot strongly support this conjecture.

While we detect positive Stokes~$Q$ toward parts of the disk, which is suggestive of mostly north-south polarization, we do not detect polarized emission over a large enough area or velocity range to accurately discern any sort of disk polarization morphology. Moreover, trying to infer the magnetic field morphology from these observations is too difficult, given the ambiguity of the GK effect (polarization can be parallel or perpendicular to the magnetic field) and possible effects from resonant scattering \citep{Houde2013}.

\section{Constraining the Polarization Percentage of the Observations} \label{sec:constrain}
%upperlimits.py
\begin{figure*}[ht!]
\begin{center}
\includegraphics[width=1.8\columnwidth]{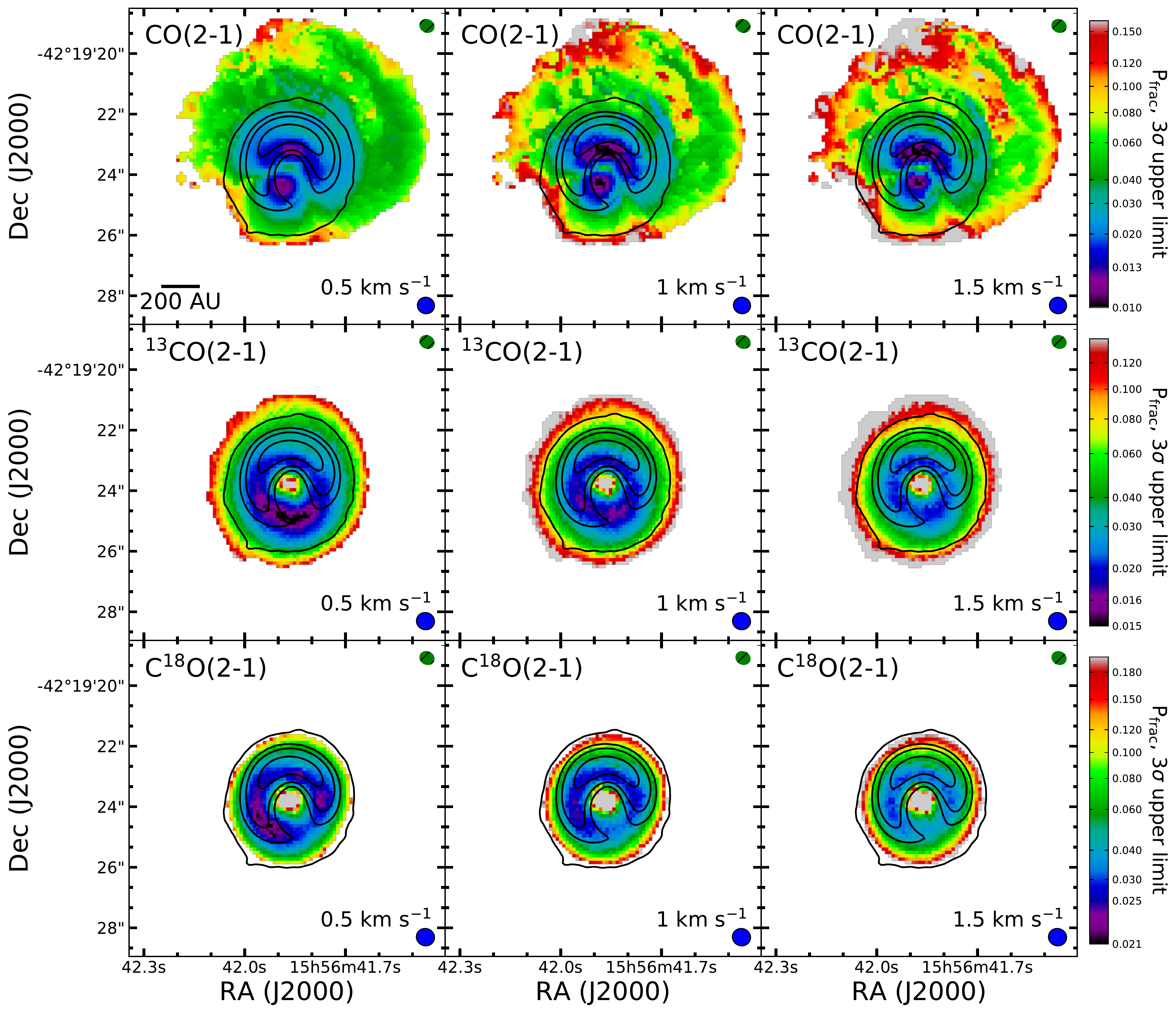}
\end{center}
\caption{\hd\ maps of the upper limit for \pfrac\ at each pixel (see Section~\ref{sec:constrain}) with 3 different spectral lines and velocity bins. From top to bottom, the rows are for \coto, \ttco, and \ceo. From left to right, the columns are for velocity bins of 0.5\,\kms, 1.0\,\kms, and 1.5\,\kms. Each row shares the same color bar. Black contours show the 1.3\,mm continuum, with levels of [4, 150, 500, 2000]~$\times$~$\sigma_I$, where $\sigma_I$~=~42\,$\mu$Jy\,bm$^{-1}$.
}
\label{fig:hd142527_upper} 
\end{figure*}

%upperlimits.py
\begin{figure*}[ht!]
\begin{center}
\includegraphics[width=1.8\columnwidth]{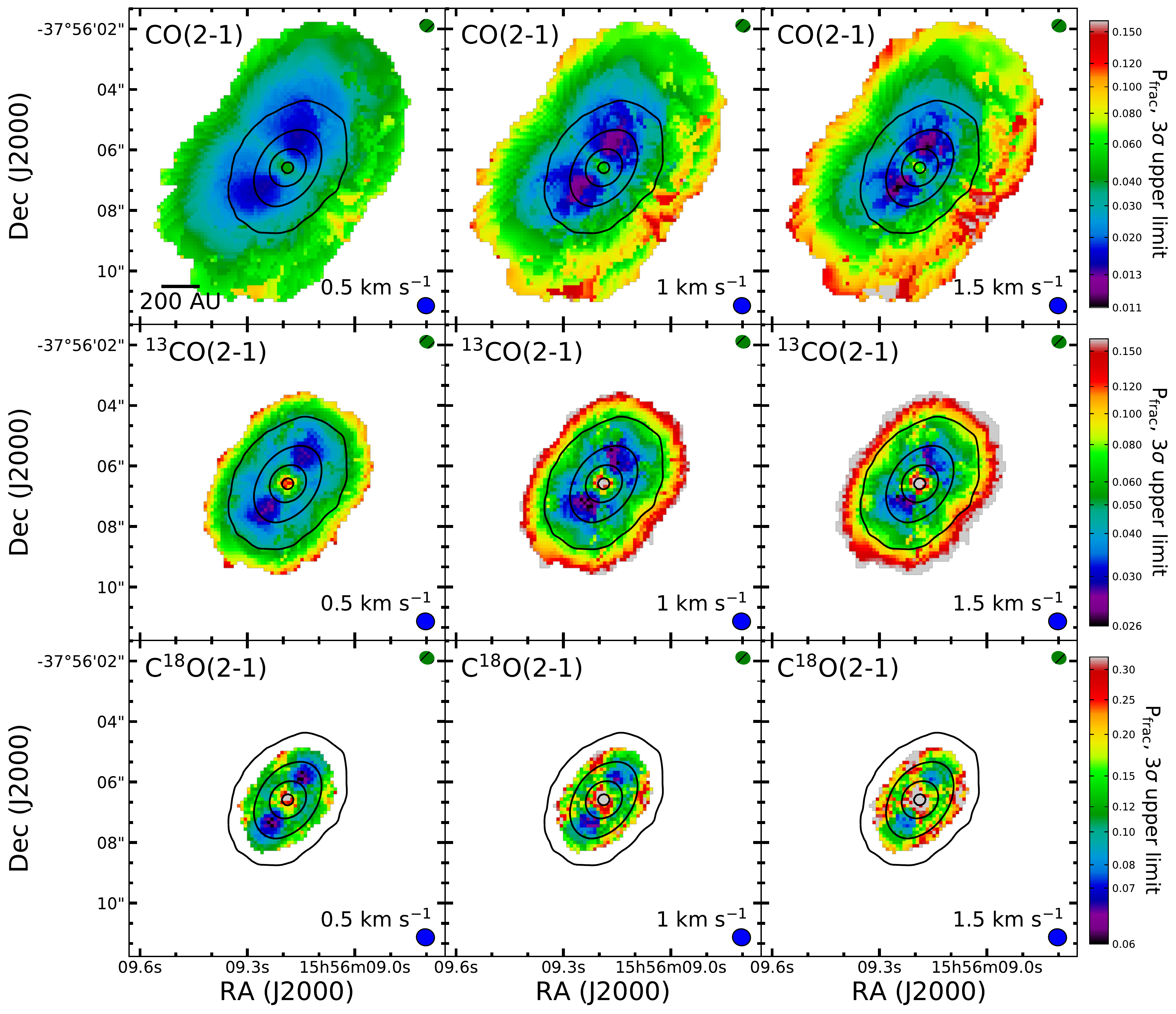}
\end{center}
\caption{Same as Figure~\ref{fig:hd142527_upper}, but now for \im\ and $\sigma_I$~=~22\,$\mu$Jy\,bm$^{-1}$.
}
\label{fig:imlup_upper} 
\end{figure*}

Although our manual inspection method detected what seems to be a marginal signal in \coto\ for \hd\ and \im, the polarization is only detected for Stokes~$Q$ and is only for select areas and velocities. To motivate future observations, it would make sense to put upper limits on the polarization percentage for each spectral line. Unlike continuum polarization, specifying the upper limit for \pfrac\ for disks is not straightforward, as it requires specifying both the integrated velocity range and location. The dominant Stokes~$I$ emission of an inclined, rotating disk as a function of velocity will (1) start on the blue-shifted side of the major axis, (2) continue toward the minor axis, and (3) finish on the red-shifted side of the major axis. Thus, integrating emission over the entire disk makes less sense than integrating over finer velocity ranges since the entire range would add noise where little to no Stokes~$I$ emission is detected.

For each spectral line, we make maps of the 3$\sigma$ upper limit for \pfrac\ by calculating these upper limits for each pixel for a variety of velocity intervals. We first mask out pixels where the moment~0 maps (Figures~\ref{fig:hd142527_mom0} and~\ref{fig:imlup_mom0}) are detected with a signal-to-noise of less than 10. This effectively masks out areas with high \pfrac\ upper limits, which aids in displaying the upper limit maps with a reasonable stretch. Then for the Stokes~$I$ spectra of each remaining pixel, we find where the summation of $N$ channels is maximum ($\Sigma I$) Taking into account the correlation between channels, we determine the 3$\sigma$ upper limit to $P_{\text{frac}}$ to be

\begin{equation}\label{eq:upper}
%P_{\text{frac},3\sigma} = 3\sigma_{\text{PI}} \frac{\sqrt{N}}{\Sigma I} ,
P_{\text{frac},3\sigma} = 3\sigma_{\text{PI}} \frac{\sqrt{1.6N - 0.6}}{\Sigma I} ,
%P_{\text{frac}, 3\sigma}} = 3\sigma_{\text{PI}} \frac{\sqrt{N}}{\Sigma I} ,
\end{equation}
%3$\sigma_{\text{PI}}\sqrt{N}/ \Sigma I$, 
where the polarized intensity sensitivity $\sigma_{\text{PI}}$ = $\sigma_Q$ = $\sigma_U$. Since we choose the maximum $\Sigma I$ at each pixel, this method is effectively the ``minimum" upper limit at each pixel for the given spatial resolution. Note that minimum upper limits could be reduced further if we did spatial smoothing. 

We generate multiple 3$\sigma$ upper limit maps, each for $N$ channels spanning velocity intervals that are a factor of 0.5\,\kms. We show the result for 0.5, 1, and 1.5\,\kms\ for \hd\ and \im\ in Figures~\ref{fig:hd142527_upper} and~\ref{fig:imlup_upper}, respectively. Any higher velocity intervals (e.g., 2, 2.5, and 3\,\kms) largely result in much worse upper limit sensitivities across the map. In general, the 0.5\,\kms\ interval performs better (i.e., lower upper limits) toward the outer part of the disk while the 1 and 1.5\,\kms\ intervals perform better for the inner part of the disk.

The upper limits vary drastically across the map for each disk, spectral line, and velocity bin. For the brightest continuum spots of \hd, the polarization 3$\sigma$ upper limits are typically less than 3\% for \coto\ and \ttco, and 4\% for \ceo. For the brightest continuum spots of \im, the polarization 3$\sigma$ upper limits are typically less than 3\% for \coto, 4\% for \ttco, and 12\% for \ceo. While these are typical values, one can see from these Figures that toward some areas of the disks, the 3$\sigma$ upper limit can be over a factor of two lower.

%could smooth over 3 arcseconds and repeat

\begin{deluxetable}{lcccc}
\label{tab:gofish}
\tablewidth{0pt}
\tablecolumns{4}
\tabletypesize{\scriptsize}
\tablecaption{\texttt{GoFish} $P_{\text{frac}}$  3$\sigma$ Upper Limits}
\tablehead{
\colhead{} & \multicolumn{2}{c}{\hd} & \multicolumn{2}{c}{\im}  \vspace{-8pt} \\
\colhead{Line} & \colhead{Whole Disk} & \colhead{Quadrants} & \colhead{Whole Disk} & \colhead{Quadrants} 
}
\startdata
\coto & 0.2\% -- 0.3\% & 0.3\% -- 0.9\% & 0.2\% -- 0.3\% & 0.3\% -- 0.5\% \\
\ttco &0.3\% -- 0.4\% & 0.4\% -- 0.9\% & 0.3\% -- 0.5\% &  0.6\% -- 1.0\% \\
\ceo &0.4\% -- 0.5\% & 0.5\% -- 1.1\% & 1.1\% -- 1.4\% & 1.9\% -- 3.3\% 
\enddata 
\tablecomments{Stacking constraints using \texttt{GoFish} (Section~\ref{sec:gofish}). Minimum and maximum 3$\sigma$ upper limits for windows of 0.5\,\kms, 1.0\,\kms, and 1.5\,\kms\ for the whole disk, and for repeating the analysis for different disk quadrants. The radii range used for the stacking is discussed in Section~\ref{sec:gofish}.
 }
\end{deluxetable}

We can also constrain the upper limit to the \pfrac\ for the \texttt{GoFish} stacking methods presented in Section~\ref{sec:gofish}. We once again use Equation~\ref{eq:upper} to calculate the ``minimum" 3$\sigma$ upper limit for each stacked \texttt{GoFish} spectra for both the entire disk (Figures~\ref{fig:hd142527_gofish_all} and~\ref{fig:imlup_gofish_all} for \coto; see Appendix for \ttco\ and \ceo\ transitions) and when stacking different disk quadrants (Figure~\ref{fig:gofish_everything} for \coto; also see Appendix for \ttco\ and \ceo\ transitions). 
We only report values for velocity intervals of 0.5, 1, and 1.5\,\kms\ because higher velocity spans do not result in better sensitivities. We take $\sigma_{\text{PI}}$ to be the noise of the Stokes~$V$ stacked spectrum. The 3$\sigma$ upper limits are given as ranges in Table~\ref{tab:gofish}. Ranges are given since the 3$\sigma$ upper limits vary significantly depending on the velocity interval and (for the quadrant analysis) the quadrant analyzed. We report the minimum and maximum 3$\sigma$ upper limit found for every permutation of velocity interval and quadrant. Compared to the pixel by pixel analysis, the stacking method performs better, having lower 3$\sigma$ upper limits by a factor of up to $\sim$10. 

For each line and source, we executed one thousand Monte Carlo runs to test how the Table~\ref{tab:gofish} upper limits change with input parameters. In these runs, we randomly add Gaussian noise to the mass, inclination, and position angles because the Keplerian profile strongly depends on these parameters. For the mass, we assume that $\sigma$ of the Gaussian noise is 15\% of the estimated masses of the disks. For the inclination and position angle, we assume that $\sigma$ of the Gaussian noise is 5$^\circ$. These runs showed that each value in Table~\ref{tab:gofish} is typical, as both the medians and means of these values over the one thousand runs were almost identical to their values. For each value in Table~\ref{tab:gofish}, the typical standard deviation for the 1000 runs is $<$10\% of its value.

Altering the radii range used for stacking slightly affects the upper limit values in Table~\ref{tab:gofish}. For example, adding or subtracting an arcsecond to $r_{\text{max}}$ affects the upper limits by zero to three tenths of a percent.

%\renewcommand{\tabcolsep}{0.1cm}

%go fish
%maybe do for entire disk too

\section{Summary and Discussion}\label{sec:summary}
We use ALMA to observe polarization toward \hd\ and \im\ in the 1.3\,mm continuum and for the spectral lines \coto, \ttco, and \ceo\ at $\sim$0.5$\arcsec$ (80\,au) resolution. We find the following results:

\begin{enumerate}
\item While the 1.3\,mm continuum images for \hd\ and \im\ show many similar features to the 870\,$\mu$m images published in other studies, there are significant differences. The polarization toward the northern dust trap in \hd\ shows a rapid change in the morphology and \pfrac\ with wavelength. Whether this wavelength dependence is consistent with the scattering interpretation previously proposed for the Band 7 polarization remains to be determined. The polarization toward the southern part of \hd\ has typically been attributed to grain alignment, but \pfrac\ is surprisingly smaller at 1.3\,mm than it is at 870\,$\mu$m. The polarization morphology toward \im\ appears more azimuthal at 1.3\,mm than 870\,$\mu$m. \pfrac\ decreases between 1.3\,mm and 870\,$\mu$m, which is the opposite of HL~Tau. This could possibly be explained by optical depth effects or grains that are 100s of $\mu$m in size.
\item For both disks, \coto\ is very optically thick ($\tau$ typically $>$30), while \ttco\ is only moderately so ($\tau \sim 1$ to a few). \ceo\ is typically optically thin, but still has significant optical depth ($\tau$~$\sim$~0.1~--~1). Based on optical depth alone, we expect \ttco\ and \ceo\ to be optimal for probing the GK effect in these disks, while \coto\ is expected to be suboptimal. However, the lines are thermalized in the midplane, suggesting the detected line polarization would come from regions above the midplane.
\item Despite the above expectations for optical depth, polarization is only detected for \coto, and only for small parts of the disk. Such a detection requires spatial averaging and integrating over a velocity range.  {\bf The polarization signal for both disks is found for Stokes~$Q$ only, and our Monte Carlo simulations suggest these are marginal detections.} \pfrac\ for \hd\ and \im\ are 1.56\,$\pm$\,0.18\% and 1.01\,$\pm$\,0.10\%, respectively. The polarization is detected for only small parts of the disks over a limited velocity range, and thus is insufficient for estimating a polarization or magnetic field pattern. %{\bf Line polarization potentially comes from beyond the midplane due to the line being thermalized within the midplane.}
\item We constrain the 3$\sigma$ upper limit for \pfrac\ at the resolution of the observations ($\sim$0$\farcs$5 or $\sim$80\,au). Toward the brightest parts of \hd, the 3$\sigma$ upper limit is typically less than 3\% for \coto\ and \ttco\ and 4\% for \ceo. Toward the brightest parts of \im\, the 3$\sigma$ upper limits are typically less than 3\%, 4\%, and 12\%, respectively. Stacking based on Keplerian rotation has 3$\sigma$ upper limits that are up to a factor of $\sim$10 lower, but we note that the stacking method can potentially average out small-scale polarization structure.
\end{enumerate}

So far, \citet{Lankhaar2020} has the only model that predicts line polarization for a disk. This paper primarily introduces PORTAL, which is a code for three-dimensional polarized line radiative transfer modeling. Their prescription for a disk is somewhat limited and it is only for \mbox{CO(3--2)}, and they defer detailed analysis to a future paper. They showed predicted polarization morphologies for both a face-on disk and a 45$^\circ$ inclined disk, making them fairly analogous with \hd\ and \im, respectively. For the face-on disk, for both toroidal and radial fields, the polarization toward the center is expected to be near 0\%, while toward the edges, it reaches values of $\sim$0.5\%. For the 45$^\circ$ inclined disk, both toroidal and radial fields, polarization is near 0\% for most of the disks. For the toroidal field, there are few locations along the major and minor axes where polarization can reach up to 0.5\%. For radial fields, the polarization can even reach up to 9\%, but the polarization is only significant for the central $\sim$40\,au (i.e., about half our beam size) and only along the minor axis of the disk. Beaming smearing would reduce these levels considerably. \citet{Lankhaar2020} already predicts low polarization levels for CO(3--2), but they do not show detailed predictions for the $J = 2 \rightarrow 1$ CO isotopologues. Future modeling of line polarization tailored for disks may show why these two disks have low levels of polarization for emission for these CO isotopologue transitions.

These observations are the first published attempt to resolve linear line polarization for circumstellar disks. Polarization is undoubtedly low. Other transitions of CO isotopologues should be observationally explored. The polarization fraction largely depends on the optical depth and radiative rates, so observations of both higher and lower frequency CO isotopologue transitions should be attempted to search for a polarized morphology. Nevertheless, because the lines are thermalized in the midplane, any detection of CO polarization either comes from beyond the midplane or is from some other (possibly unknown) effect. Other molecules could also be observationally attempted, though most molecules are much less abundant, and it has been predicted that \pfrac\ decreases with the mass for a linear molecule \citep{Deguchi1984}.

%Models of non-disks, with and without an external source of radiation, suggest that polarization for the $J = 1 \rightarrow 0$ transition is about a factor of 2 stronger than for the $J = 1 \rightarrow 0$ transition. As such, a search for polarization for the $J = 1 \rightarrow 0$ transitions, especially for the optically thinner \mbox{$^{13}$CO(1--0)} and \mbox{C$^{18}$O(1--0)} lines, could potentially result in a stronger polarization signal. 

%I looked at the Cortes paper just now.  The math is too complex for me to really follow, but I think the main difference between their calculation and one for a disk is that the optical depth will now depend on the inclination angle of the disk as well as the density, temperature, and other parameters. This should not be a show-stopper, if you are using this paper to motivate an observing proposal.

\acknowledgements
We acknowledge an anonymous referee for giving useful comments on the paper.
We thank Philip Myers, Patricio Sanhueza, and Haifeng Yang for a useful discussions, and Haifeng Yang for code that helped to generate some figures.
ZYL acknowledges support from NASA 80NSSC18K1095 and 80NSSC20K0533 and NSF AST-1716259 and AST-1815784.
LWL acknowledges support from NSF AST-1910364.
The National Radio Astronomy Observatory is a facility of the National Science Foundation operated under cooperative agreement by Associated Universities, Inc.
This paper makes use of the following ALMA data: ADS/JAO.ALMA\#2015.1.00425.S, \#2016.1.00712.S, and \#2018.1.01172.S,  ALMA is a partnership of ESO (representing its member states), NSF (USA) and NINS (Japan), together with NRC (Canada), MOST and ASIAA (Taiwan), and KASI (Republic of Korea), in cooperation with the Republic of Chile. The Joint ALMA Observatory is operated by ESO, AUI/NRAO and NAOJ. The National Radio Astronomy Observatory is a facility of the National Science Foundation operated under cooperative agreement by Associated Universities, Inc.

\facility{NSF's Atacama Large Millimiter/submillimeter Array (ALMA)}
\software{APLpy \citep{Robitaille2012}, 
CASA v5.6.0, \citep{McMullin2007},
GoFish \citep{GoFish},
PySpecKit \citep{Ginsburg2011}
}

\clearpage
%\bibliography{/Users/istephens/Documents/latex_stuff/stephens_bib}
\bibliography{stephens_bib}

%\begin{deluxetable}{cccccc} 
%\label{Teli}
%\tablewidth{0pt}
%\tablecolumns{6}
%\tabletypesize{\scriptsize}
%\tablecaption{Results of ellipse fitting in the \co velocity cube}
%\phs
%\tablehead{
%\colhead{V$_{rad}$} & \colhead{RA offset} & \colhead{Dec offset} & \colhead{B$_{maj}$} & \colhead{B$_{min}$} & \colhead{PA} \\
%\colhead{(\kms)}  & \colhead{($\arcsec$)} & \colhead{($\arcsec$)} & \colhead{($\arcsec$)}  & \colhead{($\arcsec$)} & %\colhead{($\degr$)} 
%}
%\startdata
%-10.8	& -8.54	& -2.52	& 6.776	& 5.961	& 70.1 \\
%\enddata 
%\tablecomments{}
%\tablenotetext{(a)}{Deconvolved semimajor and semiminor axis of the ring-shaped disk.}
%\tablenotetext{(b)}{Estimated from the major and minor axis lengths as the $\arccos(R_{min}/R_{maj})$.}
%\tablenotetext{(c)}{The disk mass is estimated assuming a typical gas-to-dust ratio of 100.}
%\tablenotetext{(d)}{This disk is unresolved with the present ALMA resolution.}
%\end{deluxetable}

\begin{figure}[ht!]
\begin{center}
\includegraphics[width=1\columnwidth]{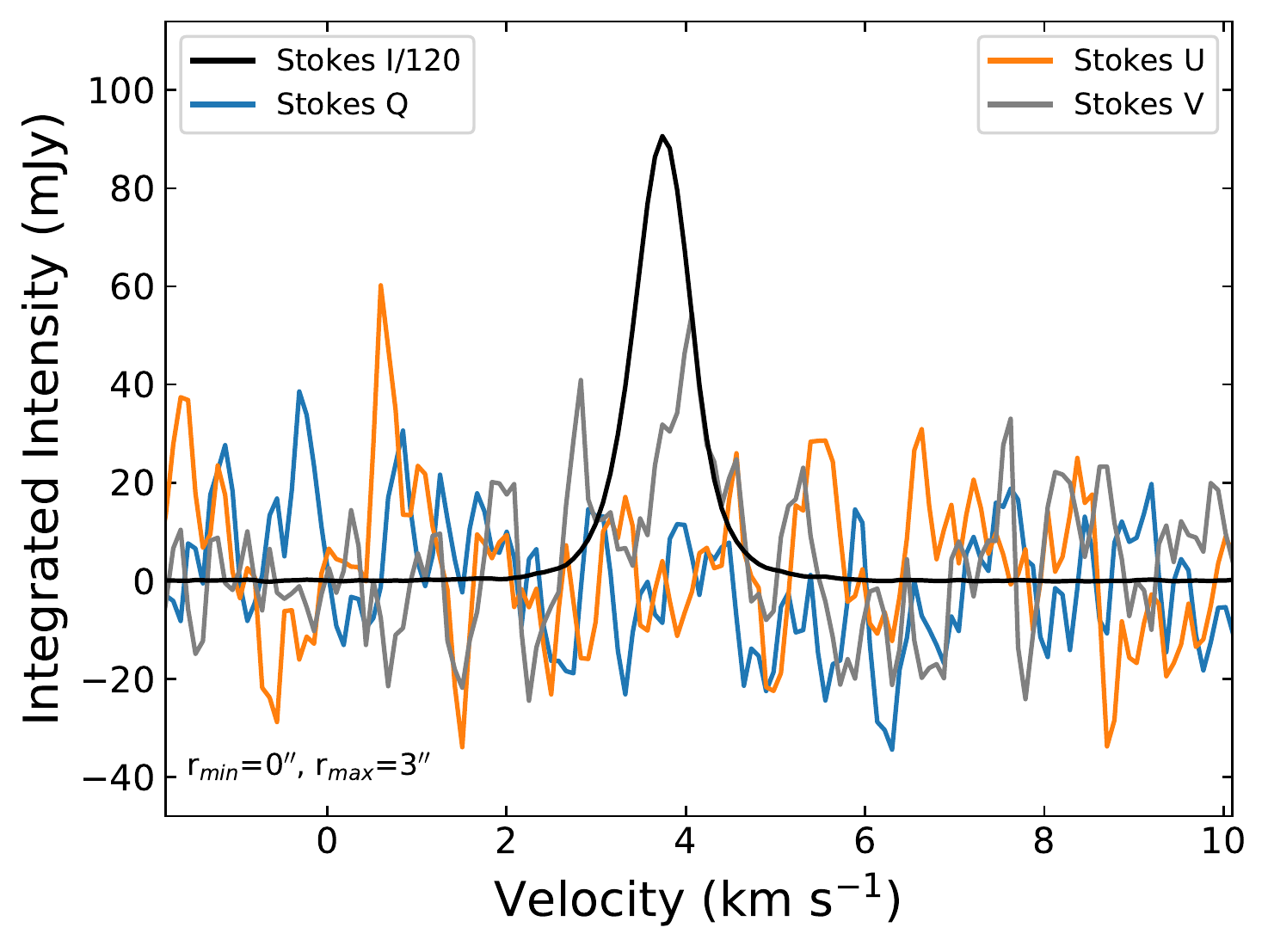}
\end{center}
\caption{\hd\ $IQUV$ \ttco\ spectra, stacking based on the Keplerian profile of the disk via \texttt{GoFish} across the entire disk out to a radius of 3$\arcsec$. The stacked Stokes~$I$ signal has been divided by 120.
}
\label{fig:hd142527_gofish_all_ttco} 
\end{figure}

\begin{figure}[ht!]
\begin{center}
\includegraphics[width=1\columnwidth]{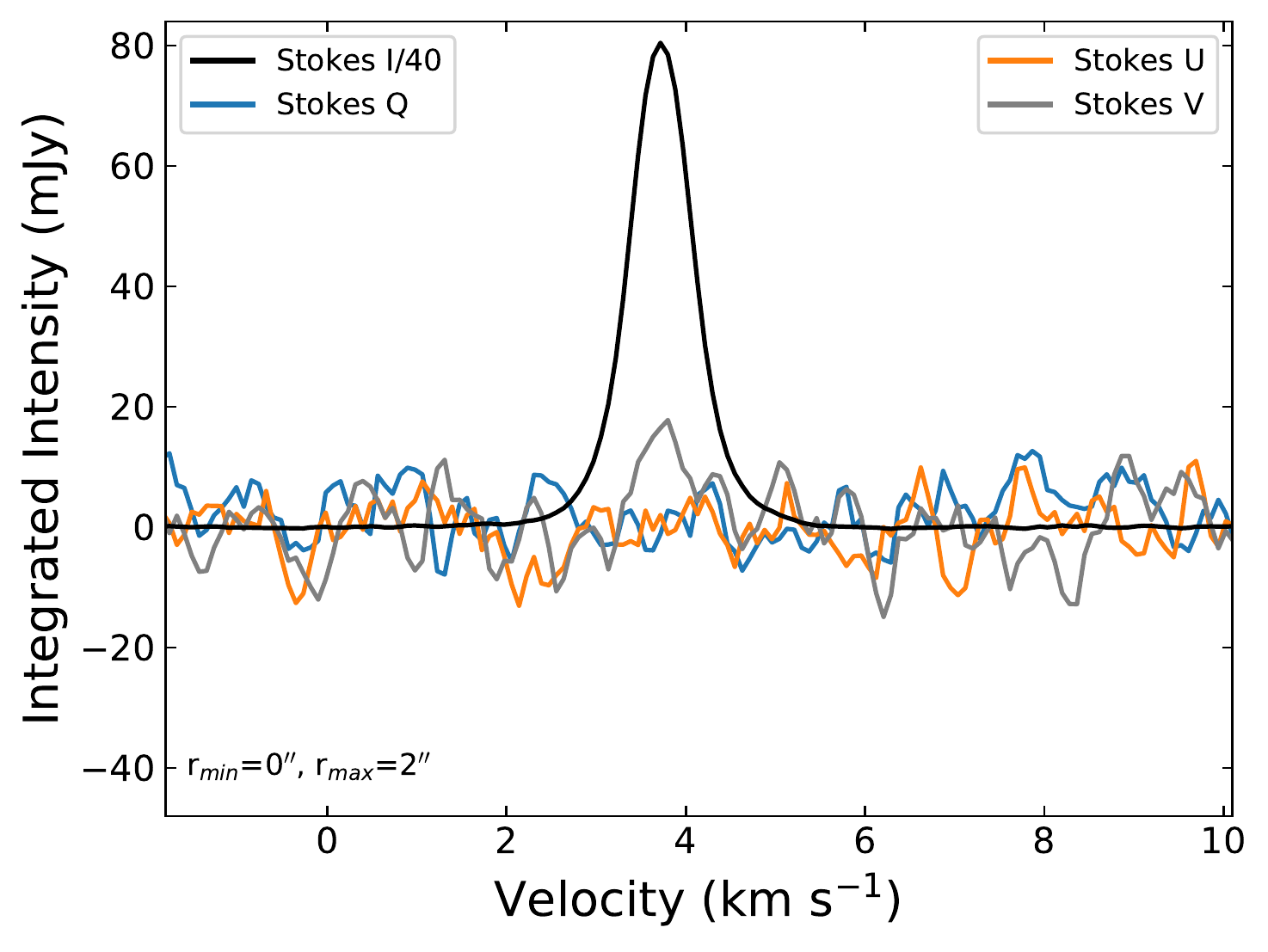}
\end{center}
\caption{\hd\ $IQUV$ \ceo\ spectra, stacking based on the Keplerian profile of the disk via \texttt{GoFish} across the entire disk out to a radius of 2$\arcsec$. The stacked Stokes~$I$ signal has been divided by 40.
}
\label{fig:hd142527_gofish_all_ceo} 
\end{figure}

\begin{figure}[ht!]
\begin{center}
\includegraphics[width=1\columnwidth]{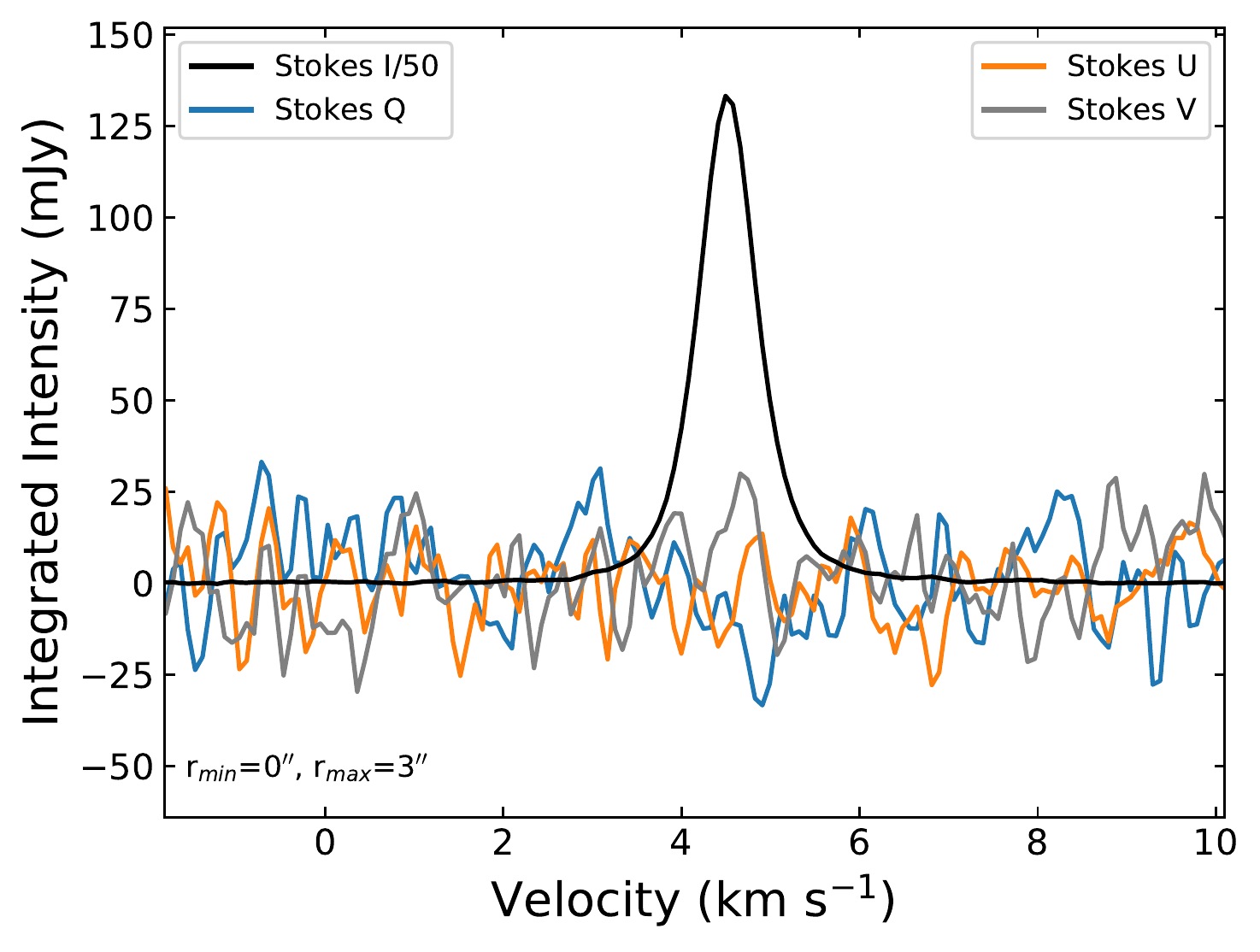}
\end{center}
\caption{\im\ $IQUV$ \ttco\ spectra, stacking based on the Keplerian profile of the disk via \texttt{GoFish} across the entire disk out to a radius of 3$\arcsec$. The stacked Stokes~$I$ signal has been divided by 50.
}
\label{fig:imlup_gofish_all_ttco} 
\end{figure}

\begin{figure}[ht!]
\begin{center}
\includegraphics[width=1\columnwidth]{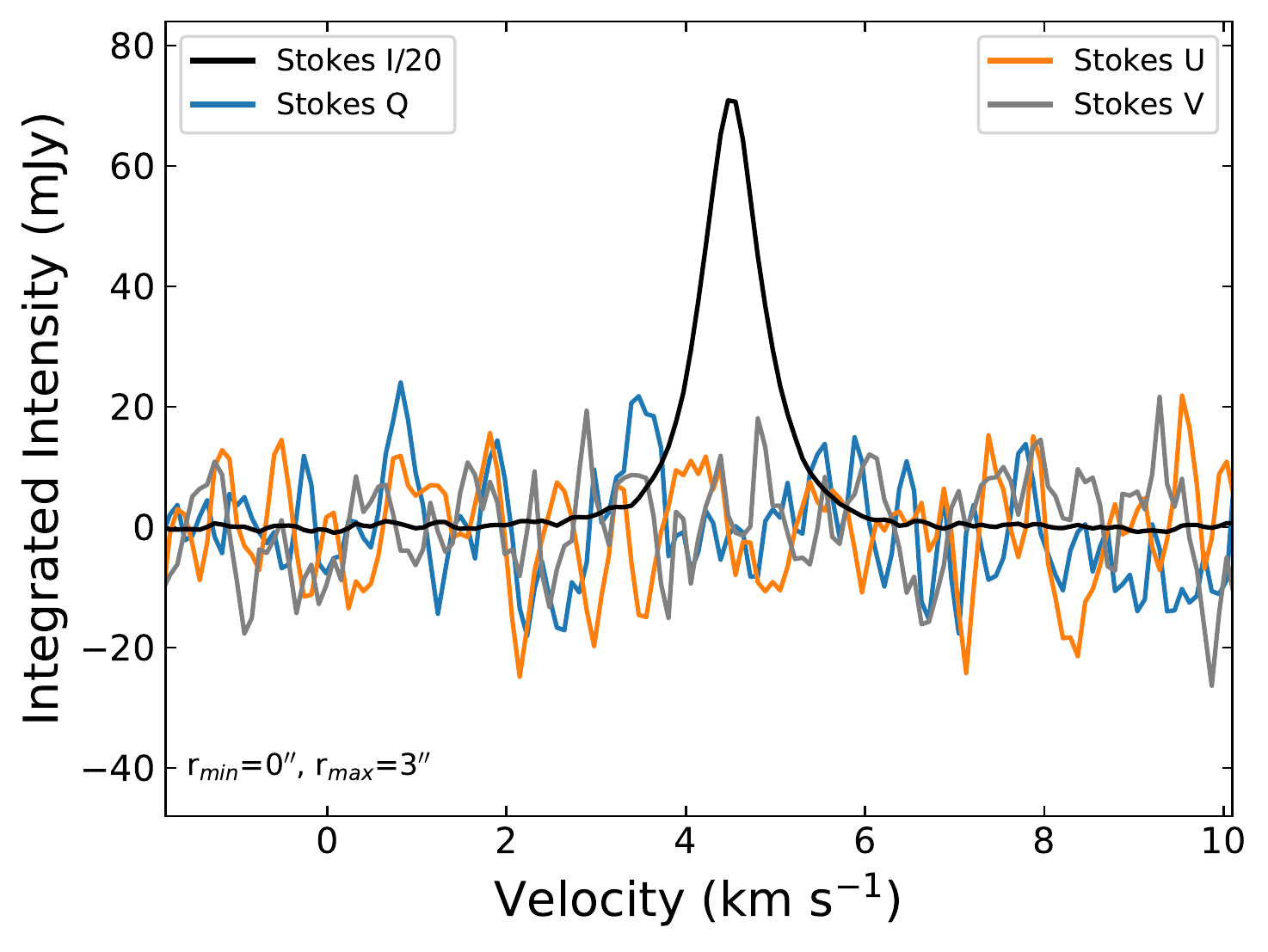}
\end{center}
\caption{\im\ $IQUV$ \ceo\ spectra, stacking based on the Keplerian profile of the disk via \texttt{GoFish} across the entire disk out to a radius of 3$\arcsec$. The stacked Stokes~$I$ signal has been divided by 20.
}
\label{fig:imlup_gofish_all_ceo} 
\end{figure}

\begin{figure*}[ht!]
\begin{center}
\includegraphics[width=0.49\columnwidth]{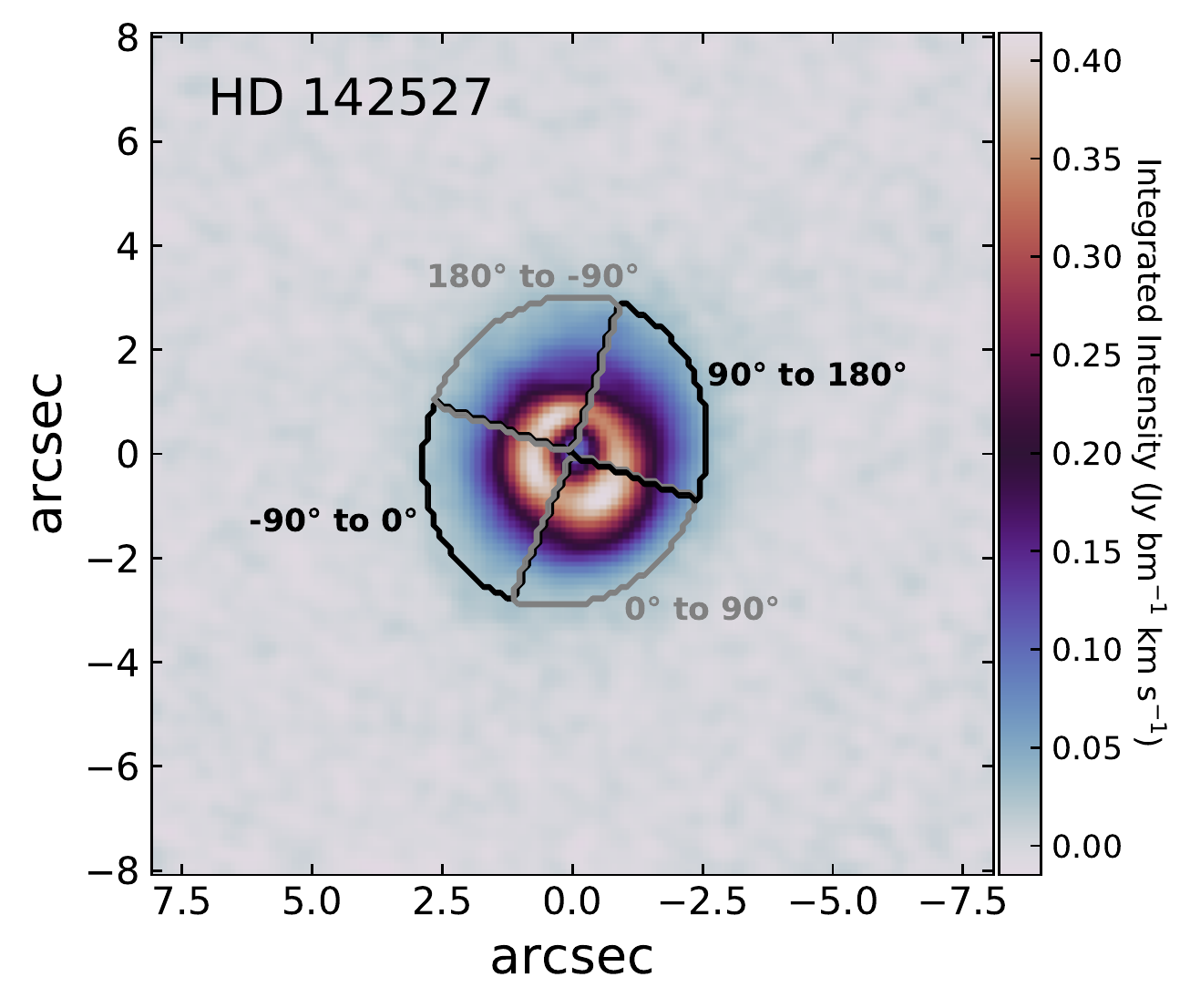}
\includegraphics[width=0.49\columnwidth]{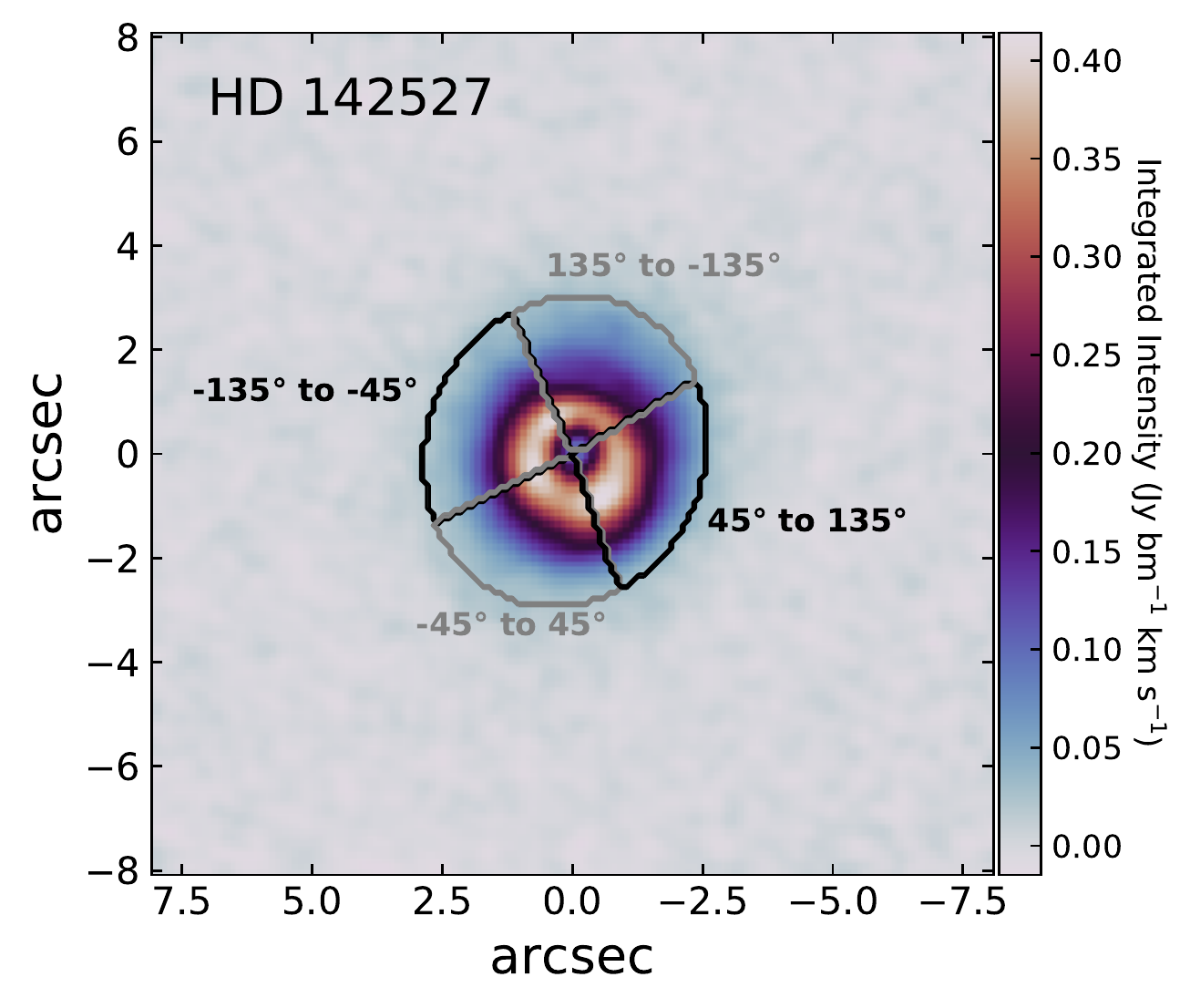}
\includegraphics[width=0.49\columnwidth]{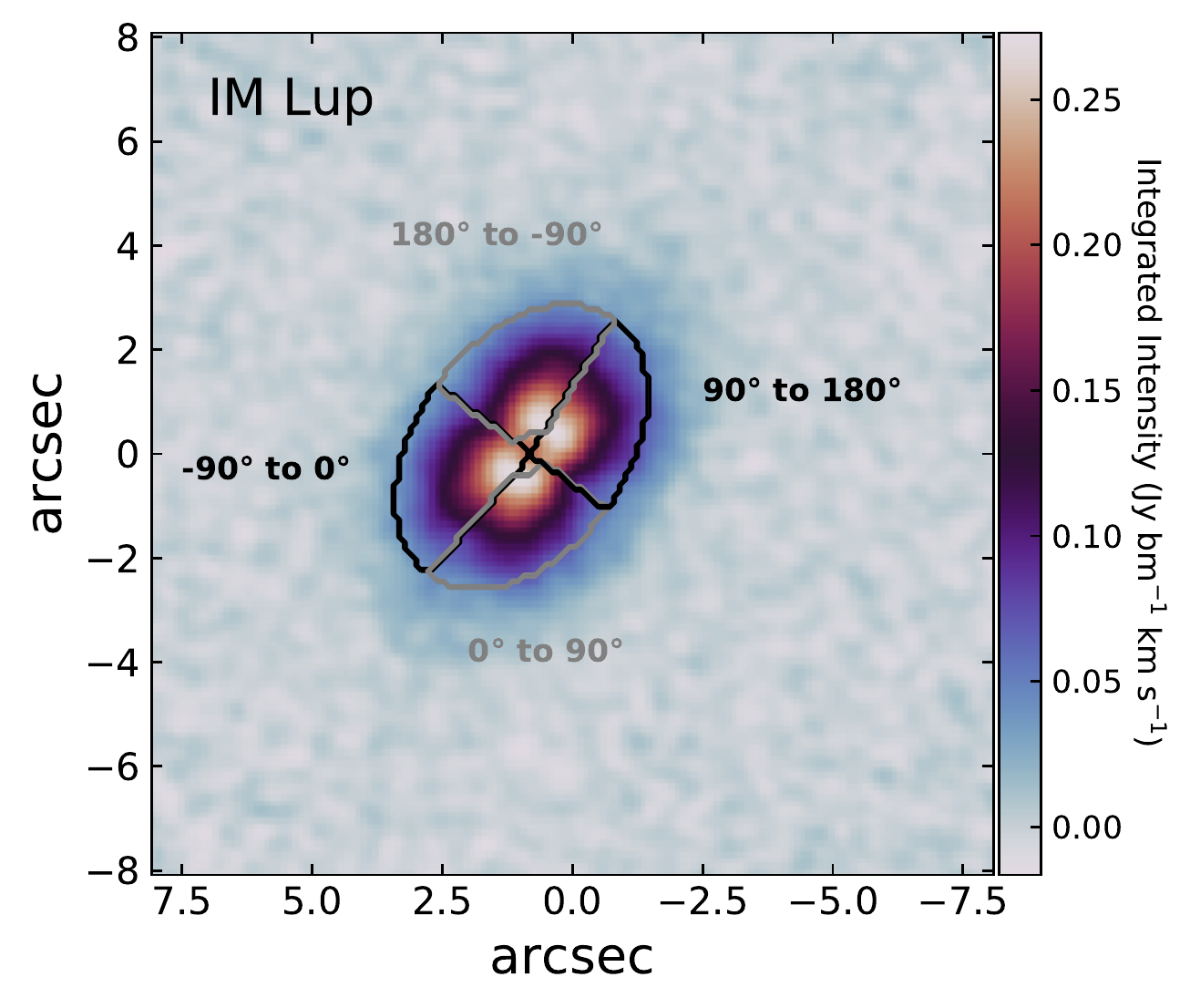}
\includegraphics[width=0.49\columnwidth]{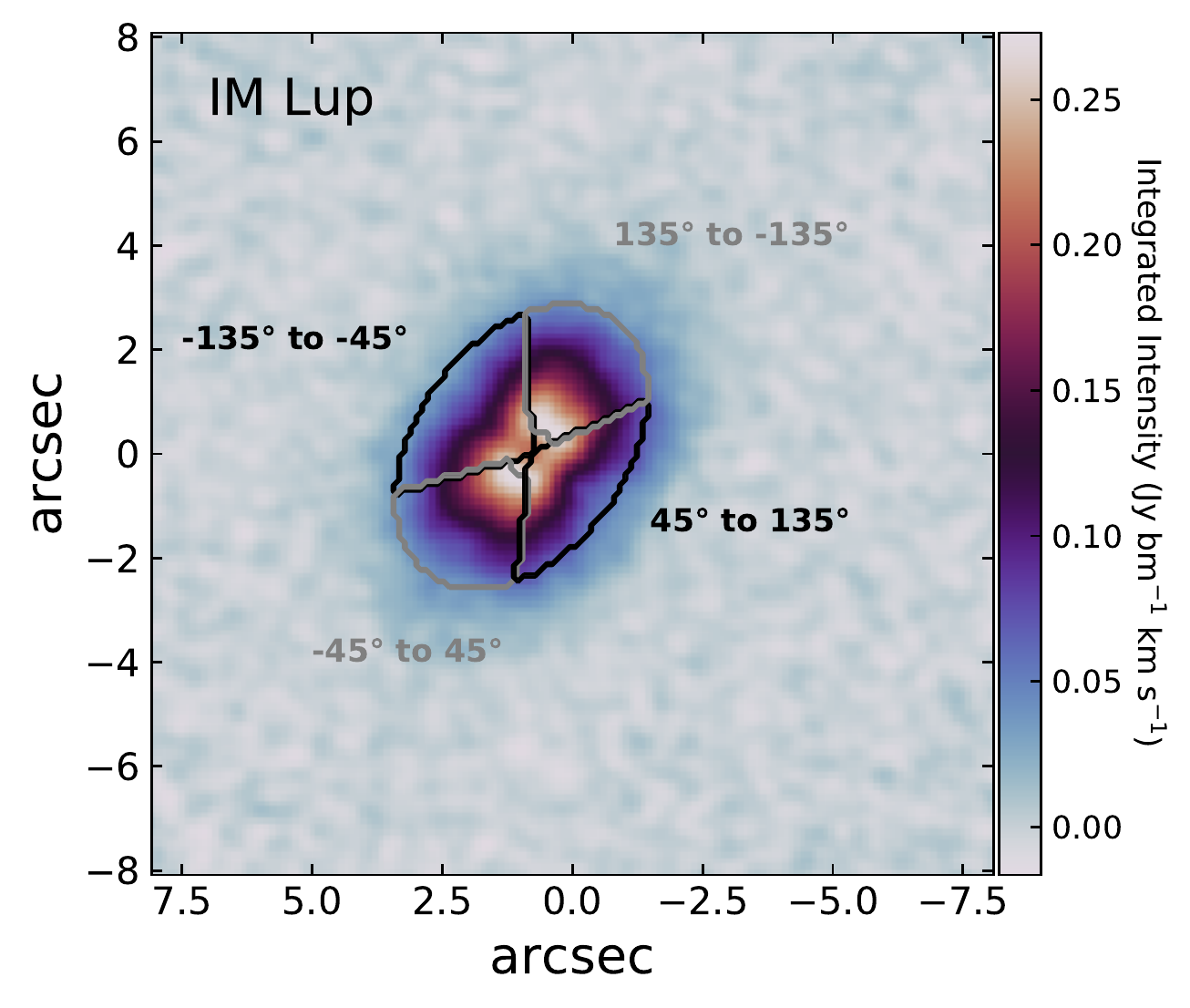}
\includegraphics[width=2\columnwidth]{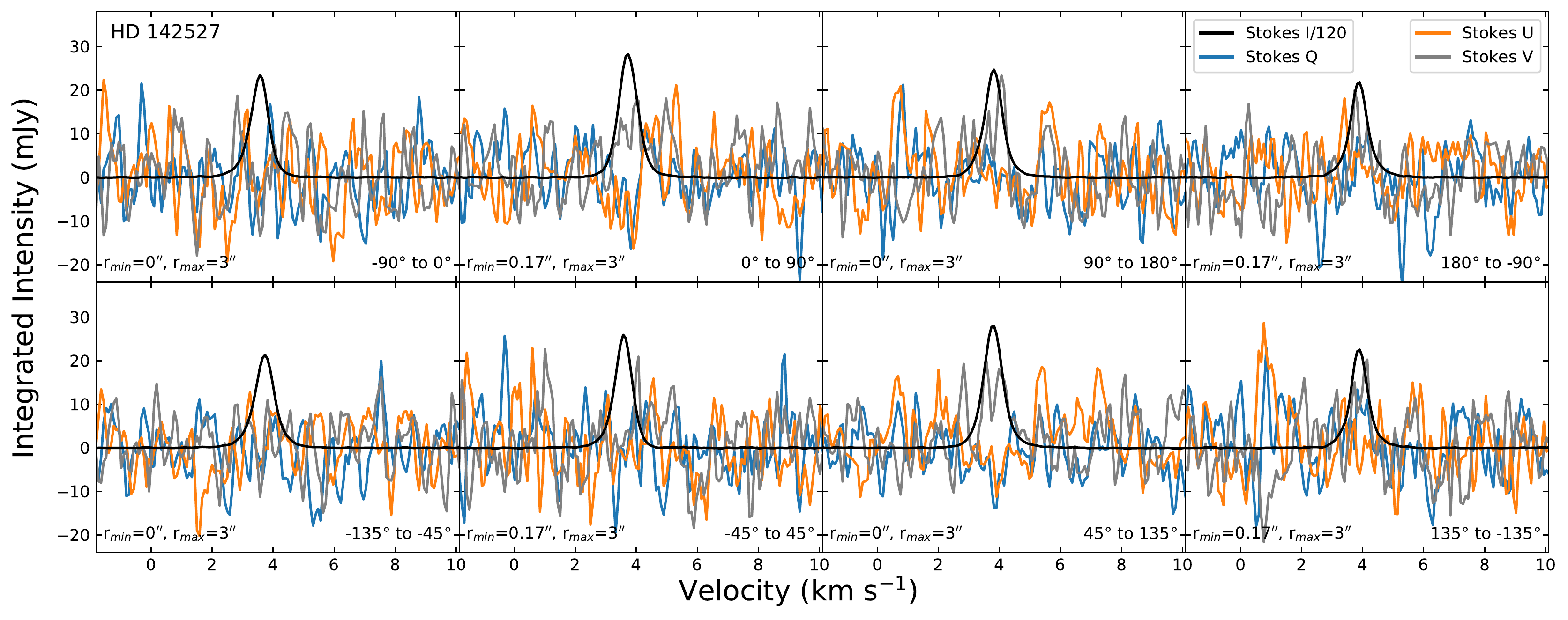}
\includegraphics[width=2\columnwidth]{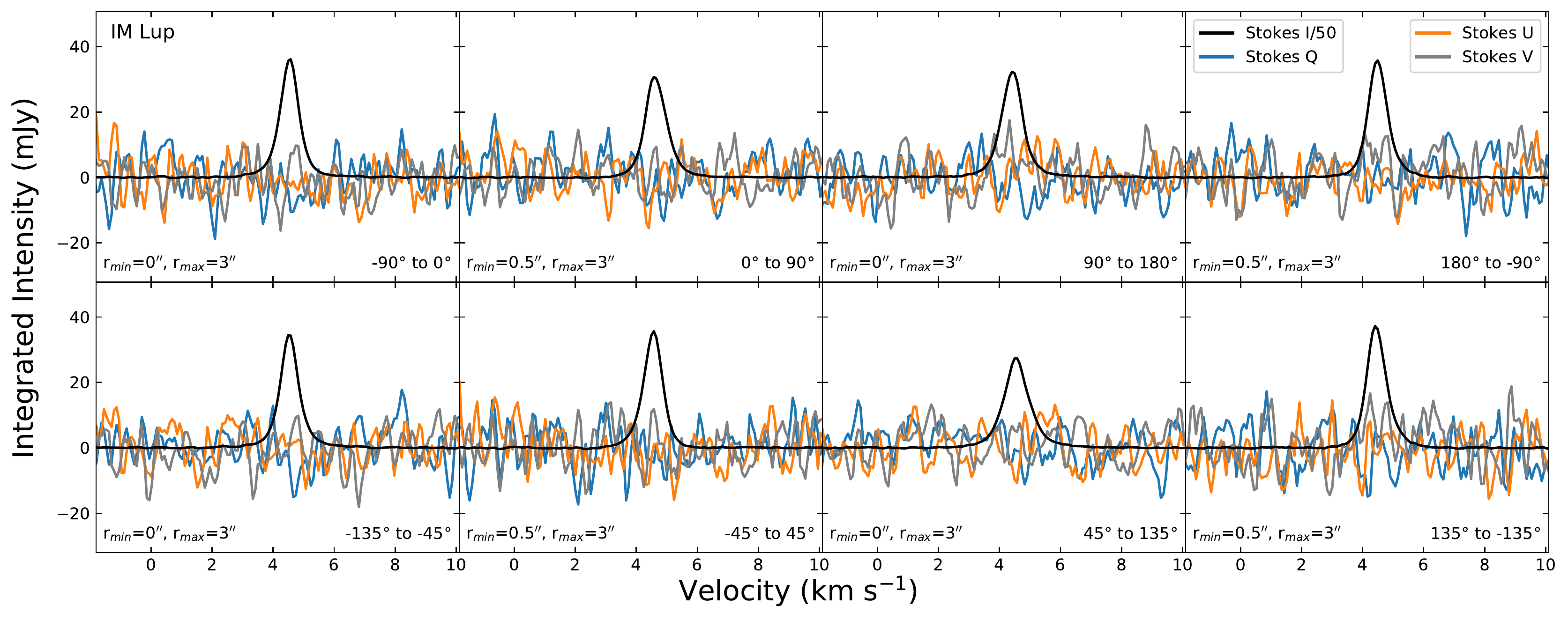}
\end{center}
\caption{Same as Figure~\ref{fig:gofish_everything}, except now for the spectral line \ttco. Stokes~$I$ for \hd\ and \im\ have been divided by 120 and 50, respectively.
}
\label{fig:gofish_everything_ttco} 
\end{figure*}

\begin{figure*}[ht!]
\begin{center}
\includegraphics[width=0.49\columnwidth]{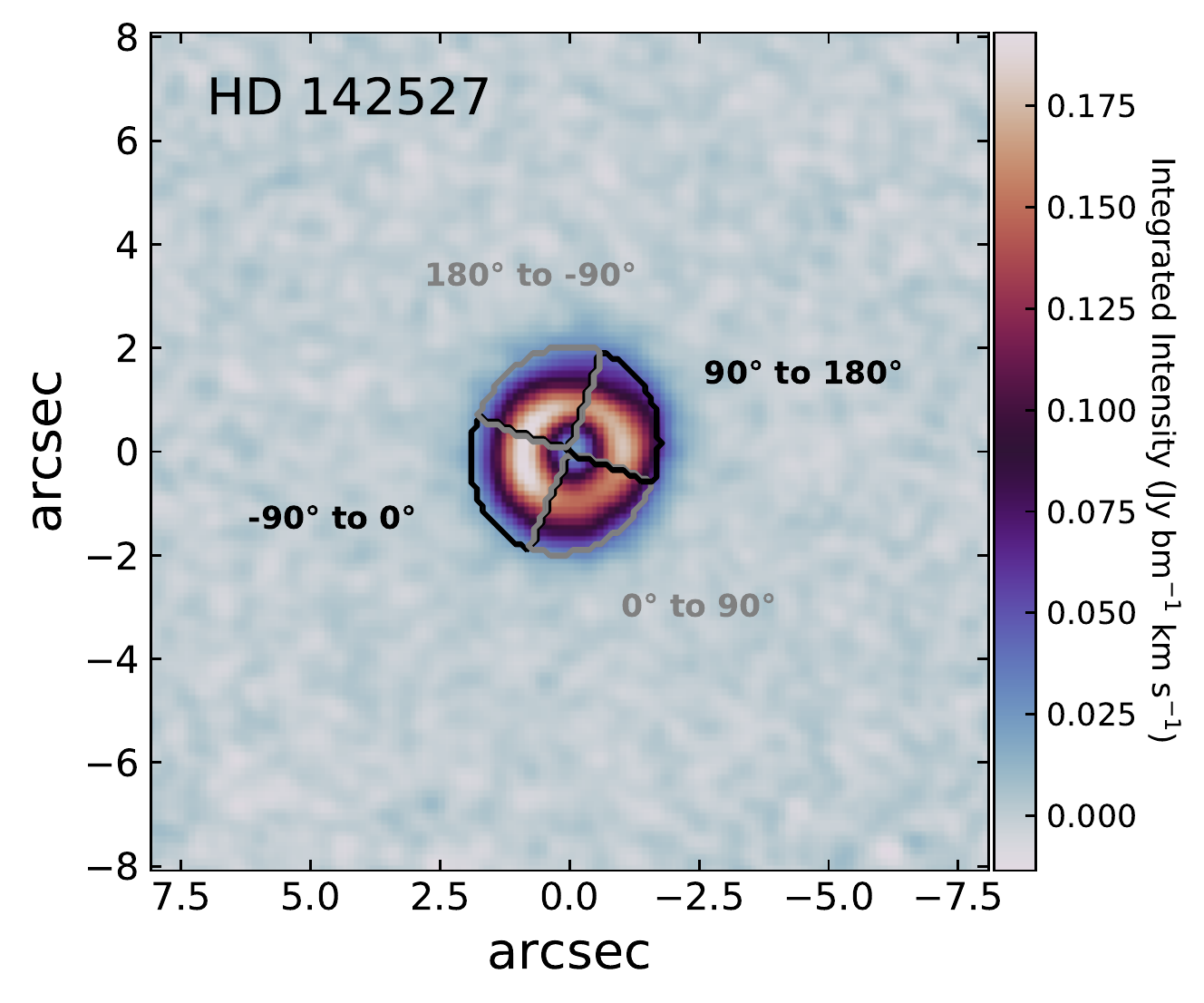}
\includegraphics[width=0.49\columnwidth]{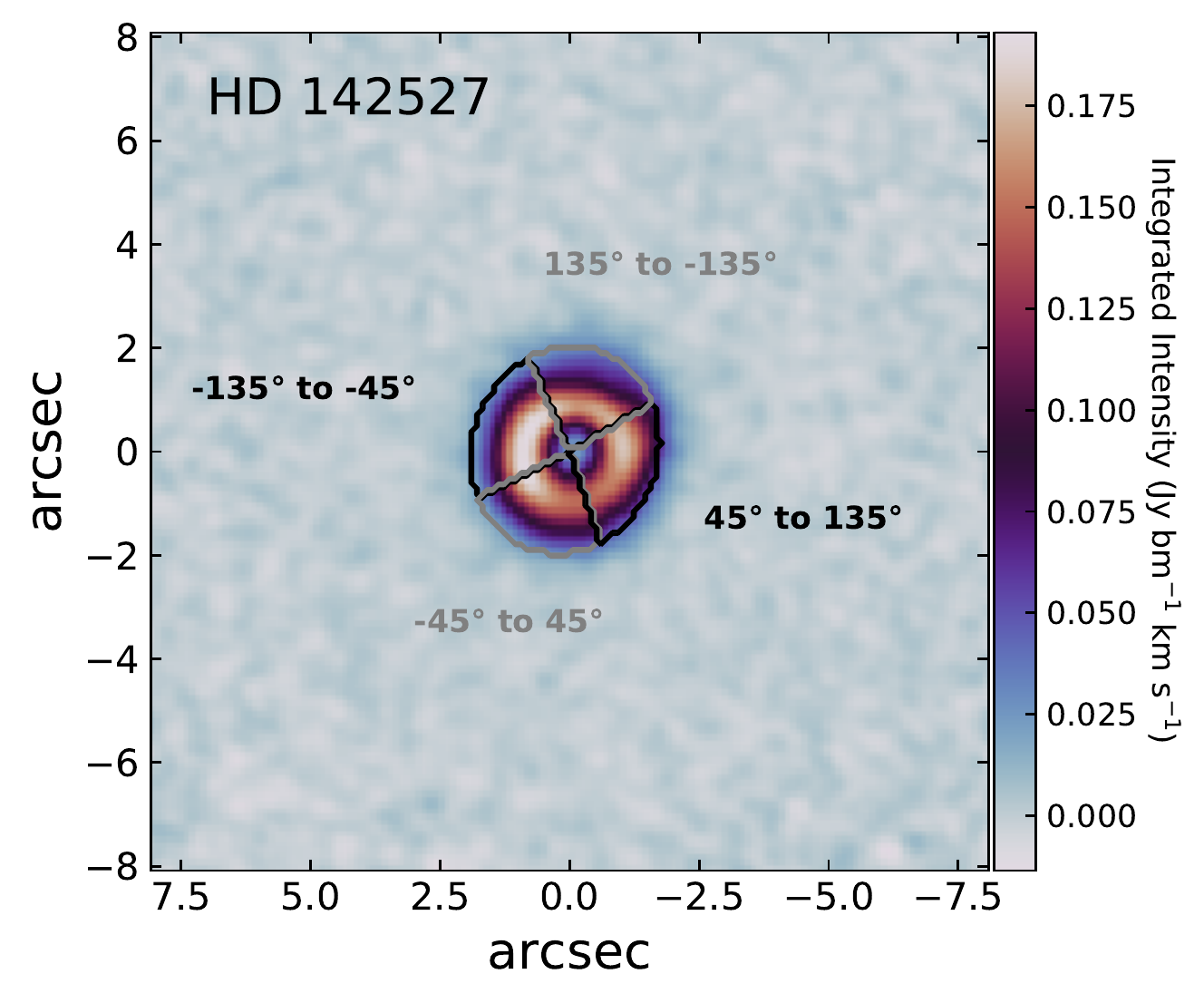}
\includegraphics[width=0.49\columnwidth]{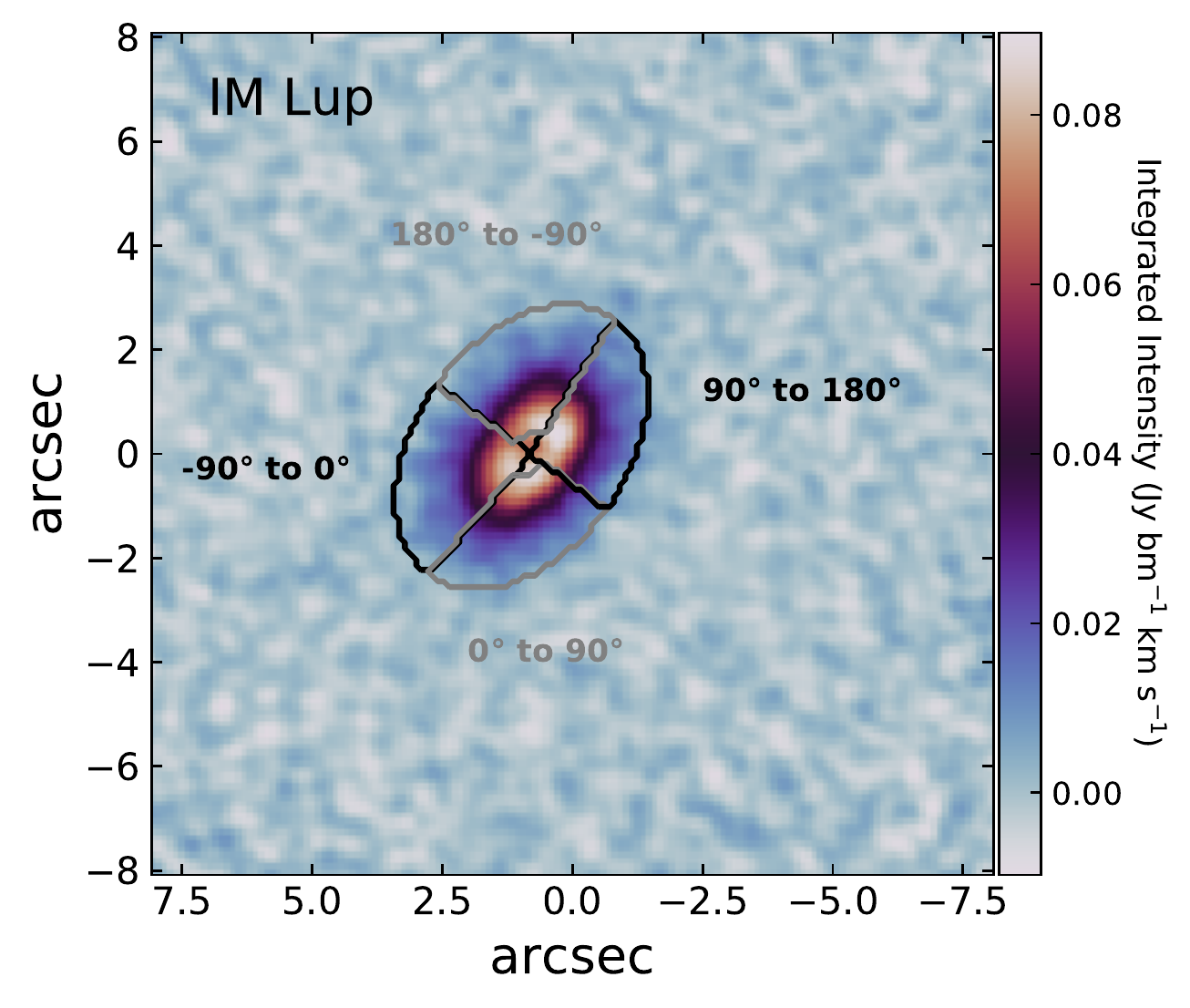}
\includegraphics[width=0.49\columnwidth]{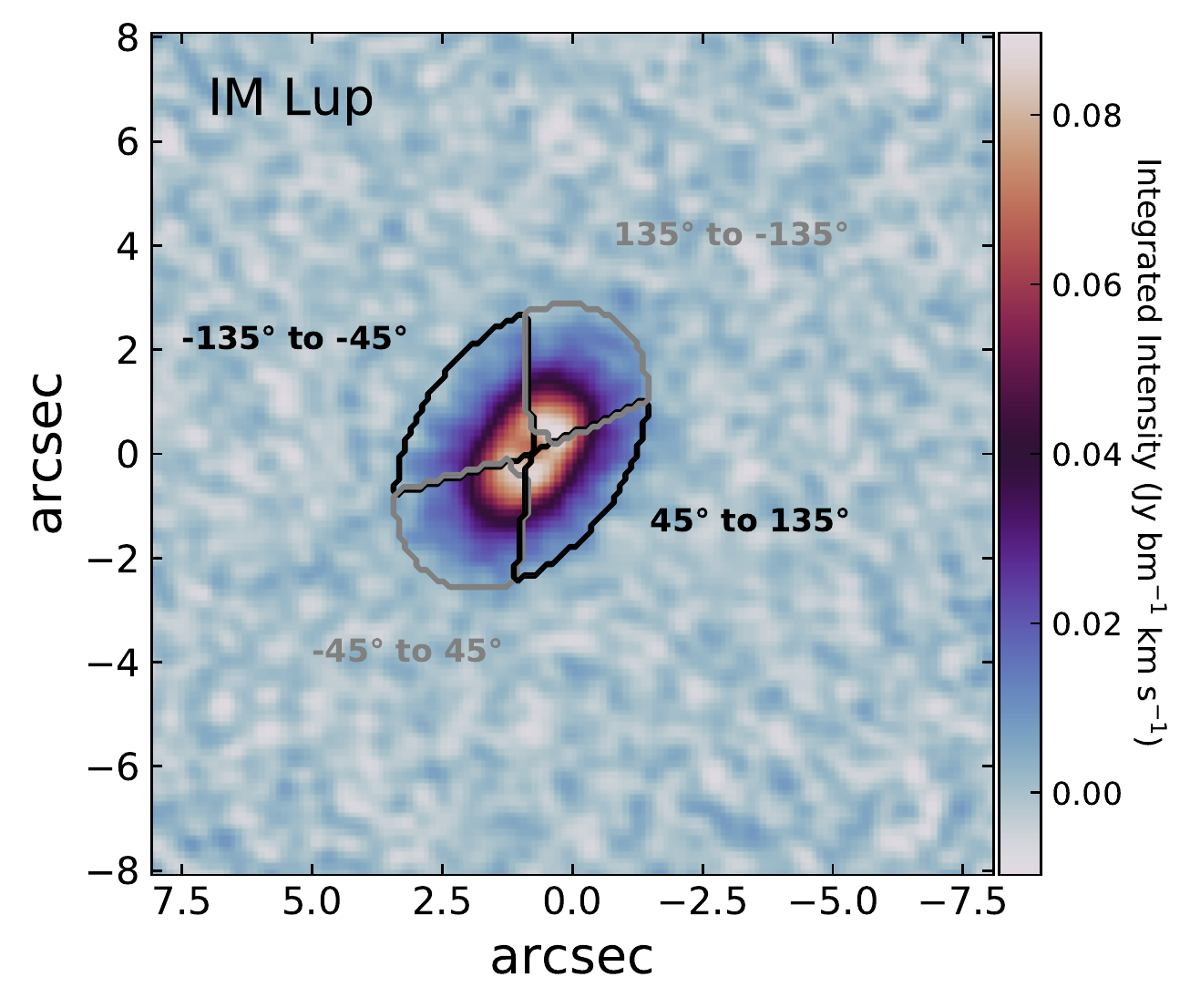}
\includegraphics[width=2\columnwidth]{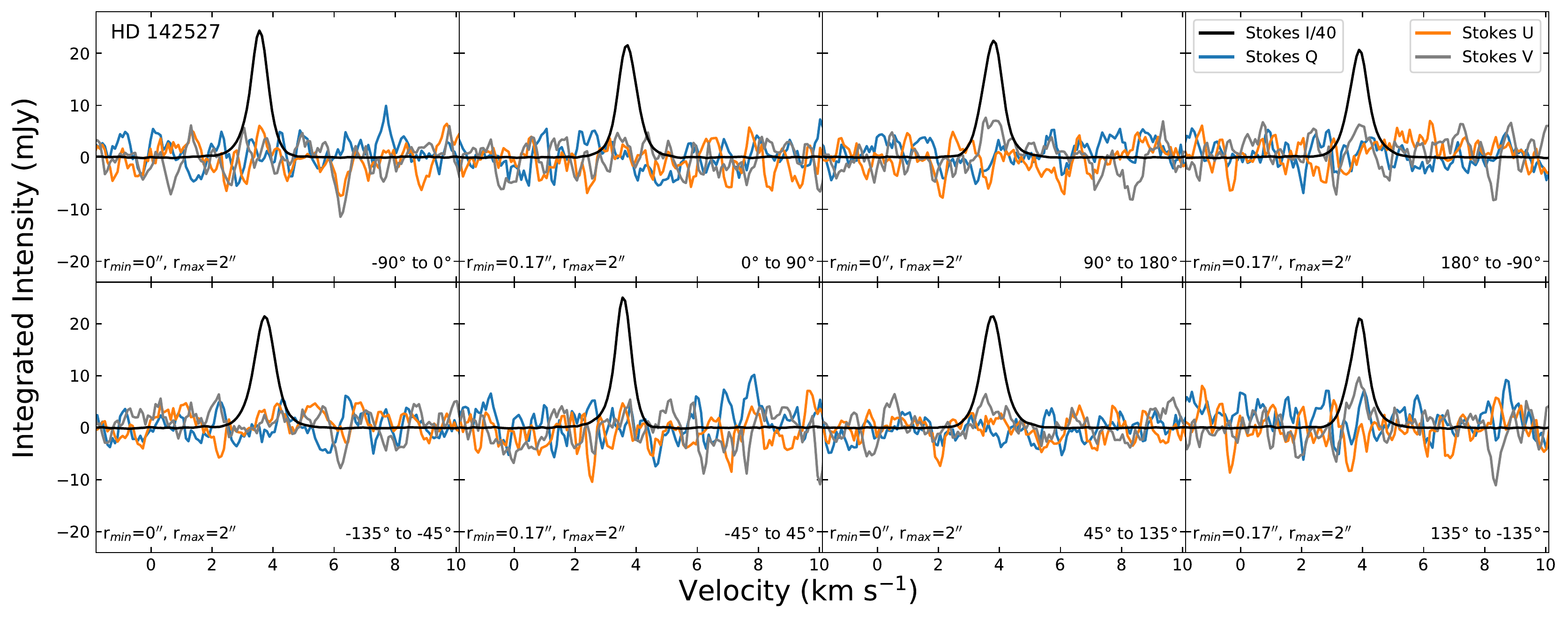}
\includegraphics[width=2\columnwidth]{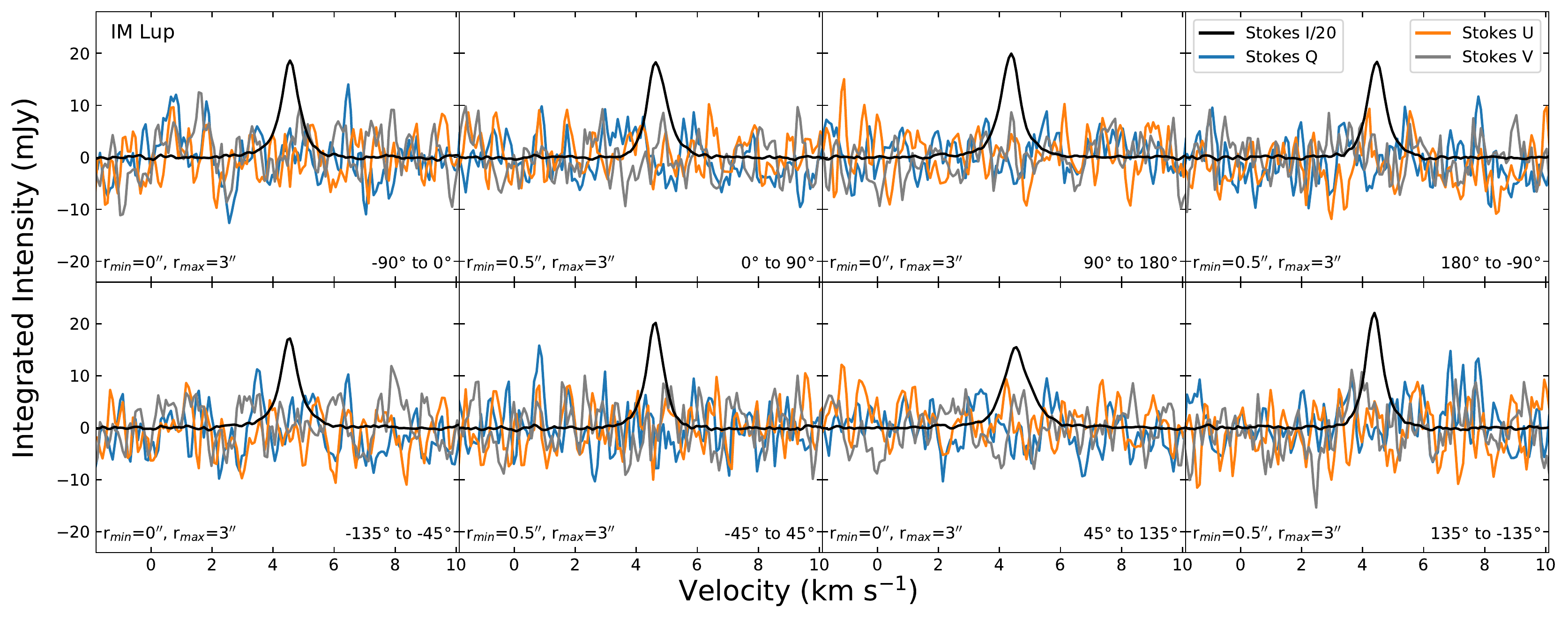}
\end{center}
\caption{Same as Figure~\ref{fig:gofish_everything}, except now for the spectral line \ceo. Stokes~$I$ for \hd\ and \im\ have been divided by 40 and 20, respectively.
}
\label{fig:gofish_everything_ceo} 
\end{figure*}

\appendix
In this appendix, we show \ttco\ and \ceo\ images made using the \texttt{GoFish} stacking techniques used in Section~\ref{sec:gofish}. Constraints on the \pfrac\ based on these stacking techniques are given in Section~\ref{sec:constrain}. Figures~\ref{fig:hd142527_gofish_all_ttco} and~\ref{fig:hd142527_gofish_all_ceo} show the \texttt{GoFish} stacking technique across the entire disk for \hd\ for \ttco\ and \ceo, respectively. Figures~\ref{fig:imlup_gofish_all_ttco} and~\ref{fig:imlup_gofish_all_ceo} show the \texttt{GoFish} stacking technique across the entire disk for \im\ for \ttco\ and \ceo, respectively. Figures~\ref{fig:gofish_everything_ttco} and~\ref{fig:gofish_everything_ceo} show the \texttt{GoFish} stacking based on disk quadrants for \ttco\ and \ceo, respectively, for both disks. %python gofish_stacking_PA_panels.py imlup 13co21
%python gofish_stacking_PA_panels.py hd142527 13co21

%python gofish_stacking_PA_panels.py imlup c18o21
%python gofish_stacking_PA_panels.py hd142527 c18o21

%python gofish_stacking_PA_panels.py imlup 13co21
%python gofish_stacking_PA_panels.py hd142527 13co21

%python gofish_stacking_PA_panels.py imlup c18o21
%python gofish_stacking_PA_panels.py hd142527 c18o21

\end{document}